\documentclass[12pt,a4paper]{article}

\PassOptionsToPackage{%
	bookmarks=false,
	bookmarksopen=false,
	colorlinks=true,
	linkcolor=red,
	citecolor=blue,
	urlcolor=black%
}{hyperref}

\PassOptionsToPackage{titles}{tocloft}

\usepackage{2025macros}

\usepackage[%
	backend=biber,
	style=numeric,
	sorting=nyt,
	giveninits=true,
	backref=true,
	backrefstyle=none,
	doi=false%
]{biblatex}
\usepackage{biblatex-shortfields}
\DeclareDelimFormat[bib,biblist]{finalnamedelim}{%
  \ifnumgreater{\value{liststop}}{2}{\finalandcomma}{}%
  \addspace\&\space
}

\ProjectPDFMetadata{
	title={Coherent manifolds and quantization},
	author={Arnold Neumaier, Phillip Josef Bachler, Arash Ghaani 
	Farashahi},
	subject={Revised draft for Coherent manifolds, arXiv:2503.09827 
	version 2},
	keywords={transformation groups, admissible half-curves, 
	infinite-dimensional Lie algebras, coherent quantization, 
	invariant domains, convenient calculus},
	producer={pdfTeX / LaTeX project build}
}

\ProjectCopyright{
	Copyright \textcopyright\ 2026 by 
	Arnold Neumaier, 
	Phillip Josef Bachler and 
	Arash Ghaani Farashahi. All rights reserved.
}

\title{%
	Coherent manifolds and  quantization
}
\subtitle{}

\addauthor{Arnold Neumaier}
	{Faculty of Mathematics, University of 
	Vienna\\Oskar-Morgenstern-Platz 1, 1090 Vienna, Austria}
	{}{Arnold.Neumaier@univie.ac.at}

\addauthor{Phillip Josef Bachler}
	{Faculty of Mathematics, University of 
	Vienna\\Oskar-Morgenstern-Platz 1, 1090 Vienna, Austria}
	{}{phillip.josef.bachler@univie.ac.at}

\addauthor{Arash Ghaani Farashahi}
	{Independent Researcher, Philadelphia, PA, USA}
	{}{ghaanifarashahi@outlook.com}

\date{November 14, 2026}

\begin{document}

\begin{Pages}
\end{Pages}

\maketitle

\begin{abstract}
We study the differential geometry of coherent manifolds modelled on convenient vector spaces and apply it to coherent quantization. 
A smooth coherent product determines geometric structures on the label manifold. 
We construct a common invariant domain for the quantizations of coherent vector fields and of coherent diffeomorphisms with smooth diffeomorphic adjoints. 
These quantizations are distinguished operators on the associated quantum space and satisfy commutator, covariance, and formal adjoint relations. 
Their coherent matrix elements can be calculated by differentiation and substitution in the coherent product. 
Extending the operators to the common invariant domain leaves their Hilbert space closures unchanged. 
Motions in subgroups of diffeomorphisms give rise to Lie algebras of generators without requiring a Lie group structure on the subgroup. Suitable coherent motions recover the infinitesimal quantization by differentiation. 
We express quantum dynamics through coherent matrix elements and obtain a coherent action principle by restricting the quantum action to coherent states. 
For Klauder spaces, the construction gives creation and annihilation operators, the canonical commutation relations, Segal fields, Weyl operators, and normal ordering on the common invariant domain.
\end{abstract}

\begin{Notes}
\end{Notes}

\vspace{1cm}


\vspace{0.5cm}


\vspace{1cm}

\newpage
\begingroup
  \hypersetup{linkcolor=black}
  \tableofcontents
\endgroup



\section{Introduction}
\label{sec:introduction}

\subsection{Scope}

Coherent spaces are a geometric way to work with concrete Hilbert 
spaces, defined as a quantum space of a coherent space. 
They combine the rich, often highly characteristic variety of 
symmetries of traditional geometric structures with the 
computational tractability of traditional tools from numerical 
analysis and statistics. 
The notion of a coherent space unifies a lot of tools from quantum mechanics into a general framework. 

\pgph


The concept of a coherent space is abstracted from 
common features of families of coherent states for Lie groups 
acting on homogeneous spaces, geometric quantization, and coherent 
states in quantum mechanics, 
see \sca{Neumaier} \cite{Neumaier2019}. 
In \sca{Neumaier  \& Ghaani Farashahi} \cite{NGF2022}, nonunitary 
transformations were considered and  many other constructions 
familiar from quantum mechanics were placed in a common framework; 
see also \sca{Zhang} et al. \cite{ZhaFG}. 

In applications to quantum physics, the infinitesimal generators of certain coherent maps serve as distinguished observables.
For example, position and momentum in nonrelativistic \QM appear as 
generators of Heisenberg groups, and the principal observables in 
relativistic single-particle \QM (energy, momentum, and angular 
momentum) appear as generators of the Poincar\'e group, the 
symmetry group of space-time.

For exactly solvable models in \QM or \QFT one can frequently find a coherent space in which  the coherent product is explicitly known and the Hamiltonian belongs to this distinguished set. 
Indeed, such an explicit coherent representation is in some 
sense equivalent to having solved the model. 

\pgph


This paper develops the differential side of coherent quantization. 
We consider coherent manifolds -- coherent spaces whose labels form 
a smooth manifold and whose coherent product is smooth. 
Differentiation then gives geometric structures on the label 
manifold and operators on the quantum space. 
Both can be studied through the same coherent product. 

One of the strengths of the coherent manifold approach is that it stays close to actual computations; computational techniques in quantum physics or related fields can be phrased profitably in terms of coherent spaces.
Once a coherent product and its compatible transformations are known, general constructions can be applied to different examples. 
In place of measures and integration, differentiation is the basic 
tool for evaluating inner products. 
This makes many calculations feasible that are difficult in Hilbert 
spaces whose inner product is defined through a measure.
Computational methods based on parametrized quantum states can 
thus be studied in a common language, retaining the information 
needed for the calculation while making its dependence on geometry 
and symmetry explicit. 

\pgph


Geometric constructions from coherent states have 
a long history, including the work of 
\sca{Rawnsley} \cite{Rawnsley1977}, 
\sca{Provost \& Vallée} \cite{ProvostVallee1980}, 
\sca{Perelomov} \cite{Perelomov1972}, and 
\sca{Odzijewicz} \cite{Odzijewicz1992}. 
Here we use the geometry of the unnormalized coherent vectors and 
allow the induced forms to be degenerate. 

A central problem is that differentiation, in general, does not 
preserve the algebraic span of the coherent states. 
To obtain a domain for repeated operator calculations, we adjoin 
their iterated derivatives along coherent vector fields. 
The resulting span is itself the quantum space of an extended 
coherent space. 
This allows the quantization of coherent maps to be used for both 
finite transformations and derivative operations. 
We obtain a common invariant domain for coherent vector fields and 
smooth coherent diffeomorphisms whose coherent adjoints are also 
smooth diffeomorphisms. 
On this domain, composition, commutators, covariance as well as 
formal adjoint relations can be treated together. 
Enlarging the domain, however, preserves the Hilbert-space closures 
of the individual operators which are originally only defined on 
the coherent span.

Domains and kernels generated by derivatives can be found in 
several earlier approaches. 
\sca{Odzijewicz \& Horowski} 
\cite[Proposition~4.2]{OdzijewiczHorowski2013}
enlarge the span of the coherent vectors by repeated application of the generator of a unitary flow. In their later paper \sca{Odzijewicz \& Horowski} \cite[paragraph following~(72)]{OdzijewiczHorowski2022}, the coherent span is a common core for the self-adjoint generators of the flows considered there. 
Here we take the span of all iterated derivatives of the coherent-state map along coherent vector fields. 
The field may be chosen independently at each differentiation. 
The resulting space is a common invariant domain for the quantizations of coherent vector fields and smooth symmetries.
\sca{Merigon} et al. \cite{MNO2015} construct common domains for kernel Lie derivatives and prove integration results under additional hypotheses. 
Jet constructions in reproducing-kernel spaces provide another related approach, see 
\sca{Douglas} et al. 
\cite{DouglasMisraVarughese2000} and
\sca{Ishikawa} et al. \cite{IshikawaHashimotoIkedaKawahara2025}. 

\pgph


The present construction organizes coherent vector fields and finite transformations through one extended coherent space, starting from a coherent manifold. 
Its purpose is to make their operator calculus available directly from coherent data, without requiring an integrating 
Lie-group action for the algebraic construction. 

To relate infinitesimal operators to transformations, we introduce 
motions in subgroups of diffeomorphisms. 
These are one-sided families with local first-order control of 
their pullbacks and inverse pullbacks. 
No smooth dependence on the motion parameter or Lie-group structure 
on the subgroup is assumed. 
Their generators form a Lie algebra and, under the stated 
compatibility hypotheses, differentiation of quantized motions 
gives the corresponding infinitesimal operators. 
This construction is related to tangent Lie algebras of 
transformation groups, see 
\sca{Hubicska \& Muzsnay} \cite{HM2019}, 
and chronological calculus, see 
\sca{Kipka \& Ledyaev} \cite{KL2014}. 
It provides a way to use transformations already available on the 
coherent label space.

Symmetries are particularly important in applications to quantum 
physics. When coherent transformations give a strongly continuous 
unitary one-parameter group, its self-adjoint generator provides a 
distinguished observable. 
Position and momentum in nonrelativistic 
quantum mechanics are familiar examples arising from the Heisenberg 
group. 
Energy, momentum and angular momentum in relativistic 
single-particle quantum mechanics are associated with the Poincaré 
group. 
A coherent description makes the action of such 
transformations and the evaluation of their matrix elements 
accessible through the labels and the coherent product.

For exactly solvable models, one can frequently choose a coherent 
family preserved by the quantum evolution. 
The evolved coherent states are then obtained from trajectories of 
their labels, and the corresponding overlaps follow from the 
coherent product. 
When both the product and the label trajectories are explicit, the 
coherent representation gives direct access to quantum dynamics. 
This mechanism already appears in \sca{Klauder} \cite{Kla.III} and 
in the coherent-state dynamics studied by 
\sca{Mehta} et al. \cite{MehCSV}. 
We formulate it within the differential coherent framework 
developed here. 

\pgph


For more general Hamiltonians, coherent families also provide trial 
states for variational dynamics. 
Restricting the quantum action to such a family expresses the 
resulting equations through coherent data. 
This connects the construction with the geometric 
time-dependent variational principle of 
\sca{Kramer \& Saraceno} \cite{KraS}. 
The distinction between action stationarity and other variational 
conditions on real trial manifolds is discussed by 
\sca{Lasser \& Su} \cite{LS2022} 
and \sca{Hackl} et al. \cite{HacklEtAl2020}. 
The coherent formulation allows that the geometric and operator 
quantities can enter these equations; 
quantitative approximation estimates require further assumptions, 
as in \sca{Lubich} \cite{Lubich}. 

The operator calculus is illustrated in an explicit setting with 
Klauder spaces: Creation and annihilation operators, number-type 
operators and Weyl operators are obtained from the coherent product 
and its transformations. 
Their familiar relations are placed on the common domain supplied 
by the differential construction. 
These spaces also connect these constructions with normal ordering. 
These calculations continue the coherent treatment in 
\sca{Neumaier \& Ghaani Farashahi} \cite{NGF2022}, and show how 
the general results can be used without first choosing a 
realization by wavefunctions. 

\pgph


\textbf{Content.}
The remainder of this section recalls the required facts about
coherent spaces and convenient calculus, and fixes our notation.
Our standard references are
\sca{Neumaier} \cite{Neumaier2019} and
\sca{Kriegl \& Michor} \cite{KM}, respectively.
Section~\ref{sec:coherent-manifolds} introduces coherent manifolds
and the differential forms determined by their coherent products.
It develops Hamiltonian and coherent vector fields and relates
infinitesimal coherent maps to their generators.
Section~\ref{sec:coherent-quantization} constructs finite and infinitesimal quantizations of coherent vector fields, coherent maps, and their infinitesimal generators. 
The resulting operators have a common invariant domain.
It then develops the associated operator algebras and their
calculus of coherent shadows.
Subsections~\ref{subsec:halfcurve-algebra} and~\ref{subsec:quantization-along-motions}
construct the Lie algebra of generators of motions and relate its coherent
generators to the derivatives of quantized motions.
Section~\ref{sec:applications} applies these constructions to Schrödinger equations defined on the quantum space of a coherent space, and their propagators and resolvents.
It relates the coherent action to the restricted quantum action
and states the conditions under which stationarity gives the
Dirac--Frenkel equations.
Finally, Klauder spaces provide explicit realizations of bosonic
operators, their commutation relations, and normal ordering. 

\pgph


\textbf{Acknowledgments.} Thanks to Waltraud Huyer and Karl-Hermann Neeb for useful remarks on an earlier version of this
paper.

\subsection{Coherent spaces}

A \bfi{Hermitian space} is a complex vector space $\Hz$ with an 
inner product $\phi^*\psi \in \Cz$ for $\phi,\psi \in \Hz$ such that it is antilinear in the first and linear in the second argument,
\[
	\ol{\phi^*\psi}=\psi^*\phi
	\AND
	\phi^*\phi > 0
	\for 
	\phi \in \Hz \setminus \{0\}.
\]
Here $\alpha > 0$ says that the complex number $\alpha$ is real and 
positive. 
We write $\|\phi\| \cq \sqrt{\phi^*\phi}$ for the \bfi{norm} on $\Hz$. 

A complex $n \* n$-matrix $G$ is \bfi{Hermitian} if 
$\ol{G_{ij}}=G_{ji}$ for $i,j=1,\ldots,n$ and 
\bfi{positive semidefinite} if $u^*Gu \geq 0$ for all 
$u \in \Cz^n$. 

\pgph

 
Let $Z$ be a nonempty set. 
A \bfi{coherent product} on $Z$ is a 
function $K \cln Z \* Z \to \Cz$ of positive type, \ie for every 
finite sequence $z_1,\ldots,z_n$ in $Z$, the $n \* n$-matrix $G$ 
with entries
\[
	G_{ij}=K(z_i,z_j)
\]
is Hermitian and positive semidefinite. 
A \bfi{coherent space} is a nonempty set $Z$ with a distinguished 
coherent product $K \cln Z \* Z \to \Cz$. 

\pgph


A \bfi{quantum space} of a coherent space $Z$ is a Hermitian space $\Qz(Z)$ spanned by a distinguished set of vectors $\ket*{z}$ with $z \in Z$ satisfying
\[
	\braket*{z}{z'} \cq K(z,z')
	\for
	z,z' \in Z
\]
with the linear functionals 
\[
	\bra*{z} \cq \ket*{z}^*
\]
acting on the \bfi{augmented quantum space} $\Qz(Z)^\*$ which is 
the algebraic antidual of $\Qz(Z)$.
By \sca{Neumaier} \cite[Theorem~5.3.2]{Neumaier2019}, the quantum space exists and is unique up to isomorphism. 
We call the Hilbert-space completion $\ol{\Qz(Z)}$ the corresponding \bfi{completed quantum space} of $Z$.
We have the canonical inclusions
\[
    \Qz(Z) \seq \ol{\Qz(Z)} \seq \Qz(Z)^\*,
\]
where $\Qz(Z)^\*$ consists of all antilinear functionals on
$\Qz(Z)$. The second inclusion identifies $\psi \in \ol{\Qz(Z)}$ with the functional $\phi \mto \phi^*\psi$ on $\Qz(Z)$. 
We equip the completed quantum space $\ol{\Qz(Z)}$ with its Hilbert norm topology and the augmented quantum space $\Qz(Z)^\*$ with the weak-$*$ topology of pointwise convergence on $\Qz(Z)$.
The \bfi{coherent-state map} $\kappa \cln Z \to \ol{\Qz(Z)}$ of 
$Z$ is defined by
\begin{equation}
\label{eq:coherent-state-map}
	\kappa(z) \cq \ket*{z}
	\for z \in Z.
\end{equation}
A coherent space is called \bfi{nondegenerate} if its coherent-state map is injective; otherwise it is called \bfi{degenerate}. 

\pgph


Let $Z$ be a coherent space. 
A map $A \cln Z \to Z$ is \bfi{coherent} if it admits a map 
$A^* \cln Z \to Z$ such that
\[
	K(Az,z')=K(z,A^*z')
	\for
	z, z' \in Z.
\]
Such a map $A^*$ is a \bfi{coherent adjoint} of $A$.
The set of coherent maps is denoted by $\Coh(Z)$.

A \bfi{symmetry} of a coherent space $Z$ is an invertible coherent map on $Z$ admitting a bijective coherent adjoint.
Its inverse is coherent with adjoint $(A^{-1})^*=(A^*)^{-1}$. 
We write $\Sym(Z)$ for the group of symmetries of $Z$.

A coherent map $A \cln Z \to Z$ is \bfi{unitary} if it is bijective 
and $A^{-1}$ is a coherent adjoint of $A$. 
In a degenerate coherent space, a different adjoint need not equal 
$A^{-1}$. 
A linear operator on $\Qz(Z)$ is called \bfi{unitary} if it is a 
surjective inner-product isometry.

An \bfi{isometry} is a coherent map 
$A \cln Z \to Z$ that preserves the coherent product:
\[
	K(Az,Az')=K(z,z')
	\for 
	z,z' \in Z.
\]
If $A^*$ is a coherent adjoint of $A$, then $K(Az,Az')=K(z,A^*Az')$.
Consequently, when $Z$ is nondegenerate, $A$ is an isometry \Iff $A^*A=1$.
A bijective isometry is a unitary coherent map, because
$K(Az,z')=K(z,A^{-1}z')$, so its inverse is a coherent adjoint.

\subsection{Lie algebras}

A \bfi{Lie algebra} $\Lz$ is a real or complex vector space with a \bfi{Lie product}, a bilinear binary operation 
$\lp \cln \Lz \* \Lz \to \Lz$, satisfying 
\[
	f \lp f=0
	\for 
	f \in \Lz
\]
and the \bfi{Jacobi identity}
\[
	f \lp (g \lp k)=(f \lp g) \lp k + g \lp (f \lp k)
	\for
	f,g,k \in \Lz.
\]
A \bfi{Lie $*$-algebra} is a real (or complex) Lie algebra $\Lz$ with Lie product $\lp$ and a mapping $*$ that assigns to every 
$X \in \Lz$ an adjoint $X^* \in \Lz$ such that, for $X,Y \in \Lz$ and $\lambda \in \Rz$ (or $\Cz$),
\[
	(X+Y)^*=X^*+Y^*,
	~~~
	(X \lp Y)^*=Y^* \lp X^*,
	~~~
	X^{**}=X,
	\AND
	(\lambda X)^*=\ol{\lambda} X^*.
\]

\subsection{Convenient calculus}
\label{subsec:convenient-calculus}

We work with Hausdorff locally convex real vector spaces which we assume to be convenient, 
see \sca{Kriegl \& Michor} \cite[Theorem~2.14]{KM}. 

\pgph


Throughout, smooth manifolds are modelled on convenient vector 
spaces and are assumed to be \bfi{smoothly Hausdorff}, i.e. 
real-valued smooth functions separate points,
see \sca{Kriegl \& Michor} \cite[Subsections~27.1 and 27.4]{KM}. 

For a smooth manifold $M$, we denote  by $TM$ the kinematic tangent bundle, and by $T_pM$ the kinematic tangent space at $p\in M$, see
\sca{Kriegl \& Michor} \cite[Sections 28 and 32]{KM}.
We usually omit the adjective “kinematic.” 

\pgph


Let $E$ be a locally convex vector space. 
A curve $c \cln \mathbb R \to E$ is \bfi{differentiable} if
\[
	\dot c(t) \cq \lim_{h\to0}\frac{c(t+h)-c(t)}{h}
\]
exists for every $t \in \Rz$, and \bfi{smooth} if all iterated 
derivatives exist. 
A curve in a smooth manifold is \bfi{smooth} if its local 
coordinate representations are smooth.
A map $f \cln M \to N$ between smooth manifolds is \bfi{smooth} if 
$f \circ c$ is smooth for every smooth curve $c \cln \Rz \to M$,
see \sca{Kriegl \& Michor} \cite[Subsection 27.2]{KM}. 

Time intervals are assumed to have nonempty interior. 
When a map is required to be smooth and its domain includes an endpoint of a time interval, we understand smoothness there in the sense of local smooth extension across the endpoint, jointly in all variables. 
At a left endpoint the time derivative is taken from the right, and at a right endpoint it is taken from the left, see 
\sca{Kriegl \& Michor} \cite[Theorem~24.8 and Proposition~24.10]{KM}.

\pgph


Let $M$ be a smooth manifold. 
The set $\Kin(M)$ of smooth kinematic vector fields, see 
\sca{Kriegl \& Michor} \cite[Section~32.1 and Theorem~32.8]{KM}, is a Lie algebra whose Lie product is the \bfi{Lie bracket} written locally as
\begin{equation}
\label{eq:Lie-bracket-convenient}
	[X,Y](p)
	\cq dY(p).X(p)-dX(p).Y(p)
	\cq dY(p)(X(p))-dX(p)(Y(p))
\end{equation}
for $p \in M, ~ X,Y \in \Kin(M)$.
The \bfi{Lie derivative} of $Y$ along $X$ is
\[
	\Lc_XY \cq [X,Y].
\] 
For a smooth map $f$ between smooth manifolds $M$ and $N$, we write
\[
	df(p) \cq T_pf \cln T_pM\to T_{f(p)}N
\]
for its differential at $p \in M$. 
Its \bfi{Lie derivative} along $X\in\Kin(M)$ is defined by
\[
	(\Lc_Xf)(p) 
	\cq df(p).X(p)
	\cq df(p)(X(p)).
\]
For a smooth manifold $M$ and a locally convex space $F$, we equip $C^\infty(M,F)$ with the initial locally convex topology, called the \bfi{mapping-space topology}, for the pullbacks
\[
	c^* \cln C^\infty(M,F) \to C^\infty(\Rz,F),
	~~~ 
	f \mto c^*f \cq f \circ c,
\]
along smooth curves $c \cln \Rz\to M$. 
Here $C^\infty(\Rz,F)$ carries the topology of uniform convergence 
on compact sets of all derivatives.
We write $L(E,F)$ for the space of bounded linear maps between 
locally convex spaces $E$ and $F$, with the topology induced by 
$C^\infty(E,F)$. 
It is a closed subspace of $C^\infty(E,F)$. 
If $F$ is convenient, then both $C^\infty(M,F)$ and $L(E,F)$ are 
convenient. 

\pgph


In Subsection \ref{subsec:halfcurve-algebra}, we will use the following technical result. 

\begin{lemma}
\label{lem:AsF}
Let $U,V,W$ be convenient manifolds. 
For every convenient vector space $F$, let
\[
	A_s^F,\ A^F \cln C^\infty(V,F) \to C^\infty(W,F)
	\AND
	B_s^F,\ B^F \cln C^\infty(U,F) \to C^\infty(V,F)
\]
be bounded linear operators, defined for $0<s<\delta$, where $\delta$ is independent of $F$.

Suppose that:
\begin{enumerate}[ref=\textup{(\roman*)}]
	\item\label{itm:natural-product-naturality} These operators commute with continuous linear maps between target spaces, \ie for every continuous linear 
	$\ell \cln F \to G$,
	   \[
	  	 A_s^G(\ell\circ g)=\ell\circ A_s^Fg,
	   \]
	   and similarly for $A,B_s,B$.
	\item\label{itm:natural-product-pointwise} For every convenient $F$ and every fixed 
	$g\in C^\infty(V,F)$, $f\in C^\infty(U,F)$,
	\[
	   \lim_{s \to 0} A_s^Fg=A^Fg,
	   ~~~
	   \lim_{s \to 0} B_s^Ff=B^Ff
	\]
	in the respective mapping-space topologies.
\end{enumerate}
Then
	\[
		\lim_{s \to 0} A_s^FB_s^F=A^FB^F 
		~~~
		\text{ in } 
		L\bigl(C^\infty(U,F),C^\infty(W,F)\bigr).
	\]
\end{lemma}

\begin{proof}
We first prove that
\begin{equation}
\label{eq:firs}
	s_n\downarrow0,
	~~~
	g_n \to  g
	~~~
	\text{ in }
	C^\infty(V,F)
	~~~ \Longrightarrow ~~~
	A_{s_n}^Fg_n  \to  A^Fg
	~~~
	\text{ in }
	C^\infty(W,F).
\end{equation}
We put $I=[0,1]$ and equip $H \cq C(I,F)$ with the seminorms
$p_I(u) \cq \sup_{t\in I}p(u(t))$,
where $p$ ranges over the continuous seminorms of $F$. 
This is a convenient vector space: By 
\sca{Kriegl \& Michor} \cite[Theorem~2.15]{KM}, the space $\ell^\infty(I,F)$ of functions with bounded image, equipped with the seminorms $p_I$, is convenient. 
Since $I$ is compact, every continuous function $I \to F$ has bounded image. 
Moreover, uniform limits with respect to all continuous seminorms preserve continuity. 
Hence, $H$ is a closed linear subspace of $\ell^\infty(I,F)$, and therefore is convenient.

Let $\gamma \cln I \to  C^\infty(V,F)$ be the continuous path that satisfy $\gamma(0)=g$, $\gamma(1/n)=g_n$, and is affine on every interval $[1/(n+1),1/n]$. 
We define $G \cln V \to  H$ by $G(v)(t) \cq \gamma(t)(v)$.
The map $G$ is smooth. 
Indeed, for every smooth curve 
$v \cln \Rz \to V$ and every $j\geq0$,
$(g_n\circ v)^{(j)} \to (g\circ v)^{(j)}$ uniformly on compact intervals, for every continuous seminorm of $F$. 
The derivatives of $G \circ v$ are obtained by interpolating these derivatives. 
To justify differentiation in the topology of $H$, we observe that the derivatives of the next two orders are uniformly bounded in $n$ on each compact interval; the usual Taylor estimate, see 
\sca{Kriegl \& Michor} \cite[Section 15.6]{KM}, therefore holds uniformly in $t$. 
Thus, $G\circ v$ is smooth.
By assumption,
\begin{equation}
\label{eq:natural-product-path-convergence}
	K_n \cq A_{s_n}^{H}G  \to  K \cq A^{H}G
	~~~
	\text { in } 
	C^\infty(W,H).
\end{equation}
For $t \in I$, evaluation
$\ev_t \cln H \to  F$, defined by $u \mto  u(t)$,
is continuous and linear. 
\ref{itm:natural-product-naturality} gives, with $t_n=1/n$,
$A_{s_n}^{F}g_n=\ev_{t_n} \circ K_n$, $A^{F}g=\ev_{0} \circ K$. 
Consequently,
\begin{equation}
\label{eq:natural-product-evaluation-split}
	A_{s_n}^{F}g_n-A^{F}g
	=\ev_{t_n} \circ(K_n-K)+(\ev_{t_n}-\ev_0) \circ K.
\end{equation}
The first term tends to zero in $C^\infty(W,F)$ by \gzit{eq:natural-product-path-convergence}, since
$p(\ev_{t_n}u) \leq p_I(u)$ uniformly in $n$.
For the second term, we fix a smooth curve $w \cln \Rz \to W$, a compact interval $J$, and $j \geq 0$. 
The map $(r,t) \mto  \bigl((K\circ w)^{(j)}(r)\bigr)(t)$
is continuous on $J\times I$. 
Hence, for every continuous seminorm $p$,
$
\sup_{r\in J}
p\!\left(
\bigl((K\circ w)^{(j)}(r)\bigr)(t_n)
-
\bigl((K\circ w)^{(j)}(r)\bigr)(0)
\right)
\to 0
$.
This proves \gzit{eq:firs}.
Now we fix $f\in C^\infty(U,F)$. 
Applying \gzit{eq:firs} to $g_n=B_{s_n}^{F}f$ and $g=B^{F}f$ gives 
$A_{s_n}^{F}B_{s_n}^{F}f \to  A^{F}B^{F}f$.
Since this holds for every sequence $s_n\downarrow0$,
\begin{equation}
\label{eq:natural-product-pointwise-convergence}
	T_s^{F}f  \to  T^{F}f
	~~~
	\text{ in }
	C^\infty(W,F),
	~~~
	T_s^{F} \cq A_s^{F}B_s^{F},
	~~~ 
	T^{F} \cq A^{F}B^{F}.
\end{equation}
Testing sequences suffices here because the parameter $s$ has a countable neighbourhood base at zero.
It remains to establish convergence in the mapping-space topology. Let $c \cln \Rz \to  C^\infty(U,F)$ be an arbitrary smooth curve, and put $E \cq C^\infty(\mathbb R,F)$.
By the smooth exponential law, the map
$\widehat c \cln U \to  E$ defined by $\widehat c(u)(t) \cq c(t)(u)$
is smooth. 
Applying \gzit{eq:natural-product-pointwise-convergence} with target space $E$ yields
\begin{equation}
\label{eq:natural-product-curve-convergence}
	R_s \cq (T_s^{E}-T^{E})\widehat c
 	 \to 0
	~~~
	\text{ in }
	C^\infty(W,E).
\end{equation}
Assumption \ref{itm:natural-product-naturality} for evaluation at $t$ gives
\begin{equation}
\label{eq:natural-product-curve-evaluation}
	R_s(w)(t)
	=\bigl((T_s^{F}-T^{F})c(t)\bigr)(w).
\end{equation}
For every smooth curve $d \cln \mathbb R\to W$, compact intervals $I_1,I_2$, nonnegative integers $j,k$, and continuous seminorm $p$ of $F$, equation \gzit{eq:natural-product-curve-convergence} implies
$	
\sup_{\substack{t\in I_1,\ r\in I_2}}
p\!\left(
\partial_t^{\,j}\partial_r^{\,k}
\bigl[R_s(d(r))(t)\bigr]
\right)
\to 0
$.
By \gzit{eq:natural-product-curve-evaluation}, these are exactly the seminorms asserting $(T_s^{F}-T^{F})\circ c \to 0$ in $C^\infty\!\left(\Rz,C^\infty(W,F)\right)$.
Since $c$ was arbitrary, $A_s^{F}B_s^{F} \to  A^{F}B^{F}$ in  $L\!\left(C^\infty(U,F),C^\infty(W,F)\right)$.
\end{proof}

A \bfi{complex vector field} on a smooth manifold $M$ is a smooth 
section of the \bfi{complexified tangent bundle}
\[
	\Cz TM \cq TM \oplus iTM.
\]
Every complex vector field has a unique representation $X+iY$, with 
$X,Y \in \Kin(M)$. 
We write
\[
	\Cz \Kin(M) \cq \Kin(M) \oplus i \Kin(M)
\]
for the space of complex vector fields.

For $f \in C^\infty(M,\Cz)$, differentiation along a complex 
vector field is defined by
\[
	\Lc_{X+iY}f \cq \Lc_Xf+i\Lc_Yf
	\for 
	X,Y \in \Kin(M).
\]
The Lie bracket extends complex-bilinearly to $\Cz \Kin(M)$:
\[
	[X+iY,U+iV] \cq [X,U]-[Y,V]+i\bigl([X,V]+[Y,U]\bigr)
	\for 
	X,Y,U,V \in \Kin(M).
\]
Complex-valued differential forms are evaluated on complex vector 
fields by complex-multi\-linear extension. 
In particular, for a complex-valued $2$-form $\omega$,
\[
	(X+iY)\omega=X\omega+iY\omega.
\]
A \bfi{complex-valued $2$-form} on 
$M$ is a smooth map $\omega \cln TM \*_M TM \to \Cz$ such that, for every $p \in M$,
\[
	\omega_p \cln T_pM \* T_pM \to  \Cz
\]
is real-bilinear and alternating.
Equivalently, $\omega=\omega_1+i\omega_2$ for some smooth real $2$-forms $\omega_1,\omega_2$.
Its \bfi{exterior derivative} is defined by
\[
	d\omega \cq d\omega_1+i \, d\omega_2.
\]
Thus, $\omega$ is closed when both $\omega_1$ and $\omega_2$ are 
closed. 

Every complex-valued differential form is extended 
complex-multilinearly to the \bfi{complexified tangent spaces}
\[
	\Cz T_pM \cq T_pM \oplus iT_pM.
\]
We denote this extension by the same symbol. We use the 
\bfi{contraction convention} 
\[
	YX\omega \cq \omega(X,Y),
\]
so that $X\omega$ is the 1-form $Y \mto \omega(X,Y)$.
Consequently, for $X,Y \in \Kin(M)$, 
$(X+iY)\omega=X\omega+i \, Y\omega$.
We define the \bfi{complex kernel} of $\omega$ at $p \in M$ by
\[
	\ker_{\Cz} \omega_p
	\cq 
	\{
		v\in \Cz T_pM
		\mid
		\omega_p(v,u)=0
		\text{ for every } u \in \Cz T_pM
	\}.
\]

\section{Coherent manifolds}
\label{sec:coherent-manifolds}

Coherent manifolds provide a geometric setting for quantization and 
for variational approximations to quantum dynamics. 
An early precursor of this approach is 
\sca{Klauder} \cite{Kla.III}, who related classical dynamics to the 
quantum action restricted to a family of states labelled by 
classical variables. 
We obtain the required geometric quantities by differentiating the 
coherent product. 
Its derivatives define differential forms and a semimetric, which 
we use to develop a coherent Poisson algebra and the coherent action 
principle. 
We then introduce coherent vector fields and relate them to 
coherent motions. 
Their quantization in Section~\ref{sec:coherent-quantization} turns these infinitesimal transformations into operators on the quantum space.

\subsection{Coherent manifolds}
\label{subsec:geometry}

A coherent product determines the scalar products of coherent 
states. When it is smooth, its derivatives also determine geometric 
structures on the space of coherent labels. We now develop these 
structures for smooth manifolds. They 
provide the geometric setting for coherent quantization and the 
coherent action principle. 

\pgph


Following \sca{Neumaier} \cite[Section~5.4, p.~77]{Neumaier2019}, we call a coherent space $Z$ a \bfi{coherent manifold} if $Z$ is a smooth manifold and its coherent product $K$ is smooth. 
From now on, $Z$ always denotes a coherent manifold unless indicated explicitly otherwise.

\begin{examples}
\begin{enumerate}
	\item Clearly, every Hilbert space is a coherent manifold.
	\item An example is the \bfi{Szegő disk}, see 
	\sca{Neumaier} \cite[Examples~5.7.3(iv)]{Neumaier2019},
	\[
		\Dz \cq \{z \in \Cz \mid |z|<1\},
	\]
	with coherent product
	\[
		K(z,z') \cq \frac{1}{1-\ol z\,z'}
		\with 
		z,z' \in \Dz.
	\]
	\item \bfi{Klauder spaces} (see 
	\sca{Neumaier} \cite[Example 5.2.2]{Neumaier2019}), discussed in more detail in Subsection \ref{subsec:klauder}, are another class of coherent manifolds.\qedhere
\end{enumerate}
\end{examples}

\begin{proposition}
\label{prop:smooth-kappa}
The coherent-state map \gzit{eq:coherent-state-map} is 
smooth.
\end{proposition}

\begin{proof}
Let $c \cln \Rz \to Z$ be a smooth curve on $Z$.
Then $K(c(s),c(t))$ is smooth on $\Rz^2$. 
By \sca{Neeb} \cite[Theorem~7.1]{Neeb2010}, $\kappa \circ c$ is 
a smooth curve on $\ol{\Qz(Z)}$. 
Thus, $\kappa$ is smooth.
\end{proof}

Let $M$ be a smooth manifold. 
For $f \in C^\infty(M  \*  M,\Cz)$, we denote its 
partial differentials at $(p,q) \in M \* M$ by
\[
	D_Lf(p,q) \cln T_pM \to \Cz,
	~~~
	D_Rf(p,q) \cln T_qM \to \Cz.
\]
For $X\in\Kin(M)$, we put
\[
	L_Xf(p,q) \cq  D_Lf(p,q).X(p),
	~~~
	R_Xf(p,q) \cq  D_Rf(p,q).X(q).
\]
Thus, $L_X$ and $R_X$ differentiate along $X$ in the first and 
second variables, respectively. 

\pgph


Let $Z$ be a coherent manifold. 
We define the \bfi{coherent $1$-form} $\theta$ by 
\begin{equation}
\label{eq:theta}
	\theta(z) \cq  -i D_RK(z,z)
	\for 
	z \in Z.
\end{equation}
Its associated \bfi{coherent $2$-form} $\omega$ is defined by
\begin{equation}
\label{eq:omega}
	\omega(z) \cq d\theta(z)
	\for 
	z \in Z.
\end{equation}
Note that $\theta$ and $\omega$ are complex-valued forms. 

\pgph


Furthermore, the \bfi{coherent semimetric} is
\[
	g_z(v,u)
	\cq \re  D_L D_RK(z,z).(v,u)
	\for 
	z \in Z, ~ u,v \in T_zZ.
\]
A coherent manifold $Z$ is called 
\bfi{infinitesimally nondegenerate} if its coherent-state map 
\gzit{eq:coherent-state-map} is \bfi{tangent-injective}, that is, if
\[
	d\kappa(z) \cln T_zZ  \to \ol{\Qz(Z)}
\]
is injective for every $z\in Z$.

A \bfi{smooth Riemannian semimetric} on a smooth manifold $M$ is 
a smooth map
\[
	g \cln TM \* _M TM \to \Rz
\]
whose restriction $g_p$ to each fibre is symmetric, real-bilinear 
and positive semidefinite. 

\begin{proposition}
\label{prop:can-tens-formulas}
The following hold:
\begin{enumerate}[label=\textup{(\roman*)}, ref=\textup{(\roman*)}]
	\item\label{itm:kernel-two-form} The coherent $2$-form $\omega$ is smooth and closed, 
	and
	\[
		\omega_z(v,u)
		=2\im  D_L D_RK(z,z).(v,u)
		\for 
		z \in Z, ~ v,u \in T_zZ.
	\]
	\item\label{itm:kernel-semimetric} The coherent semimetric $g$ is a smooth Riemannian 
	semimetric, and
	\[
		g_z(v,v)=\|d\kappa(z).v\|^2
		\for 
		z \in Z, ~ v \in T_zZ.
	\]
	\item\label{itm:kernel-tangent-nullspace} For every $z\in Z$,
	\[
		\ker g_z=\ker d\kappa(z)
		=\bigl\{v \in T_zZ\mid  D_RK(z',z).v=0
		\text{ for every } z' \in Z\bigr\}.
	\]
	Thus, $K$ separates tangent vectors at $z$ when 
	$d\kappa(z)$ is 
	injective.
	\item\label{itm:kernel-two-form-nullspace} For every $z\in Z$,
	\[
		\ker d\kappa(z) \seq \ker\omega_z
		=\bigl\{v\in T_zZ\mid \omega_z(v,u)=0
		\text{ for every }u\in T_zZ\bigr\}.
	\]
	Hence, if $\ker\omega_z=\{0\}$ for every $z\in Z$, then 
	$Z$ is infinitesimally nondegenerate.
\end{enumerate}
\end{proposition}

\begin{proof}
By Proposition~\ref{prop:smooth-kappa}, the coherent-state map
$\kappa$ is smooth. 
Differentiating $K(z,z')=\braket*{z}{z'}$ gives
$D_RK(z,z').u=\bra*{z}d\kappa(z').u$ and
$D_L D_RK(z,z')(v,u)=(d\kappa(z).v)^*d\kappa(z').u$.
In particular, $\theta_z(v)=-i(\bra*{z}d\kappa(z).v)$,
so $\theta$ is smooth.
To compute $d\theta$, work in a chart and extend $v,u$ as 
constant coordinate vector fields. 
We obtain
$d(\theta(u))(z).v
=-i((d\kappa(z).v)^*d\kappa(z).u+\bra*{z}d^2\kappa(z).(v,u))$.
The constant coordinate vector fields commute. 
Thus, $d\theta(z).(v,u)=d(\theta(u))(z).v-d(\theta(v))(z).u$.
The second-derivative terms cancel by symmetry of $d^2\kappa$. 
Therefore,
\[
\begin{aligned}
	\omega_z(v,u)
	&=-i((d\kappa(z).v)^*d\kappa(z).u-(d\kappa(z)u)^*d\kappa(z).v)
	=2 \,\im (d\kappa(z).v)^*d\kappa(z).u \\
	&=2 \,\im  D_L D_RK(z,z).(v,u).
\end{aligned}
\] 
Smoothness follows immediately, and closedness follows from
$d\omega=d^2\theta=0$.
This proves \ref{itm:kernel-two-form}.
Furthermore, $g_z(v,u)=\re (d\kappa(z).v)^*d\kappa(z).u$.
Since the tangent map of $\kappa$ and the Hilbert-space inner 
product are smooth, $g$ is smooth. 
Hermitian symmetry and real linearity of $d\kappa(z)$ show that 
$g_z$ is symmetric and real-bilinear. 
Moreover, $g_z(v,v)=\|d\kappa(z).v\|^2 \geq 0$.
This proves \ref{itm:kernel-semimetric}.
By definition,
$\ker g_z
\cq \bigl\{v\in T_zZ \mid g_z(v,u)=0 \text{ for every }u
\in T_zZ\bigr\}$.
If $d\kappa(z).v=0$, the formula for $g_z$ gives 
$v \in \ker g_z$. 
Conversely, if $v \in \ker g_z$, taking $u=v$ now gives
$0=g_z(v,v)=\|d\kappa(z).v\|^2$,
hence $d\kappa(z).v=0$. 
Thus, $\ker g_z=\ker d\kappa(z)$.
The identity $D_RK(z',z).v=\bra*{z'}d\kappa(z).v$
shows that $d\kappa(z).v=0$ implies $ D_RK(z',z).v=0$ for every 
$z' \in Z$. 
Conversely, if all these pairings vanish, then $d\kappa(z)v$ is 
orthogonal to the complex span of the coherent vectors. 
This span is dense in $\ol{\Qz(Z)}$, so $d\kappa(z).v=0$. 
This proves \ref{itm:kernel-tangent-nullspace}.
Finally, \ref{itm:kernel-two-form} shows that $d\kappa(z).v=0$ implies $\omega_z(v,u)=0$ for every $u \in T_zZ$. 
Hence, $\ker d\kappa(z) \seq \ker\omega_z$.
If $\ker\omega_z=\{0\}$, then $d\kappa(z)$ is injective. 
This proves \ref{itm:kernel-two-form-nullspace}.
\end{proof}

\begin{proposition}
\label{prop:open-coherent-subsets}
Let $U  \seq  Z$ be nonempty and open. 
Then the restricted atlas of $Z$ on $U$ and the kernel
\[
	K_U \cq K|_{U\times U}
\]
turn $U$ into a coherent manifold. 
The inclusion $\inj \cln U \hto Z$ 
is smooth and preserves the coherent product. 
Moreover, $TU=TZ|_U$.
\end{proposition}

\begin{proof}
Restricting the charts of $Z$ to their intersections with $U$ 
gives charts with $c^\infty$-open images. 
Their transition maps are restrictions of smooth transition maps, hence smooth. 
The resulting natural topology on $U$ is the subspace topology. 
The separation properties required of the ambient manifold are 
inherited.
In these charts, $\inj$ is represented by the identity. 
Thus, it is smooth. 
Hermitian symmetry restricts to $K_U$. 
For $z_1,\ldots,z_n \in U$ and $c_1,\ldots,c_n\in\Cz$,
\[
	\sum_{j,k=1}^{n}\ol{c_j}c_kK_U(z_j,z_k)
	=\sum_{j,k=1}^{n}\ol{c_j}c_kK(z_j,z_k)
	\geq 0.
\]
Finally, $K_U=K\circ(\inj\times\inj)$ is smooth, and $K(\inj(u),\inj(v))=K_U(u,v)$ for $u,v\in U$.
\end{proof}

\begin{proposition}
\label{prop:restricted-coherent-adjoints}
Let $U  \seq  Z$ be nonempty and open, and let $A$ be a coherent map on $Z$ with coherent adjoint $A^*$ such that 
\[
	K(Az,z')=K(z,A^*z')
	\for 
	z,z'\in Z.
\]
If
\[
	A(U) \seq  U,
	~~~ 
	A^*(U) \seq  U,
\]
then $A^*|_U$ is a coherent adjoint of $A|_U$ for $K_U$. 
Smoothness of either map is preserved by restriction.
\end{proposition}

\begin{proof}
The invariance assumptions make both restrictions mappings from $U$ to $U$.

For $z,z'\in U$,
$K_U(A|_U(z),z')=K(Az,z')=K(z,A^*z')=K_U(z,A^*|_U(z'))$.
Smoothness is preserved by restriction to an open domain and 
corestriction to an open range.
\end{proof}

\subsection{Hamiltonian vector fields and coherent Poisson algebras}

We may associate with the coherent 2-form of a coherent manifold a
coherent Lie product by means of a construction taken from
\sca{Neumaier \& Westra} \cite[Section 18.1]{NeuWest.CQMvLie}. 

\pgph


For a complex closed 2-form $\omega$ on a smooth manifold $M$, we 
call $f \in C^\infty(M,\Cz)$ \bfi{compatible with $\omega$} if there 
is a complex vector field $X_f \in \Cz \Kin(M)$, called a
\bfi{Hamiltonian vector field associated with $f$}, satisfying
\begin{equation}
\label{eq:dfhom}
	df=X_f\omega.
\end{equation}
We write $\Ez(\omega)$ for the \eh{set of all functions 
$f \in C^\infty(M,\Cz)$ that are compatible with $\omega$}.

\begin{proposition}
\label{prop:coh2form}
Let $\omega$ be a complex closed 2-form on a smooth manifold $M$, 
and let $X_f$ be a Hamiltonian vector field associated with
$f \in \Ez(\omega)$. 
Then $X_f$ is unique up to addition of an arbitrary smooth vector 
field $Y$ with $Y\omega=0$.
In particular, if $\ker_{\Cz} \omega_p=\{0\}$ for every $p \in M$, 
then
\begin{equation}
\label{eq:assHam}
	X_f=df \, \omega^{-1}
\end{equation}
is uniquely determined by $f$. 
Here $\omega^{-1}$ denotes the inverse of $v \mto v\omega$ on its 
image.
\end{proposition}

\begin{proof}
$X$ is another Hamiltonian vector field associated with 
$f$ \Iff $X\omega=df=X_f\omega$, hence \Iff $Y\cq X-X_f$ satisfies 
$Y\omega=0$.
\end{proof}

We follow the Lie-product convention of 
\sca{Neumaier \& Westra} \cite{NeuWest.CQMvLie}. 
The operator $\lp$ defined on $\Ez(\omega)$ by 
\[
	f \lp g
	\cq X_f\,dg
	=X_fX_g\omega
	=-X_g\,df
	\for 
	f,g \in \Ez(\omega)
\]
is a Lie product related to the usual Poisson bracket by
\[
	\{f,g\} \cq g \lp f
	=-f \lp g.
\]
Since the Leibniz identity $f \lp (gh)=(f \lp g)h+g(f \lp h)$ holds, this turns $\Ez(\omega)$ into a commutative Poisson algebra, the \bfi{coherent Poisson algebra} of $Z$. 

Note the ordering of the arguments: In $f \lp g$, the first argument generates differentiation by $f$.
Thus, the first argument determines the Hamiltonian vector field 
that differentiates the second argument. 
In particular, for a Hamiltonian $H$,
\[
	H \lp f
	=df.X_H
	=\Lc_{X_H}f.
\]
Therefore, Hamilton’s equation for an observable $f$, 
conventionally written $\dot f=\{f,H\}$, takes the form 
$\dot f=H \lp f$. 

\begin{theorem}[%
	\sca{Neumaier \& Westra}
	{\cite[Theorem~18.1.3, p.~380]{NeuWest.CQMvLie}}%
]\label{lem:clo2form}
For every complex closed 2-form $\omega$ on a smooth manifold 
$M$, the set $\Ez(\omega)$ is a Poisson algebra with pointwise 
multiplication and Lie product given by
\[
	f \lp g \cq X_fdg =-df.X_g=X_fX_g\omega,
\]
which is independent of the choice of $X_f$ and $X_g$.
A Hamiltonian vector field associated with $f \lp g$ is given by
\[
	X_{f \lp g} \cq [X_f,X_g].
\]
\end{theorem}

\begin{corollary}
\label{prop:coherent-poisson}
For the coherent $2$-form \gzit{eq:omega}, $\Ez(\omega)$, with pointwise multiplication and the coherent Lie product
\[
	f \lp g \cq dg.X_f,
\]
is a unital complex Poisson algebra.
\end{corollary}

\begin{proof}
Proposition~\ref{prop:can-tens-formulas} shows that $\omega$ is 
smooth and closed. 
Theorem~\ref{lem:clo2form} applies.
\end{proof}

\subsection{Coherent vector fields}
\label{subsec:coherent-vector-fields}

A vector field $X \in \Kin(Z)$ is called \bfi{coherent} if there 
exists $X^* \in \Kin(Z)$ such that
\begin{equation}
\label{eq:vector-fields-coherent-adjoint}
	L_XK=R_{X^*}K.
\end{equation}
Any such $X^*$ is called a \bfi{coherent adjoint} of $X$. 
We write $\vect(Z)$ for the 
set of coherent vector fields on $Z$. 
Equivalently, on $Z \* Z$, put
$V(z,z') \cq \bigl(X(z),-X^*(z')\bigr)$.
Then $X^*$ is a coherent adjoint of $X$ \Iff
$dK(z,z').V(z,z')=0$ for $z,z' \in Z$.

\pgph

If $\beta$ is an action of a symmetric Lie algebra $(\mathfrak g,\tau)$, see \sca{Mérigon} et al. \cite[Definition~5.1]{MNO2015}, by vector fields and $K$ is compatible with $\beta$ in the sense of \sca{Mérigon} et al. \cite[Definitions~4.1 and~5.3]{MNO2015}, then $-\beta(\tau x)$ is a coherent adjoint of $\beta(x)$.

\begin{proposition}
\label{prop:coherent-algebra}
Equation  
\gzit{eq:vector-fields-coherent-adjoint} is 
symmetric, and $\vect(Z)$ is a real Lie subalgebra of $\Kin(Z)$.
If $X^*,Y^*$ are coherent adjoints of $X,Y$, respectively, and 
$a,b \in \Rz$, then $aX^*+bY^*$ is a coherent adjoint of
$aX+bY$, and $[Y^*,X^*]$ is a coherent adjoint of $[X,Y]$.
Furthermore, the subspace 
\[
	\Null(Z) \cq 
	\{
		X \in \Kin(Z) \mid d\kappa(z).X(z)=0 
		\text{ for every } z \in Z
	\} 
\]
is a real Lie ideal in $\vect(Z)$, and any two coherent adjoints of the same vector field differ by an element of $\Null(Z)$. 
The quotient $\vect(Z)/\Null(Z)$ carries a well-defined 
real-linear involution
\[
	(X+\Null(Z))^* \cq X^*+\Null(Z)
\]
satisfying
\[
	[X+\Null(Z),Y+\Null(Z)]^*
	=[Y^*+\Null(Z),X^*+\Null(Z)].
\]
If $\kappa$ is tangent-injective, then $\Null(Z)=\{0\}$, and 
coherent adjoints are unique. In this case, the mapping 
$X \mto X^*$ turns $\vect(Z)$ into a real Lie $*$-algebra.
\end{proposition}

\begin{proof}
The Hermitian symmetry of $K$, differentiated along a vector field 
$X$, gives $\ol{(L_XK)(z',z)}=(R_XK)(z,z')$.
Hence, conjugating $L_XK=R_{X^*}K$ and interchanging the 
variables gives $L_{X^*}K=R_XK$.
Thus, $X$ is a coherent adjoint of $X^*$. 
This proves symmetry.
For $a,b \in \Rz$, we have
$L_{aX+bY}K=aL_XK+bL_YK=aR_{X^*}K+bR_{Y^*}K=R_{aX^*+bY^*}K$.
Furthermore,
\[
	\begin{aligned}
		L_{[X,Y]}K
		&=L_XL_YK-L_YL_XK
		=L_XR_{Y^*}K-L_YR_{X^*}K
		=R_{Y^*}L_XK-R_{X^*}L_YK \\
		&=R_{Y^*}R_{X^*}K-R_{X^*}R_{Y^*}K 
		=R_{[Y^*,X^*]}K.
	\end{aligned}
\]
Thus, $[Y^*,X^*]$ is a coherent adjoint of $[X,Y]$. 
Since the zero field is coherent, these identities show that 
$\vect(Z)$ is a real Lie subalgebra of $\Kin(Z)$.
By Proposition~\ref{prop:smooth-kappa}, the coherent-state map
is smooth. 
After differentiating $K(z',z)=\braket*{z'}{z}$
in the second variable, we get
$(R_XK)(z',z)=\bra*{z'}d\kappa(z).X(z)$.
The coherent vectors have dense complex linear span in 
$\ol{\Qz(Z)}$. 
Therefore, using also Hermitian symmetry, we have
$X \in \Null(Z)$ \Iff $R_XK=0$ \Iff $L_XK=0$.
In particular, every element of $\Null(Z)$ is a coherent vector field, with zero as a coherent adjoint. 
Real linearity of $d\kappa(z)$ shows that $\Null(Z)$ is a real 
vector subspace.
Let $N \in \Null(Z)$ and $Y \in \vect(Z)$. 
We choose a coherent adjoint $Y^*$. 
Then
$L_{[N,Y]}K=L_NL_YK-L_YL_NK=L_NR_{Y^*}K=R_{Y^*}L_NK=0$.
Hence, $[N,Y] \in \Null(Z)$, proving that $\Null(Z)$ is an ideal.
If $X_1^*,X_2^*$ are coherent adjoints of $X$, then
$R_{X_1^*-X_2^*}K=R_{X_1^*}K-R_{X_2^*}K=L_XK-L_XK=0$.
Thus, $X_1^*-X_2^* \in \Null(Z)$. 
Furthermore, replacing $X$ by $X+N$, with $N \in \Null(Z)$, 
leaves 
$L_XK$, and hence its set of coherent adjoints, unchanged. 
The operation on the quotient is therefore independent of both 
the representative and the chosen adjoint.
Its real linearity and bracket reversal follow from the 
formulas already proved. 
Symmetry of the coherent-adjoint relation makes it involutive. 
Finally, if $\kappa$ is tangent-injective, then
$d\kappa(z)N(z)=0$ implies $N(z)=0$ at every $z \in Z$. 
Hence, $\Null(Z)=\{0\}$, and coherent adjoints are unique.
\end{proof}

\subsection{Infinitesimal coherent maps}
\label{subsec:inf-coh-maps}

A symmetry of $Z$ is called \bfi{smooth} if it is a diffeomorphism which admits a coherent adjoint that is also a diffeomorphism. 
We write $\sSym(Z)$ for the group of smooth symmetries of $Z$.

\phantomsection\label{not:coherent-motion} 
Let $I \seq \Rz$ be a closed interval and let
$A \cln I \to \sSym(Z)$ be a curve of smooth symmetries.
For each $t \in I$, choose a bijective coherent adjoint $A^*(t)$
of $A(t)$. Thus,
\begin{equation}
\label{eq:adjoint-identity}
    K(A(t)z,z')=K(z,A^*(t)z')
    \for z,z' \in Z.
\end{equation}

The following two theorems relate curves of smooth symmetries to their infinitesimal generators. Differentiating an infinitesimal coherent map, together with a sufficiently regular family of adjoints, yields a coherent vector field. 
Conversely, a time-dependent coherent vector 
field generates a curve of smooth symmetries if both the field and the negative of a chosen adjoint field admit global evolutions with the required regularity. 
The adjoint of the resulting curve is obtained by reversing the latter evolution.

Suppose that every $A(t)$ is unitary and that the maps
$(t,z) \mto A(t)z$ and $(t,z) \mto A(t)^{-1}z$ are smooth.
Then $-X(t)$ is a coherent adjoint of the logarithmic derivative
$X(t) \cq \dot A(t) \circ A(t)^{-1}$, and
\[
    L_{X(t)}K+R_{X(t)}K=0
\]
by Proposition~\ref{prop:quantized-unitary-motion}.
This is the infinitesimal condition for preservation of the 
coherent product. 
Conversely, under the stated existence and regularity assumptions, this condition ensures that the evolution consists of invertible isometries, hence of unitary coherent maps.
Their quantizations, see 
Propositions~\ref{prop:newprop} 
and~\ref{prop:quantized-unitary-evolution}, give unitary evolution 
on the completed quantum space and carry coherent states into coherent states.

\begin{theorem}\label{thm:coherent-motion-generator}
Let $A \cln I \to \sSym(Z)$ be a curve of smooth symmetries with coherent adjoint $A^*$.
Suppose that
\[
	(t,z) \mto A(t)^{\pm1}z
	\AND
	(t,z) \mto (A^*(t))^{\pm1}z
	\for
	t \in I, ~ z \in Z
\]
are smooth.

Then
\[
	X(t)(z) \cq \dot A(t)\bigl(A(t)^{-1}z\bigr)
\]
defines a smooth time-dependent coherent vector field. 
A coherent adjoint is
\[
	Y(t)(z') \cq 
	d\bigl(A^*(t)^{-1}\bigr)(A^*(t)z').\dot A^*(t)(z')
	\for
	z' \in Z.
\]
In particular,
\[
	L_{X(t)}K=R_{Y(t)}K.
\]
\end{theorem}

\begin{proof}
Smoothness and the chain rule show that $X(t)(z)$ and $Y(t)(z')$ 
are smooth. 
Their values belong to $T_zZ$ and $T_{z'}Z$, respectively.
We fix $t \in I$. 
Differentiating $K(A(t)p,z')=K(p,A^*(t)z')$
with respect to $t$, while keeping $p,z' \in Z$ fixed, gives
\[
	D_LK(A(t)p,z').\dot A(t)(p)= D_RK(p,A^*(t)z').\dot A^*(t)(z').
\]
We set $p=A(t)^{-1}z$. 
Then
$L_{X(t)}K(z,z')= D_RK(A(t)^{-1}z,A^*(t)z').\dot A^*(t)(z')$.
Equation \gzit{eq:adjoint-identity} also gives
$K(A(t)^{-1}z,z')=K(z,A^*(t)^{-1}z')$.
Differentiating with respect to $z'$ yields
\[
	D_RK(A(t)^{-1}z,q).v
	= D_RK(z,A^*(t)^{-1}q)(d\bigl(A^*(t)^{-1}\bigr)(q).v)
	\for 
	q \in Z, ~ v \in T_{q}Z.
\]
Taking $q=A^*(t)z'$ and $v=\dot A^*(t)(z')$, we obtain
\[
\begin{aligned}
	L_{X(t)}K(z,z')
	&=D_RK(z,z')(
	d\bigl(A^*(t)^{-1}\bigr)(A^*(t)z').\dot A^*(t)(z')) 
	=D_RK(z,z').Y(t)(z') \\
	&=R_{Y(t)}K(z,z').
\end{aligned}
\]
Thus, $Y(t)$ is a coherent adjoint of $X(t)$.
\end{proof}

Suppose that, for every initial point and 
every initial time $s \in I$, where $I \seq \Rz$ is a closed interval, the equations
\[
	\dot z(t)=X(t)(z(t))
	\AND
	\dot{z^\prime}(t)=-Y(t)(z'(t))
	\for 
	X(t),Y(t) \in \Kin(Z), ~ t \in I
\]
have unique solutions on all of $I$.
We denote their \bfi{transition maps} $U(t,s)$ and $V(t,s)$ by
\[
	U(t,s)z(s) \cq z(t),
	~~~
	V(t,s)z'(s) \cq z'(t),
\] 
We assume that the maps
\[
    (t,s,z) \mto U(t,s)z,
    ~~~
    (t,s,z) \mto V(t,s)z
\]
are smooth on $I \* I \* Z$. 
By definition of the transition maps,
\[
	\begin{aligned}
		\partial_tU(t,s)z&=X(t)(U(t,s)z),&
		~~~
		U(s,s)z&=z,\\
		\partial_tV(t,s)z'&=-Y(t)(V(t,s)z'),&
		~~~
		V(s,s)z'&=z'.
	\end{aligned}
\]

\begin{theorem}
\label{thm:coherent-evolution}
Let $I \seq  \Rz$ be a closed interval, and let $X(t)$ and $Y(t)$, $t \in I$, be smooth time-dependent vector fields satisfying
\begin{equation}
\label{eq:K-preserved-infinitesimal}
	L_{X(t)}K=R_{Y(t)}K.
\end{equation}

Then, for every fixed $s\in I$, the family $t \mto U(t,s)$ is 
a curve of smooth symmetries, with coherent adjoint
\begin{equation}
	\label{eq:coherent-adjoint-U}
	U(t,s)^*=V(s,t).
\end{equation}
More precisely,
\[
	K(U(t,s)z,z')=K(z,V(s,t)z')
	\for 
	s,t\in I, ~ z,z' \in Z.
\]
Its logarithmic derivative is $X(t)$:
\begin{equation}
\label{eq:logaritmic-derivative-U}
	\partial_tU(t,s) \circ U(s,t)=X(t).
\end{equation}
\end{theorem}

\begin{proof}
Uniqueness of solutions gives $U(t,r) U(r,s)=U(t,s)$ and $V(t,r)V(r,s)=V(t,s)$ for $r \in I$.
Indeed, either side gives the value at time $t$ of the solution 
with the prescribed initial value at time $s$. 
Together with $U(s,s)=V(s,s)=\id_Z$, these identities imply
$U(t,s)^{-1}=U(s,t)$ and $V(t,s)^{-1}=V(s,t)$.
Smoothness therefore makes all transition maps smooth 
diffeomorphisms.
Fix $s,t \in I$ and $z,z' \in Z$. 
We put $p(r) \cq U(r,s)z$ and $q(r) \cq V(r,t)z'$.
Differentiating with respect to $r$ gives
$\dot p(r)=X(r)(p(r))$ and $\dot q(r)=-Y(r)(q(r))$.
Hence, by \gzit{eq:K-preserved-infinitesimal},
\[
\begin{aligned}
	\frac{d}{dr}K(p(r),q(r))
	&=D_LK(p(r),q(r)).X(r)(p(r))
		-D_RK(p(r),q(r)).Y(r)(q(r)) \\
	&=L_{X(r)}K(p(r),q(r))
		-R_{Y(r)}K(p(r),q(r))
	=0.
\end{aligned}
\]
Thus, $K(p(r),q(r))$ is constant on $I$. 
For $r=s$ and $r=t$, we get $K(z,V(s,t)z')=K(U(t,s)z,z')$.
Therefore, $V(s,t)$ is an invertible coherent adjoint of 
$U(t,s)$, proving \gzit{eq:coherent-adjoint-U}.
Thus, $U(t,s)$ is a symmetry. 
Its smoothness shows that, for fixed $s$, it defines a curve of smooth symmetries.
Finally,
$\partial_tU(t,s)\bigl(U(s,t)z\bigr)=X(t)\bigl(U(t,s)U(s,t)z\bigr)
=X(t)(z)$.
This proves \gzit{eq:logaritmic-derivative-U}.
\end{proof}

Note that for fixed $s$, writing $B(t) \cq V(s,t)$, the adjoint family satisfies
\[
	\partial_tB(t)(z')=dB(t)(z').Y(t)(z').
\]
Indeed, differentiating $V(t,s)(B(t)z')=z'$ gives
\[
	-Y(t)(z')+dV(t,s)(B(t)z').\partial_tB(t)(z')=0.
\]
Applying $dB(t)(z')$ proves the asserted equation. 
It also gives
\[
	d\bigl(B(t)^{-1}\bigr)(B(t)z').\partial_tB(t)(z')
	=Y(t)(z'),
\]
in accordance with Theorem \ref{thm:coherent-motion-generator}. 

\pgph


A curve of smooth symmetries $A \cln I \to \sSym(Z)$ is called \bfi{unitary} if every $A(t)$ is a \bfi{unitary coherent map}.

\begin{corollary}
\label{cor:unitary-coherent-evolution}
Let $I \seq  \Rz$ be a closed interval, and let $X(t)$, $t \in I$, be a smooth time-dependent vector 
field satisfying
\begin{equation}
\label{eq:unitary-K-preserved-infinitesimal}
	L_{X(t)}K+R_{X(t)}K=0.
\end{equation}
Suppose that the equation 
\[
	\dot{z}(t)=X(t)(z(t))
\]
has unique global solutions on $I$, with smooth transition maps 
$U(t,s)$ with $s,t \in I$. 
Then
\[
	K(U(t,s)z,U(t,s)z')=K(z,z')
	\for 
	z,z' \in Z,
\]
and $t \mto U(t,s)$ is a curve of smooth symmetries for every 
fixed $s$, with
\[
	U(t,s)^*=U(s,t)=U(t,s)^{-1}.
\]
\end{corollary}

\begin{proof}
Set $Y(t)=-X(t)$. 
Then $L_{X(t)}K=R_{Y(t)}K$, and the equation generated by $-Y(t)$ is the equation generated by $X(t)$. 
Thus, $V=U$. Theorem~\ref{thm:coherent-evolution} now gives
\[
	K(U(t,s)z,z')=K(z,U(s,t)z').
\]
Replacing $z'$ by $U(t,s)z'$ yields $K(U(t,s)z,U(t,s)z')=K(z,z')$.
The inverse identity proves unitarity.
\end{proof}


\section{Coherent quantization}
\label{sec:coherent-quantization}

We recall the quantization theorem for coherent maps due to \sca{Neumaier \& Ghaani Farashahi} 
\cite[Theorems~3.11 and 3.12]{NGF2022}. 
To quantize a coherent vector field, we append the field to each derivative label. 
This defines a coherent map of the extended coherent space by Theorem \ref{thm:coherent-map-extended}. 
Its quantization gives the operator associated with the field in Theorem \ref{thm:intrinsic-representation}.

\begin{theorem}
\label{thm:finite-quantization}
Let $Z$ be a coherent space, and let $A$ be a coherent map with coherent adjoint $A^*$. 
Then there is a unique linear map
$\Gamma(A) \cln \Qz(Z) \to \Qz(Z)$ satisfying
\[
	\Gamma(A)\ket*{z}=\ket*{Az}
	\for 
	z \in Z.
\]
As an operator in $\ol{\Qz(Z)}$, it is densely defined and 
closable.
We have
\[
	\begin{aligned}
		\Gamma(AB)&=\Gamma(A)\Gamma(B)
		\for 
		A,B \in \Coh(Z), \\
		\Gamma(A^{-1})&=\Gamma(A)^{-1}
		\qquad
		\text{ if }
		A^{-1} \in \Coh(Z), \\
		\Gamma(A^*)&=\Gamma(A)^* 
		\qquad
		\text{ on } 
		\Qz(Z)
		\for 
		A \in \Coh(Z),\\
		\Gamma(1)&=1.
	\end{aligned}
\]
\end{theorem}

\subsection{Extended coherent spaces and quantization}
\label{subsec:extended-coherent-quantization}

Sums and products of quantum operators require a common invariant domain. 
In representation theory of Lie groups, the space of smooth vectors 
provides such a domain for the derived representation. 
\sca{Neeb} \cite{Neeb2010} develops this approach for 
infinite-dimensional Lie groups and studies its relation to the 
domains of iterated infinitesimal operators. 

\pgph


Under the kernel compatibility and smooth action hypotheses of 
\sca{Mérigon} et al. \cite[Definitions~5.3 and~5.4]{MNO2015}, the kernel vectors belong to the common domain of all iterated Lie derivatives \sca{Merigon} et al. \cite[Proposition 5.6]{MNO2015}. 
Their span is a core for the selfadjoint operators of 
\sca{Merigon} et al.\cite[Corollary 5.7]{MNO2015} and for the generators of the unitary one-parameter groups of 
\sca{Merigon} et al.\cite[Lemma 5.8]{MNO2015}. 
Theorem~\hyperref[itm:common-domain-core]{\ref*{thm:common-domain-finite-quantization}\ref*{itm:common-domain-core}} gives a core criterion for a quantized coherent map of the extended space. 
If the image of a coherent vector under this operator depends smoothly on its label, then the span of the original coherent vectors is a core for the operator.

\pgph


We write elements of $\vect(Z)^n$ as 
\[
	X_{1:n} \cq (X_1,\ldots, X_n).
\] 

\begin{proposition}
With the coherent product
\[
\wh{K}([z;X_{1:n}],[z';Y_{1:m}])
	\cq
		\bigl(
		L_{X_1}\cdots L_{X_n}
		R_{Y_1}\cdots R_{Y_m}
		K
		\bigr)(z,z'),
\]
the set 
\[
	\wh{Z}
	\cq 
	Z \* \bigcup_{n\geq0}\vect(Z)^n
	\with 
	\vect(Z)^0 \cq \{\emptyset\}
\]
is a coherent space, the \bfi{extended coherent space} of $Z$.
The space $\Qz(\wh{Z})$ of all complex linear combinations of $|z;X_{1:n}\>$ with $z \in Z$, $n \in \Nz_0$ and $X_i \in \vect(Z)$ for $i=1,\ldots,n$ is a quantum space of $\wh{Z}$, the 
\bfi{extended quantum space} of $Z$.
Furthermore,
\[
	\Qz(Z) \seq \Qz(\wh{Z}) \seq \ol{\Qz(Z)}.
\]
\end{proposition}

\begin{proof}
For $z=[z;X_{1:n}] \in \wh Z$, we put
$\ket*{z} \cq (\Lc_{X_1}\cdots\Lc_{X_n}\kappa)(z) \in \ol{\Qz(Z)}$.
These derivatives exist by Proposition~\ref{prop:smooth-kappa}.
For $n=0$, the empty product gives
$\ket*{[z;\emptyset]}=\ket*{z}$.
Differentiating the coherent product $K(z,z')=\kappa(z)^*\kappa(z')$ in its two arguments gives
$\wh K(z,z')=\braket*{z}{z'}$ for $z,z' \in \wh Z$.
Thus, $\wh K$ is Hermitian. 
For arbitrary $z_1,\ldots,z_N \in \wh Z$ and 
$c_1,\ldots,c_N \in \Cz$, we have
\[
    \sum_{a,b=1}^N \ol{c_a}c_b\,\wh K(z_a,z_b)
    =\left\|\sum_{a=1}^N c_a\ket*{z_a}\right\|^2 \geq 0.
\]
Hence, $\wh K$ is of positive type, and the complex linear span of
the vectors $\ket*{z}$ in $\ol{\Qz(Z)}$ is a quantum space of
$\wh Z$. The labels $[z;\emptyset]$ give
$\Qz(Z)\seq\Qz(\wh Z)$.
\end{proof}

For $A \in \Sym^{\infty}(Z)$, we define the pushforward by $A$ as
\begin{equation}
\label{eq:pushforward-A}
	A_*X \cq TA \circ X \circ A^{-1}
\end{equation}
where $TA$ is the tangential map of $A$.

\begin{theorem}
\label{thm:coherent-map-extended}
Let $A \in \Sym^{\infty}(Z)$ and $X\in\vect(Z)$. 
Then
\begin{enumerate}[ref=\textup{(\roman*)}]
	\item\label{itm:extended-coherent-map} 
	the maps $\wh A$ and $\wh X$ defined by
	\[
		\wh A[z;X_{1:n}] \cq [Az;A_*X_1,\ldots,A_*X_n],
	\]
	\[
		\wh X[z;X_{1:n}] \cq [z;X_{1:n},X] 
		\for 
		[z;X_{1:n}] \in \wh{Z}.
	\]
	are coherent maps on $\wh{Z}$;
	\item\label{itm:extended-pushforward} $A_{*}\vect(Z)=\vect(Z)$.
\end{enumerate}
\end{theorem}

\begin{proof}
We choose a smooth diffeomorphic coherent adjoint $A^*$. 
Differentiating
\[
	K(Az,z')=K(z,A^*z')
	\with 
	z,z' \in Z
\]
shows that $A_*X$ is coherent with coherent adjoint
$((A^*)^{-1})_*X^*$ for $X \in \vect(Z)$.
Applying the same argument to $A^{-1}$ gives $A_*\vect(Z)=\vect(Z)$, which proves \ref{itm:extended-pushforward}. 

Hence, $\wh A[z;X_{1:n}]\cq [Az;A_*X_1,\ldots,A_*X_n]$
defines a bijection of $\wh{Z}$. 
By iterative use of the chain rule,
$\wh K( \wh A[z;X_{1:n}],[z';Y_{1:m}])
=\wh K([z;X_{1:n}], \wh{A^*}[z';Y_{1:m}])$ for 
$[z;X_{1:n}],[z';Y_{1:m}] \in \wh{Z}$ so $\wh A$ is a coherent map on $\wh{Z}$.
Finally, 
$\wh K(\wh X[z;X_{1:n}],[z';Y_{1:m}])
	=
	\bigl(
		L_{X_1}\cdots L_{X_n}L_X
		R_{Y_1}\cdots R_{Y_m}K
	\bigr)(z,z')
$.
Left and right differentiations commute. 
Using $L_XK=R_{X^*}K$, we therefore obtain
\[
	\wh  K(\wh X[z;X_{1:n}],[z';Y_{1:m}])
	=
	\bigl(
		L_{X_1}\cdots L_{X_n}
		R_{Y_1}\cdots R_{Y_m}R_{X^*}K
	\bigr)(z,z')
	=\wh K([z;X_{1:n}],\wh{X^*}[z';Y_{1:m}]).
\]
This proves \ref{itm:extended-coherent-map}.
\end{proof}
We write $\wh{\Gamma}(C)$ for the quantization of
$C \in \Coh(\wh Z)$. 
By Theorem~\ref{thm:finite-quantization},
\[
    \wh{\Gamma}(1)=1,
    \qquad
    \wh{\Gamma}(CD)=\wh{\Gamma}(C)\wh{\Gamma}(D)
    \for C,D \in \Coh(\wh Z).
\]
If $C$ has a coherent inverse, then
\[
    \wh{\Gamma}(C)^{-1}=\wh{\Gamma}(C^{-1}).
\]

\begin{theorem}
\label{thm:common-domain-finite-quantization}
Let $C$ be a coherent map on $\wh Z$.
\begin{enumerate}[ref=\textup{(\roman*)}]
	\item\label{itm:common-domain-quantization} There is a unique complex-linear endomorphism $\wh{\Gamma}(C)$ of $\Qz(\wh Z)$ satisfying
	\[
	 	\wh{\Gamma}(C)|u\>=|Cu\> \for u \in \wh Z.
	\]
	It is densely defined and closable in $\ol{\Qz(Z)}$. 
	For every coherent adjoint $C^*$ of $C$,
	\[
		 \Qz(\wh Z)\seq \Dom\wh{\Gamma}(C)^*,
		 \qquad
		 \left.\wh{\Gamma}(C)^*\right|_{\Qz(\wh Z)}=\wh{\Gamma}(C^*).
	\]
	\item\label{itm:common-domain-core} Suppose that $\wh{\Gamma}(C) \circ \kappa$ is smooth as a map from $Z$ to $\ol{\Qz(Z)}$. 
	Then, for all $n\geq0$, $z \in Z$, and $X_1,\ldots,X_n\in\vect(Z)$,
	\[
		 \wh{\Gamma}(C)|z;X_1,\ldots,X_n\>
		 =\bigl(L_{X_1}\cdots L_{X_n}(\wh{\Gamma}(C)\circ\kappa)\bigr)(z).
	\]
	Moreover, $\Qz(Z)$ is a core for $\ol{\wh{\Gamma}(C)}$.
	\item\label{itm:common-domain-group-laws} Let 
	$A \in \Sym^{\infty}(Z)$. 
	Then the quantization of the coherent map $\wh A$ defined in
	Theorem~\hyperref[itm:extended-coherent-map]{\ref*{thm:coherent-map-extended}\ref*{itm:extended-coherent-map}} satisfies
	\[
	 	\wh{\Gamma}(\wh A)|z;X_{1:n}\>=|Az;A_*X_1,\ldots,A_*X_n\>
	\]
	and $\wh \Gamma(\wh A)|_{\Qz(Z)}=\Gamma(A)$.
\end{enumerate}
\end{theorem}

\begin{proof}
\ref{itm:common-domain-quantization} We apply Theorem~\ref{thm:finite-quantization} to $\wh Z$.
For $\phi,\psi \in \Qz(\wh Z)$, the adjoint pairing now gives
\[
	(\wh{\Gamma}(C)\phi)^*\psi=\phi^*\wh{\Gamma}(C^*)\psi.
\]
Since $\Qz(\wh Z)$ contains the dense subspace $\Qz(Z)$ of $\ol{\Qz(Z)}$, both $\wh{\Gamma}(C)$ and its adjoint are densely defined. 
Hence, $\wh{\Gamma}(C)$ is closable.

\ref{itm:common-domain-core} The smooth map
$z \mto (\kappa(z),\wh{\Gamma}(C)\kappa(z))$
takes values in the graph of $\ol{\wh{\Gamma}(C)|_{\Qz(Z)}}$.
This graph is a closed linear subspace, so it contains all iterated
derivatives of the displayed map. 
Applying $L_{X_1}\cdots L_{X_n}$ gives
$((L_{X_1}\cdots L_{X_n}\kappa)(z),\,
(L_{X_1}\cdots L_{X_n}(\wh{\Gamma}(C)\circ\kappa))(z))$.
The first component is $|z;X_{1:n}\>$ since pairing either
vector with $|w\>$ gives $(R_{X_1}\cdots R_{X_n}K)(w,z)$, and the vectors $|w\>$ have dense span. 
Thus, the first component belongs to the domain of
$\wh{\Gamma}(C)$. 
Since the pair also belongs to the graph of
$\ol{\wh{\Gamma}(C)}$, its second component equals
$\wh{\Gamma}(C)|z;X_1,\ldots,X_n\>$. 
This proves the differentiation formula.
The vectors $|z;X_1,\ldots,X_n\>$ span $\Qz(\wh Z)$. 
Hence,
$\wh{\Gamma}(C)|_{\Qz(Z)} \seq \wh{\Gamma}(C)
\seq \ol{\wh{\Gamma}(C)|_{\Qz(Z)}}$.
\ref{itm:common-domain-group-laws} By Theorem~\hyperref[itm:extended-coherent-map]{\ref*{thm:coherent-map-extended}\ref*{itm:extended-coherent-map}}, $\wh A$ acts coherently on $\wh Z$. 
Part \ref{itm:common-domain-quantization} gives the stated action on extended coherent vectors.
For $n=0$ it gives $\wh{\Gamma}(\wh A)|z\>=|Az\>=\Gamma(A)|z\>$.
Thus, the two operators agree on $\Qz(Z)$. 
Since $\wh{\Gamma}(\wh A) \circ \kappa=\kappa \circ A$ is smooth by
Proposition~\ref{prop:smooth-kappa}, \ref{itm:common-domain-core} gives the equality of closures. 
Finally, $(AB)_*X=A_*(B_*X)$, so the lifts preserve composition and the identity. 
\end{proof}

\begin{theorem}
\label{thm:intrinsic-representation}
The quantization $d\Gamma(X) \cq \wh{\Gamma}(\wh X)$ of $\wh X$ defined in Theorem~\ref{thm:coherent-map-extended} determines a real-linear map $X \mto d\Gamma(X)$ satisfying
\[
 [d\Gamma(X),d\Gamma(Y)]=d\Gamma(X \lp Y), 
\]
where 
\begin{equation}
\label{eq:brackets}
	X \lp Y \cq-[X,Y]=dX.Y-dY.X
	\AND
	[d\Gamma(X),d\Gamma(Y)] 
	\cq d\Gamma(X)d\Gamma(Y)-d\Gamma(Y)d\Gamma(X).
\end{equation}
For every coherent adjoint $X^*$ of $X$,
\begin{equation}
\label{eq:domain-adjoint}
 \Qz(\wh Z)\seq \Dom d\Gamma(X)^*,\qquad
 \left.d\Gamma(X)^*\right|_{\Qz(\wh Z)}=d\Gamma(X^*).
\end{equation}
The operator $d\Gamma(X)$ is densely defined and closable in $\ol{\Qz(Z)}$, with
\begin{equation}\label{eq:dGamma-closed}
 d\Gamma(X)|z\>=d\kappa(z)X(z),\qquad
 \ol{\left.d\Gamma(X)\right|_{\Qz(Z)}}=\ol{d\Gamma(X)}
\end{equation}
and
\begin{equation}
\label{eq:dGamma-restriction}
 d\Gamma(X)=\left.\ol{\left.d\Gamma(X)\right|_{\Qz(Z)}}
                   \right|_{\Qz(\wh Z)}.
\end{equation}
Thus, $\Qz(Z)$ is a core for $\ol{d\Gamma(X)}$.

Finally, $\ker d\Gamma=\Null(Z)$. Hence, $d\Gamma$ induces a faithful
representation of $\vect(Z)/\Null(Z)$ with the bracket $\lp$.
If $\kappa$ is tangent-injective, $d\Gamma$ is faithful on $\vect(Z)$.
\end{theorem}

\begin{proof}
By Theorem~\hyperref[itm:extended-coherent-map]{\ref*{thm:coherent-map-extended}\ref*{itm:extended-coherent-map}}, $X$ defines the coherent map
\[
 (z;X_1,\ldots,X_n) \mto (z;X_1,\ldots,X_n,X)
\]
on $\wh Z$, and every coherent adjoint $X^*$ of $X$ defines a coherent adjoint of this map. 
We put
\begin{equation}\label{eq:dGamma-Gamma}
 d\Gamma(X)\cq \wh{\Gamma}(\wh X).
\end{equation}
Theorem~\hyperref[itm:common-domain-quantization]{\ref*{thm:common-domain-finite-quantization}\ref*{itm:common-domain-quantization}} gives existence, uniqueness, closability, and \gzit{eq:domain-adjoint}. 
In particular, all coherent adjoints of $X$ induce the same operator on $\Qz(\wh Z)$.

For $n=0$, we get
$\wh{\Gamma}(\wh X)\kappa(z)=|z;X\>=(L_X\kappa)(z)=d\kappa(z)X(z)$.
Since $X$ and $\kappa$ are smooth, $L_X\kappa$ is smooth. 
Thus, Theorem~\hyperref[itm:common-domain-core]{\ref*{thm:common-domain-finite-quantization}\ref*{itm:common-domain-core}} gives
\gzit{eq:dGamma-closed} and, by restriction to $\Qz(\wh Z)$,
\gzit{eq:dGamma-restriction}. 
It also gives
$d\Gamma(X)|z;X_1,\ldots,X_n\>=(L_{X_1}\cdots L_{X_n}L_X\kappa)(z)$.
The right-hand side is real-linear in $X$. 
Since the extended coherent vectors span 
$\Qz(\wh Z)$, $X\mto d\Gamma(X)$ is real-linear. 
Applying the formula twice gives
\[
\begin{aligned}
 [d\Gamma(X),d\Gamma(Y)]|z;X_1,\ldots,X_n\>
   &=\bigl(L_{X_1}\cdots L_{X_n}(L_YL_X-L_XL_Y)\kappa\bigr)(z)
   =(L_{X_1}\cdots L_{X_n}L_{X \lp Y}\kappa)(z)\\
   &=d\Gamma(X \lp Y)|z;X_1,\ldots,X_n\>.
\end{aligned}
\]

Finally, $d\Gamma(X)=0$ implies $L_X\kappa=0$ by \gzit{eq:dGamma-closed}.
Conversely, if $L_X\kappa=0$, then every $L_{X_1}\cdots L_{X_n}L_X\kappa$
vanishes, so $d\Gamma(X)=0$. Hence,
\[
 \ker d\Gamma
 =\{X\in\vect(Z)\mid d\kappa(z)X(z)=0\text{ for every }z\in Z\}
 =\Null(Z).
\]
By Proposition~\ref{prop:coherent-algebra}, $\Null(Z)$ is a real Lie ideal.
The quotient representation is therefore faithful. If $\kappa$ is
tangent-injective, $d\kappa(z)X(z)=0$ for every $z$ implies $X=0$.
\end{proof}

\begin{corollary}
\label{cor:covariance}
Let $A \in \sSym(Z)$. 
Then, for every $X \in \vect(Z)$,
\begin{equation}
\label{eq:covariance-Gamma-dGamma}
	\wh{\Gamma}(\wh A)d\Gamma(X)\wh{\Gamma}(\wh A)^{-1}
	=d\Gamma(A_*X)
	\qquad
	\text{ on }
	\Qz(\wh{Z}).
\end{equation}
\end{corollary}

\begin{proof}
By Theorem~\hyperref[itm:extended-pushforward]{\ref*{thm:coherent-map-extended}\ref*{itm:extended-pushforward}}, $A_*X \in \vect(Z)$. 
All the operators in \gzit{eq:covariance-Gamma-dGamma} preserve 
$\Qz(\wh{Z})$. 
For $z \in Z$ and $X_1,\ldots,X_n \in \vect(Z)$,
\[
	\wh{\Gamma}(\wh A)d\Gamma(X)|z;X_{1:n}\>
	=|Az;A_*X_1,\ldots,A_*X_n,A_*X\>
	=d\Gamma(A_*X)\wh{\Gamma}(\wh A)|z;X_{1:n}\>.
\]
Extending by linearity and composing on the right with 
$\wh{\Gamma}(\wh A)^{-1}$ proves \gzit{eq:covariance-Gamma-dGamma}.
\end{proof}

\begin{corollary}
\label{cor:unitary}
Let $G$ be a subgroup of $\Diff(Z)$ such that
\[
	K(Az,Az')=K(z,z')
	\for 
	A \in G, ~ z,z'\in Z.
\]
Then, for every $A\in G$, the quantization $\Gamma(A)$ extends uniquely to a unitary operator $U_A$ on $\ol{Q(Z)}$. Moreover,
\[
	U_A Q(\wh Z)=Q(\wh Z),
	\qquad
	U_A|_{Q(\wh Z)}=\wh\Gamma(\wh A),
	\qquad
	\ol{\wh\Gamma(\wh A)}=U_A.
\]
The map $A \mto  U_A$ is a unitary representation of $G$.
\[
	U_{AB}=U_AU_B,
	\qquad
	U_{\id_Z}=\id_{\ol{\Qz(Z)}}.
\]
\end{corollary}

\begin{proof}
Substituting $A^{-1}w$ for $w$ in the kernel identity gives
$K(Az,z')=K(z,A^{-1}z')$.
Thus, $A^{-1}$ is a diffeomorphic coherent adjoint of $A$. 
Since $A$ preserves $K$ and is bijective, $\Gamma(A)$ is a surjective isometry of $Q(Z)$. 
It therefore extends uniquely to a unitary operator $U_A$ on $\ol{Q(Z)}$.

By Theorem~\hyperref[itm:common-domain-group-laws]{\ref*{thm:common-domain-finite-quantization}\ref*{itm:common-domain-group-laws}}, $\wh\Gamma(\wh A)$ extends $\Gamma(A)$, and by part~\ref{itm:common-domain-core} these operators have the same closure. 
Hence, $\wh\Gamma(\wh A)=U_A|_{Q(\wh Z)}$. 
Applying this also to $A^{-1}$ gives $U_AQ(\wh Z)=Q(\wh Z)$. 
Finally,
$U_AU_B|z\>=|ABz\>=U_{AB}|z\>$ and
$U_{\id_Z}|z\>=|z\>$.
The coherent vectors have dense linear span, so these identities hold on $\ol{Q(Z)}$.
\end{proof}

\subsection{Quantization of infinitesimal coherent maps}

We now apply the above results to curves of smooth symmetries considered in Subsection~\ref{subsec:inf-coh-maps}, page \pageref{not:coherent-motion}. 
Proposition~\ref{prop:quantized-unitary-motion} shows that $-X_t$ is a coherent adjoint of $X_t$.
Proposition~\ref{prop:newprop} quantizes the given family, and Proposition~\ref{prop:quantized-unitary-evolution} quantizes the evolution generated by a vector field that infinitesimally preserves the coherent product. 

\pgph


Let $A \cln I \to \sSym(Z)$ be a curve of unitary smooth symmetries on a closed interval $I \seq \Rz$ such that
\begin{equation}
\label{eq:at}
	(t,z)  \mto   A_tz,
	\qquad
	(t,z)  \mto   A_t^{-1}z
	\for 
	z \in Z, ~ t \in I
\end{equation}
are smooth.
We define the time-dependent smooth vector fields
\[
	X_t \cq \partial_tA_t\circ A_t^{-1}
	\for 
	t \in I.
\]

\begin{proposition}
\label{prop:quantized-unitary-motion}
Let $A \cln I \to \sSym(Z)$ be a curve of unitary smooth symmetries on a closed interval $I \seq \Rz$ such that the maps in \gzit{eq:at} are smooth. 
Then $-X_t$ is a coherent adjoint of $X_t$, and we have
\begin{equation}
\label{eq:LXtRXt}
	L_{X_t}K+R_{X_t}K=0.
\end{equation}
\end{proposition}

\begin{proof}
The assumed smoothness implies that
$X_t=\dot A_t\circ A_t^{-1}$ is a smooth time-dependent vector field. 
Since $A_t$ preserves the coherent product,
$K(A_tp,A_tq)=K(p,q)$ for $p,q\in Z$.
Differentiating with respect to $t$, while keeping $p,q$ fixed, and using $\dot A_t(p)=X_t(A_tp)$,
gives $0=(L_{X_t}K+R_{X_t}K)(A_tp,A_tq)$.
As $A_t$ is surjective, $L_{X_t}K+R_{X_t}K=0$.
Thus, $L_{X_t}K=R_{-X_t}K$, so $-X_t$ is a coherent adjoint of $X_t$. 
\end{proof}

\begin{proposition}
\label{prop:newprop}
For every $t\in I$, the quantization of $A_t$ extends uniquely to a unitary operator $U_t$ on $\ol{\Qz(Z)}$ satisfying
\[
	U_t|z\>=|A_tz\>,\qquad
	U_t\Qz(Z)=\Qz(Z),\qquad
	U_t\Qz(\wh Z)=\Qz(\wh Z).
\]
The family $t\mto U_t$ is strongly continuous. 
For every $\psi\in\Qz(\wh Z)$, the curve $t\mto U_t\psi$ is smooth in $\ol{\Qz(Z)}$ and satisfies
\begin{equation}
\label{eq:diff-equation-U-t}
	\frac{d}{dt}U_t\psi=d\Gamma(X_t)U_t\psi,
\end{equation}
with one-sided derivatives at the endpoints of $I$.
\end{proposition}

\begin{proof}
The $A_t$ generate a group of smooth coherent isometries, with coherent involution given by inversion. 
Corollary~\ref{cor:unitary} gives the unitary extensions $U_t$ and invariance of $\Qz(\wh Z)$.

Since $U_t|z\>=|A_tz\>$ and $A_t$ is bijective, $U_t\Qz(Z)=\Qz(Z)$.

By Proposition~\ref{prop:smooth-kappa} and \gzit{eq:at}, the map $F(t,z)\cq\kappa(A_tz)$ is smooth. From Proposition~\ref{prop:quantized-unitary-motion}, we conclude $X_t\in\vect(Z)$. 
Since $\partial_t A_tz=X_t(A_tz)$, the chain rule and Theorem~\ref{thm:intrinsic-representation} give
\[
	\partial_tF(t,z)=(L_{X_t}\kappa)(A_tz)=d\Gamma(X_t)U_t|z\>.
\]

For $Y_1,\ldots,Y_n\in\vect(Z)$, we put $P=L_{Y_1}\cdots L_{Y_n}$, $Y_j^t=(A_t)_*Y_j$, and $P_t=L_{Y_1^t}\cdots L_{Y_n^t}$. Theorem~\ref{thm:common-domain-finite-quantization} gives
$U_t|z;Y_{1:n}\>=(P_zF)(t,z)=|A_tz;Y^t_{1:n}\>$, 
where $P_z$ acts on the $Z$ variable. 
Since $F$ is smooth and $P$ is independent of $t$, $\partial_t$ commutes with $P_z$. 
At fixed $t$, the chain rule gives $P(f\circ A_t)=(P_t f)\circ A_t$. Thus,
\[
	\frac{d}{dt}U_t|z;Y_{1:n}\>
	=(P_z\partial_tF)(t,z)
	=(P_t L_{X_t}\kappa)(A_tz)
	=d\Gamma(X_t)|A_tz;Y^t_{1:n}\>,
\]
where the last equality is the action of $d\Gamma(X_t)$ on extended coherent vectors from Theorem~\ref{thm:intrinsic-representation}. These vectors $|z;Y_{1:n}\>$ span $\Qz(\wh Z)$, which proves \gzit{eq:diff-equation-U-t} on the stated domain.

In particular, $t \mto U_t\phi$ is continuous for every 
$\phi\in\Qz(Z)$. 
For $\psi\in\ol{\Qz(Z)}$, unitarity now gives
$\|(U_t-U_{t_0})\psi\| \leq 2\|\psi-\phi\|+\|(U_t-U_{t_0})\phi\|$.
Since $\Qz(Z)$ is dense in $\ol{\Qz(Z)}$, this proves strong continuity.
\end{proof}

\begin{proposition}
\label{prop:quantized-unitary-evolution}
Let $I \seq \Rz$ be a closed interval, and let $X_t$ be a smooth time-dependent vector field, satisfying \gzit{eq:LXtRXt}.
Suppose that, for every $s \in I$ and $z \in Z$, the curve
$t \mto U(t,s)z$ is the unique solution on $I$ of
\[
    \dot u(t)=X_t(u(t)), \qquad u(s)=z.
\]
Assume that $(t,s,z) \mto U(t,s)z$ is smooth on $I \* I \* Z$.
Thus,
\begin{equation}
\label{eq:diff-phi}
	U(s,s)=1,
	\qquad
	\partial_tU(t,s)z=X_t(U(t,s)z)
	\for
	z \in Z, ~ s,t \in I.
\end{equation}
Then, for fixed $s$, the family $t \mto U(t,s)$ is a curve of unitary smooth symmetries.
For each $t,s$, the quantization of $U(t,s)$ extends uniquely to a unitary operator $V(t,s)$ on $\ol{\Qz(Z)}$, satisfying
\[
	V(t,s)|z\>=|U(t,s)z\>,
	\qquad
	V(t,s)\Qz(\wh Z)=\Qz(\wh Z).
\]
These operators form a strongly continuous unitary evolution:
\[
	V(s,s)=1,
	\qquad
	V(t,r)V(r,s)=V(t,s),
	\AND
	V(t,s)^*=V(s,t).
\]
For every $\psi\in\Qz(\wh Z)$,
\begin{equation}
\label{eq:diff-U-t}
	\partial_tV(t,s)\psi
	=d\Gamma(X_t)V(t,s)\psi.
\end{equation}
\end{proposition}

\begin{proof}
By Corollary~\ref{cor:unitary-coherent-evolution}, for fixed $s$ the family $t\mto U(t,s)$ is a curve of unitary smooth symmetries, 
with $U(t,s)^{-1}=U(s,t)$.
For fixed $s\in I$, the family $A_t \cq U(t,s)$ and its inverse 
act smoothly. 
Its logarithmic derivative is $\partial_tU(t,s)\circ U(s,t)=X_t$.
Proposition~\ref{prop:newprop} therefore supplies the unique unitary operators $V(t,s)$, satisfying $V(t,s)|z\>=|U(t,s)z\>$ and $\partial_tV(t,s)\psi=d\Gamma(X_t)V(t,s)\psi$ for 
$\psi\in\Qz(\wh Z)$.
This proves \gzit{eq:diff-U-t}.
For fixed $r,s \in I$ and $z \in Z$, the curves
$t \mto U(t,r)U(r,s)z$ and $t \mto U(t,s)z$ solve the same
equation and have the same value $U(r,s)z$ at time $r$.
Uniqueness therefore gives $U(t,r)U(r,s)=U(t,s)$.
Multiplicativity of quantization and unitarity consequently yield
$V(s,s)=1$, $V(t,r)V(r,s)=V(t,s)$, and $V(t,s)^*=V(s,t)$. 

It remains to prove continuity. 
For every $z\in Z$, the map
$(t,s)\mto V(t,s)|z\>=\kappa(U(t,s)z)$ is smooth. 
Hence, $(t,s)\mto V(t,s)\phi$ is continuous for every $\phi\in\Qz(Z)$. 
For $\psi\in\ol{\Qz(Z)}$, unitarity gives
$\|(V(t,s)-V(t_0,s_0))\psi\|
\leq 2\|\psi-\phi\|+\|(V(t,s)-V(t_0,s_0))\phi\|$.
Since $\Qz(Z)$ is dense in $\ol{\Qz(Z)}$, this proves continuity in $\ol{\Qz(Z)}$.
\end{proof}

\subsection{Enveloping algebras and shadow operators}
\label{subsec:enveloping-shadow-operators}

Quantum observables and Hamiltonians often involve sums and 
products of elementary operators. 
Subsection~\ref{subsec:extended-coherent-quantization} constructs finite and infinitesimal 
coherent quantizations on a common invariant domain. 
We now describe the algebra generated by these operators and 
develop a calculus for their coherent matrix elements. 
This allows composite operators to be studied directly 
through the coherent product.
A description of operators by shadows was developed by 
\sca{Neumaier \& Ghaani Farashahi} \cite{NGF2022}, who characterize 
shadows by admissibility and describe their  behaviour under 
adjunction. 
The transfer of operator multiplication to functions goes back to 
\sca{Berezin} \cite{Berezin1974}. 

\pgph


Here we combine these ideas with the common invariant domain 
constructed above, allowing products of possibly unbounded 
operators on a smooth coherent manifold. 
The universal enveloping algebra records the infinitesimal 
commutator relations, and a crossed product incorporates finite 
transformations and their covariance. 
Their representation gives a unital operator algebra with formal 
adjoints. 
We show that each operator in this algebra is uniquely determined 
on the whole common domain by its shadow. 
Operator composition and formal adjunction therefore induce an 
algebra structure with involution on the space of these smooth 
kernels.
The resulting calculus has a direct computational use. 
Finite coherent transformations act on shadows by substitution, 
infinitesimal operators act by differentiation, and formal 
adjunction acts by conjugation and interchange of the arguments. 
Matrix elements of products can thus be evaluated by successive 
operations on the coherent product. 

\pgph


We write $\Cz\Lz \cq \Lz \oplus i \Lz$ for the complexification of a real Lie algebra $\Lz$.
The \bfi{universal enveloping algebra} of $\Lz$ is the complex algebra
\[
	\Uz(\Lz) \cq U\bigl(\Cz\Lz\bigr)\big/ I(\Lz),
\]
where
\[
	U\bigl(\Lz\bigr) \cq \bigoplus_{n \geq 0}\Lz^{\otimes n}
\]
and $I(\Lz)$ is the two-sided ideal generated by
\[
    X \otimes Y-Y \otimes X-(X \lp Y)
    \for X,Y \in \Cz\Lz,
\]
where $\lp$ is extended complex-bilinearly to $\Cz\Lz$.
We write $X$ also for its canonical image in $\Uz(\Lz)$. 

\pgph


Throughout this section, we consider the Lie algebra $\Lz=\vect(Z)$ with the Lie product opposite to the Lie bracket,
\[
	X \lp Y \cq -[X,Y] 
	\for 
	X,Y \in \vect(Z).
\]

\begin{proposition}
\label{prop:enveloping-representation}
Let $X \in \vect(Z)$. 
Then the map $X \mto d\Gamma(X)$ extends uniquely to a unital algebra homomorphism
\[
	\pi \cln \Uz(\vect(Z)) \to \End_{\Cz}\bigl(\Qz(\wh{Z})\bigr).
\]
For $X_1,\ldots,X_n \in \vect(Z)$ and $z \in Z$,
\begin{equation}
\label{eq:Pi-Gamma}
	\pi(X_1\cdots X_n)
	=d\Gamma(X_1)\cdots d\Gamma(X_n)
	\AND
	\pi(X_1\cdots X_n)|z\>
	=(\Lc_{X_n}\cdots\Lc_{X_1}\kappa)(z).
\end{equation}
In particular,
\begin{equation}
\label{eq:ext-span}
	\Qz(\wh{Z})
	=\Span_{\Cz}
	\{\pi(u)|z\> \mid u \in \Uz(\vect(Z)),\ z \in Z\}.
\end{equation}
\end{proposition}

\begin{proof}
Theorem~\ref{thm:intrinsic-representation} gives a real-linear 
map $d\Gamma \cln  \vect(Z) \to \End_{\Cz}(\Qz(\wh{Z}))$
satisfying
\begin{equation}
\label{eq:Gamma-Lie}
	[d\Gamma(X),d\Gamma(Y)]
	=-d\Gamma([X,Y])
	=d\Gamma(X \lp Y)
	\for
	X,Y \in \vect(Z).
\end{equation}
Its complex-linear extension to $\Cz\vect(Z)$ satisfies 
\gzit{eq:Gamma-Lie} for complexified vector fields as well.
We extend $d\Gamma$ multiplicatively to the tensor algebra. 
By \gzit{eq:Gamma-Lie}, each generator of the ideal is mapped to
$d\Gamma(X)d\Gamma(Y)-d\Gamma(Y)d\Gamma(X)+d\Gamma([X,Y])=0$.
Hence, the extension vanishes on the ideal and factors 
through a unital complex algebra homomorphism $\pi$. 
Since $\Uz(\vect(Z))$ is generated by $\Cz\vect(Z)$, this 
homomorphism is unique.
From Theorem~\ref{thm:intrinsic-representation}, we know
$d\Gamma(X)|z;X_1,\ldots,X_n\>=|z;X_1,\ldots,X_n,X\>$.
Iterating proves \gzit{eq:Pi-Gamma} on coherent 
vectors. 
These iterated derivatives span $\Qz(\wh{Z})$, which proves 
\gzit{eq:ext-span}.
\end{proof}

We call $G$ a \bfi{coherent $*$-group} of $Z$ if it is equipped with an involution $A \mto A^*$ satisfying
\[
	1^*=1,
	\qquad
	A^{**}=A,
	\qquad
	(AB)^*=B^*A^*,
	\AND
	K(Az,z')=K(z,A^*z')
\]
for all $A,B \in G$ and $z,z' \in Z$. 
If $G$ is a subgroup of $\Diff(Z)$, we call it a 
\bfi{smooth coherent $*$-group}.

\begin{proposition}
A smooth coherent $*$-group $G$ of $Z$ is a subgroup of $\sSym(Z)$.
\end{proposition}

\begin{proof}
Let $A\in G$. 
By definition, $K(Az,z')=K(z,A^*z')$ for $z,z' \in Z$.
Hence, $A$ is coherent with coherent adjoint $A^*$. 
Since $G$ is a $*$-group,
$A^*(A^{-1})^*=(A^{-1}A)^*=1$ and $(A^{-1})^*A^*=(AA^{-1})^*=1$.
Thus, $(A^{-1})^*=(A^*)^{-1}$ and
$K(A^{-1}z,z')=K\bigl(z,(A^*)^{-1}z'\bigr)$.
Therefore, $A^{-1}$ is coherent as well. 
Since the group operation in $G$ is composition, the assertion follows.
\end{proof}

Let $G$ be a smooth coherent $*$-group of $Z$. 
By Theorem~\hyperref[itm:extended-pushforward]{\ref*{thm:coherent-map-extended}\ref*{itm:extended-pushforward}},
\[
	A_*\vect(Z)=\vect(Z)
	\for 
	A \in G.
\]
The pushforward preserves Lie brackets and therefore induces an 
automorphism
\[
	\alpha_A \in \Aut_{\Cz}(\Uz(\vect(Z))),
	\qquad
	\alpha_A(X) \cq A_*X.
\]
Moreover,
\[
	\alpha_{AB}=\alpha_A\alpha_B
	\for
	B \in G.
\]

Let $\Cz[G]$ be the \bfi{group algebra} of $G$ consisting of the finitely supported complex-valued functions on $G$. 
For $A \in G$, let
\[
\delta_A(B) \cq
	\begin{cases}
		1,&B=A,\\
		0,&B\ne A.
	\end{cases}
\]
We equip the vector space
\[
	\Exp(G)
	\cq \Uz(\vect(Z)) \otimes_{\Cz}\Cz[G]
\]
with the multiplication obtained by bilinear extension of 
\[
	(u\otimes\delta_A)(v\otimes\delta_B)
	\cq u\,\alpha_A(v) \otimes \delta_{AB}.
\]
It is a unital algebra called the \bfi{exponomial algebra} of $G$.
The canonical embeddings
\[
	\iota(u) \cq u \otimes \delta_{\id_Z},
	\qquad
	j(A) \cq 1_{\Uz(\vect(Z))} \otimes \delta_A
\]
satisfy the covariance relation
\[
	j(A)\,\iota(u)\,j(A)^{-1}
	=\iota\bigl(\alpha_A(u)\bigr).
\]

For $A \in G$, let $\wh A$ be the coherent map on $\wh Z$
defined in Theorem~\hyperref[itm:extended-coherent-map]{\ref*{thm:coherent-map-extended}\ref*{itm:extended-coherent-map}}.
We write 
\[
	\wh{\Gamma}(A) \cq \wh{\Gamma}(\wh A).
\]

\begin{proposition}
\label{prop:expolynomial-representation}
Let $G$ be a smooth coherent $*$-group of $Z$. 
Then there is a unique unital complex algebra homomorphism
\[
	\Pi \cln \Exp(G) \to \End_{\Cz}(\Qz(\wh{Z}))
\]
such that
\begin{equation}
	\label{eq:Pi-pi-Gamma}
	\Pi(u \otimes \delta_A)=\pi(u)\wh{\Gamma}(A)
	\for
	A \in G, ~ u \in \Uz(\vect(Z)).
\end{equation}
Its image is the unital complex algebra $\Az(G)\cong\Exp(G)/\ker\Pi$ generated by
$d\Gamma(X)$ with $X \in \vect(Z)$ and $\wh{\Gamma}(A)$ with 
$A \in G$.
Every $T \in \Az(G)$ has a unique adjoint $T^* \in \Az(G)$, characterized by
\[
	(T\phi)^*\psi
	=\phi^*(T^*\psi)
	\for 
	\phi,\psi \in \Qz(\wh{Z}).
\]
The operation $*$ is a conjugate-linear involution satisfying
\begin{equation}
\label{eq:T-U}
	(TU)^*=U^* T^*
	\for
	U \in \Az(G),
\end{equation}
\begin{equation}
\label{eq:Gamma-dGamma}
	\wh{\Gamma}(A)^*=\wh{\Gamma}(A^*),
	\qquad
	d\Gamma(X)^*=d\Gamma(X^*)
	\for 
	X \in \vect(Z), ~ A \in G.
\end{equation}
These identities are independent of the chosen coherent 
adjoints. 
Every $T \in \Az(G)$, regarded as an operator in $\ol{\Qz(Z)}$ with 
domain $\Qz(\wh{Z})$, is closable.
\end{proposition}

\begin{proof}
Complex linearity and multiplication extend equation
\gzit{eq:covariance-Gamma-dGamma} to
$\wh{\Gamma}(A)\pi(v)\wh{\Gamma}(A)^{-1}=\pi(\alpha_A(v))$
for $v \in \Uz(\vect(Z))$.
Consequently,
$
\pi(u)\wh{\Gamma}(A)\pi(v)\wh{\Gamma}(B)
=\pi(u)\pi(\alpha_A(v))\wh{\Gamma}(AB)
=\pi(u\alpha_A(v))\wh{\Gamma}(AB)
$.
Thus, \gzit{eq:Pi-pi-Gamma} defines a unital algebra 
homomorphism. 
Its values on the generators determine it uniquely and identify 
its image with $\Az(G)$. 

Theorems~\ref{thm:common-domain-finite-quantization} and
\ref{thm:intrinsic-representation} supply adjoints for 
the generators. 
These adjoints belong to $\Az(G)$, since $A^*\in G$ and 
$X^* \in \vect(Z)$. 
Reversing products and conjugating coefficients therefore supplies 
an adjoint for every polynomial expression in the generators.
Suppose $S_1,S_2 \in \Az(G)$ are formal adjoints of the same operator $T$. 
Then $\phi^*(S_1-S_2)\psi=0$ for $\phi,\psi \in \Qz(\wh{Z})$.
Density of $\Qz(\wh{Z})$ in $\ol{\Qz(Z)}$ gives 
$S_1\psi=S_2\psi$ for every $\psi \in \Qz(\wh{Z})$. 
Hence, the adjoint is unique. 
Now \gzit{eq:T-U} and \gzit{eq:Gamma-dGamma} follow from the same 
uniqueness.

Finally, 
$\Qz(\wh{Z}) \seq \operatorname{Dom}(T^*)$ and
$T^*|_{\Qz(\wh{Z})}=T^*$, where $T^*$ on the left side is the Hilbert-space adjoint. 
Its domain is dense, so $T$ is closable.
\end{proof}

The \bfi{shadow} of $T \in \Az(G)$ is the map 
$S_T \cln Z \* Z \to \Cz$ defined by
\[
	S_T(z,z')
	\cq 
	\bra*{z}T\ket*{z'}
	\for z,z' \in Z.
\]

\begin{proposition}
\label{prop:shadow-algebra}
Let $G$ be a smooth coherent $*$-group of $Z$.  
Then the map $\Sh \cln \Az(G) \to C^\infty(Z \* Z,\Cz)$ defined by $\Sh T \cq S_T$ is complex-linear and injective. 
The operations
\[
	S_TS_U \cq S_{TU},
	\qquad
	S_T^* \cq S_{T^*}
	\for 
	T,U \in \Az(G)
\]
turn $\Sh \Az(G)$ into a unital complex algebra with involution, 
and $\Sh$ is an isomorphism onto this algebra. 
Its identity is the shadow $S_1=K$, and we have
\begin{equation}
\label{eq:S-t}
	S_T^*(z,z')=\ol{S_T(z',z)}
	\for
	z,z' \in Z.
\end{equation}
\end{proposition}

\begin{proof}
Every $T \in \Az(G)$ is a finite complex linear combination of 
operators $\pi(X_1\cdots X_n)\wh{\Gamma}(A)$ for 
$X_1,\ldots,X_n \in \vect(Z)$, and $A \in G$.
For such an operator, 
Proposition~\ref{prop:enveloping-representation} gives, for 
$z \in Z$, $T\ket*{z}=(L_{X_n}\cdots L_{X_1}\kappa)(Az)$.
The right-hand side is a smooth $\ol{\Qz(Z)}$-valued map. 
Hence, the map
$z \mto T\ket*{z}$ is smooth for every $T \in \Az(G)$. 
Since $z \mto  |z\>$ and $z' \mto  T|z'\>$ are smooth and the inner product is continuous and real-bilinear, their pairing 
$(z,z') \mto \< z|T|z'\>$ is smooth.
Thus, $S_T \in C^\infty(Z \* Z,\Cz)$. 
Complex linearity follows from linearity of the inner product in 
its second argument.
Suppose $S_T=0$. 
Then, for $z,z'\in Z$,
$0=S_T(z,z')
=\<z|T|z'\>
=(T^*\ket*{z})^*\ket*{z'}
$.
The coherent vectors have a dense complex linear span in 
$\ol{\Qz(Z)}$. 
Therefore, $T^*\ket*{z}=0$ for $z \in Z$.
For every $\psi \in \Qz(\wh{Z})$, it follows that
$\bra*{z}(T\psi)=(T^*\ket*{z})^*\psi=0$ for $z \in Z$.
Since $\Qz(Z)$ is dense in $\ol{\Qz(Z)}$, we get $T\psi=0$. 
Hence, $T=0$, proving injectivity.
The algebra and involution of $\Az(G)$ can therefore be 
transported uniquely to $\Sh \Az(G)$. 
The identity is $S_{1_{\Az(G)}}=K$, and we also have
$
S_{T^*}(z,z')
=\bra*{z}T^*\ket*{z'}
=\ol{\<z'|T|z\>}
=\ol{S_T(z',z)}
$
for $z,z' \in Z$.
This proves \gzit{eq:S-t}. 
\end{proof}

Let $G$ be a smooth coherent $*$-group of $Z$. 
For $T \in \Az(G)$, we define the right and left 
\bfi{shadow operators} by
\begin{equation}
	\label{eq:shadow-actions}
	\Oc(T)S_B \cq S_{BT},
	\qquad
	\Oc_L(T)S_B \cq S_{TB}
	\for 
	B \in \Az(G).
\end{equation}
They are well-defined complex-linear endomorphisms of $\Sh \Az(G)$.
For $f\in C^\infty(Z  \* Z,\Cz)$, $A \in G$, and $X \in \vect(Z)$, 
we put
\begin{equation}
	\label{eq:P-A}
	(P_Af)(z,z') \cq f(z,Az')
	\for z,z' \in Z.
\end{equation}

\begin{theorem}
	\label{thm:kernel-realization}
	Let $G$ be a smooth coherent $*$-group of $Z$.
	Let $T,U \in \Az(G)$, $X \in \vect(Z)$, $A \in G$, and let 
	$f \in C^\infty(Z \* Z, \Cz)$. 
	Then the maps $\Oc$ and $\Oc_L$ are faithful and 
	complex-linear, 
	with
	\begin{equation}
		\begin{gathered}
			\Oc(TU)=\Oc(U)\Oc(T),
			\qquad
			\Oc_L(TU)=\Oc_L(T)\Oc_L(U),\\
			\Oc_L(T)\Oc(U)=\Oc(U)\Oc_L(T),
			\qquad
			\Oc(T)K=\Oc_L(T)K=S_T.
		\end{gathered}
		\label{eq:shadow-composition}
	\end{equation}
	On $\Sh(\Az(G))$, for $z,z' \in Z$,
	\begin{equation}
		\begin{aligned}
			\Oc(\wh{\Gamma}(A))f(z,z')&=f(z,Az'),
			&\Oc(d\Gamma(X))f&=R_Xf,\\
			\Oc_L(\wh{\Gamma}(A))f(z,z')&=f(A^*z,z'),
			&\Oc_L(d\Gamma(X))f&=L_{X^*}f.
		\end{aligned}
		\label{eq:shadow-generators}
	\end{equation}
	In particular, for $A \in G$, $z,z' \in Z$, and 
	$X_1,\ldots,X_n \in \vect(Z)$,
	\begin{equation}
		\Sh\!\left(\wh{\Gamma}(A)d\Gamma(X_1)\cdots d\Gamma(X_n)\right)
		=R_{X_n}\cdots R_{X_1}K(z,Az').
		\label{eq:mixed-shadow}
	\end{equation}
\end{theorem}

\begin{proof}
Associativity in $\Az(G)$ gives the first three identities in
\gzit{eq:shadow-composition}. Taking $B=1_{\Az(G)}$ in
\gzit{eq:shadow-actions} gives the last one. It also proves
faithfulness by injectivity of $\Sh$.
For a finite transformation and $z,z' \in Z$,
$
(P_AS_B)(z,z')
=S_B(z,Az')
=\mel{z}{B}{Az'}
=\<z|B \wh{\Gamma}(A)|z'\>
=S_{B\wh{\Gamma}(A)}(z,z')
$.
For differentiation, using $S_B(z,z')=(B^*|z\>)^*|z'\>$, 
we obtain
\[
	(R_XS_B)(z,z')
	=(B^*|z\>)^*d\kappa(z')X(z')
	=(B^*|z\>)^*d\Gamma(X)|z'\>
	=\< z|B\,d\Gamma(X)|z'\> 
	=S_{B\,d\Gamma(X)}(z,z').
\]
For the left action, Theorem 
\ref{thm:common-domain-finite-quantization} gives
\[
	S_{\wh{\Gamma}(A)B}(z,z')
	=(\wh{\Gamma}(A^*)|z\>)^*B|z'\>
	=\< A^*z|B|z'\>
	=S_B(A^*z,z').
\]
Similarly, Theorem \ref{thm:intrinsic-representation} gives
\[
	S_{d\Gamma(X)B}(z,z')
	=(d\Gamma(X^*)|z\>)^*B|z'\>
	=(d\kappa(z)X^*(z))^*B|z'\>
	=(L_{X^*}S_B)(z,z').
\]
This proves \gzit{eq:shadow-generators}.

Finally, the antihomomorphism property of $\mathcal O$ implies
\[
	\operatorname{Sh}\!\left(
	\wh{\Gamma}(A)d\Gamma(X_1)\cdots d\Gamma(X_n)
	\right)
	=
	\mathcal O(d\Gamma(X_n))\cdots
	\mathcal O(d\Gamma(X_1))
	\mathcal O(\wh{\Gamma}(A))K
	=R_{X_n}\cdots R_{X_1}(P_AK).
\]
Since $(P_AK)(z,z')=K(z,Az')$, this is 
\gzit{eq:mixed-shadow}, with differentiation applied to the 
whole composite function.
\end{proof}

By \gzit{eq:shadow-composition},
\[
    [\Oc(T),\Oc(U)]=\Oc(UT-TU)
    \for T,U \in \Az(G).
\]
Thus, $\Oc$ is a representation of $\Az(G)$ whose Lie product is $(T,U) \mto UT-TU$. 
In particular, for $X \in \vect(Z)$
\[
	X \mto \Oc(d\Gamma(X))=R_X|_{\Sh \Az(G)}
\] 
is a representation of $\vect(Z)$ with bracket $[\cdot,\cdot]$:
\[
	[\Oc(d\Gamma(X)),\Oc(d\Gamma(Y))]
	=\Oc(d\Gamma([X,Y]))
	\for 
	X,Y \in \vect(Z).
\]
For finite transformations and covariance the corresponding 
identities on $\Sh \Az(G)$ are, using \gzit{eq:P-A} and 
\gzit{eq:pushforward-A},
\[
	P_AP_B=P_{BA},
	\qquad
	P_A^{-1}R_XP_A=R_{A_*X}
\]
for $A,B \in G$ and $X \in \vect(Z)$.

\begin{corollary}
\label{cor:formal-shadow-action}
Let $G$ be a smooth coherent $*$-group of $Z$.
Then there is a unique unital algebra antihomomorphism
\[
	\wt{\Oc} \cln \Exp(G) \to \End_{\Cz}(C^\infty(Z \* Z,\Cz))
\]
such that 
\begin{equation}
	\label{eq:O-wt}
	\wt{\Oc}(X)=R_X 
	\AND 
	\wt{\Oc}(j(A))=P_A
	\for
	X \in \vect(Z), ~ A \in G.
\end{equation}
It satisfies, for $e \in \Exp(G)$ and $B \in \Az(G)$,
\begin{equation}
	\begin{gathered}
		\wt{\Oc}(e)K=\Sh \Pi(e),
		\qquad
		\wt{\Oc}(e)S_B=S_{B\Pi(e)}, \\
		\Sh \Az(G)=\wt{\Oc}(\Exp(G))K,
		\qquad
		\wt{\Oc}(e)|_{\Sh \Az(G)}=\Oc(\Pi(e)).
	\end{gathered}
\label{eq:formal-shadow-factorization}
\end{equation}
Moreover,
\begin{equation}
\label{eq:kernel-Pi}
	\ker\Pi
	=\{e \in \Exp(G) \mid \wt{\Oc}(e)K=0\}
	=\{e \in \Exp(G) \mid \wt{\Oc}(e)|_{\Sh \Az(G)}=0\}.
\end{equation}
\end{corollary}

\begin{proof}
Lie differentiation and the chain rule on $C^\infty(Z \* Z,\Cz)$ give
\begin{equation}
\label{eq:relations}
	[R_X,R_Y]=R_{[X,Y]},
	\qquad
	P_AP_B=P_{BA},
	\AND
	R_XP_A=P_AR_{A_*X}
\end{equation}
for $X,Y \in \vect(Z)$ and $A,B \in G$.
We define $\Phi$ on $U(\Cz\vect(Z))$ by complex linearity, $\Phi(1)=1$, and
$\Phi(X_1\otimes\cdots\otimes X_n) \cq R_{X_n}\cdots R_{X_1}$.
Then $\Phi(uv)=\Phi(v)\Phi(u)$. 
By the first identity in \gzit{eq:relations},
$\Phi\bigl(X\otimes Y-Y\otimes X+[X,Y]\bigr)
=R_YR_X-R_XR_Y+R_{[X,Y]}=0$.
Hence, the two-sided ideal defining $U(\vect(Z))$ lies in $\ker\Phi$, and $\Phi$ induces a unital antihomomorphism on $U(\vect(Z))$.

To extend $\Phi$ to $\Exp(G)$, assign $P_A$ to $j(A)$. Since $\Phi(uv)=\Phi(v)\Phi(u)$, $j(A)j(B)=j(AB)$ requires
$P_BP_A=P_{AB}$,
and $j(A)X=(A_*X)j(A)$ requires $R_XP_A=P_AR_{A_*X}$.
Both identities follow from \gzit{eq:relations}. 
Thus, the assignment extends uniquely to a unital antihomomorphism $\wt \Oc$ on $\Exp(G)$, given on monomials by
$\wt\Oc\bigl(X_1\cdots X_nj(A)\bigr)=P_AR_{X_n}\cdots R_{X_1}$.
Theorem~\ref{thm:kernel-realization} gives $\wt{\Oc}(e)S_B=S_{B\Pi(e)}$ for all $e \in \Exp(G)$. 
Taking $B=1$ proves the first identity in 
\gzit{eq:formal-shadow-factorization}; surjectivity of
$\Pi \cln \Exp(G) \to \Az(G)$ gives the equality of shadow spaces. 
Finally, \gzit{eq:kernel-Pi} follows from injectivity of $\Sh$ 
and from $K \in \Sh \Az(G)$.
\end{proof}

\begin{proposition}
\label{ex:shadow-space-example}
The map $\Sh$ in Proposition~\ref{prop:shadow-algebra} 
need not be surjective. 
\end{proposition}

\begin{proof}
Let $Z \cq \Rz$ with $K(z,z')=zz'$ for $z,z' \in Z$, and $G=\{1\}$.
Then $\Qz(Z)=\Qz(\wh{Z})=\ol{\Qz(Z)}=\Cz$ and $\ket*{z}=z$ for 
$z \in Z$. 
The smooth vector field $X(z)=z\partial_z$ is a coherent vector field with coherent adjoint $X^*=X$, since $L_XK=K=R_XK$.
Moreover, $d\Gamma(X)=1$, and $(R_Xf)(z,z')=z'D_2f(z,z')$.
Thus,
\[
	\Pi(X-1)=0,
\]
whereas, for the constant function $1 \in C^\infty(Z \* Z,\Cz)$,
\[
	\wt O(X-1)1=-1.
\]
Consequently, the operations \gzit{eq:O-wt} on 
$C^\infty(Z \* Z,\Cz)$ need not descend to the algebra $\Az(G)$. 
Here $\Az(G)=\Cz$, $\Sh \Az(G)=\Cz K$, and $R_XK=K$.
Hence, $R_X$ acts as the identity on $\Sh \Az(G)$, as required 
by $d\Gamma(X)=1$. 
This also shows that a relation in $\Az(G)$ may identify a 
quantized vector field with a scalar operator.
\end{proof}

\subsection{A Lie algebra for generators of motions}
\label{subsec:halfcurve-algebra}

In this Subsection, we associate a real Lie algebra with a subgroup 
$G \seq \Diff(M)$ of the diffeomorphism group of a smooth manifold $M$. 
The construction uses motions which are one-sided families through the identity whose forward and inverse actions on smooth functions 
admit compatible first-order expansions. 
We show that their generators form a real Lie subalgebra of 
$\Kin(M)$. 
This requires no Lie group structure on $G$ and provides the 
infinitesimal framework for the coherent quantization developed 
below. 

\pgph


Let $E$ be a locally convex space, and let $F$ be a convenient vector space, and let $R_s\in L(E,F)$ for $0<s<\delta$, where 
$\delta>0$. 
For $\alpha \ge 0$, we define the \bfi{Landau symbol} $o(s^\alpha)$ by
\[
	R_s=o(s^\alpha)
	\qquad\text{if}\qquad
	s^{-\alpha}R_s \to 0
	\qquad
	\text{ in }L(E,F)
	\for 
	s\downarrow0.
\]
In particular, $R_s=o(1)$ means $R_s\to0$ in $L(E,F)$ as 
$s\downarrow0$. 

\pgph


Let $M$ be a smooth manifold, and let $G$ be a subgroup of 
$\Diff(M)$. 
Let $\eps > 0$.
A map
\[
	h \cln [0,\eps) \to G,
	\with
	h_0=\id_M,
\]
is called a \bfi{motion} if there exists a smooth vector field 
$X \in \Kin(M)$ with the following property.
For every open set $U \seq M$ and every $x \in U$, there exist an 
open neighbourhood $x \in V \seq U$ and $0 < \delta \leq \eps$ 
such that
\begin{equation}
\label{eq:trapping}
	h_s^{\pm1}(V) \seq U
	\for 
	0 \leq s < \delta,
\end{equation}
and, for $f \in C^\infty(U,F)$, where $F$ is an arbitrary 
convenient vector space,
\begin{equation}
\label{eq:forward-expansion}
	(h_s^{\pm 1})^*f(p) 
	\cq f\bigl(h_s^{\pm1}(p)\bigr)
	=f(p) \pm s \, \Lc_Xf(p)+o(s)
	\for
 	p \in V, ~ 0 < s < \delta.
\end{equation}
The vector field $X$ is called the \bfi{generator} of $h$. 
We write $\Half(G)$ for the \eh{set of all motions} on $G$. 

Let $\eps_h, \eps_k > 0$. 
We call two motions $h \cln [0,\eps_h) \to G$ and 
$k \cln [0,\eps_k) \to G$ \bfi{equivalent} and write $h \sim k$ if 
there is a  $0 < \delta \leq \min\{\eps_h,\eps_k\}$ such that 
$h_s=k_s$ for $s \in [0,\delta)$.
The \bfi{germ} $\germ(h)$ of a motion $h$ at $0$ is the equivalence 
class consisting of all $k \in \Half(G)$ which satisfy $h \sim k$. 

\pgph


Before we proceed, some remarks are in order. 
\begin{enumerate}[label=\textup{(\arabic*)}]
	\item\label{rem:motion-expansion-restriction} In \gzit{eq:forward-expansion}, $o(s)$ is understood in $L(C^\infty(U,F),C^\infty(V,F))$, and both $f$ and $\Lc_Xf$ on the right-hand side are restricted to $V$.
	\item \label{rem:motion-restriction} 
	The restriction to $V$ in \gzit{eq:trapping} is essential. 
	Requiring $V=U$ for every open $U$ would make each open set 
	invariant under $h_s$ for sufficiently small $s$, and would force $X=0$. 
	Indeed, choose a chart $u \cln U \to F$ around $p$ and a continuous 
	linear functional $\ell \cln F \to \Rz$, and put
	\[
		f(q) \cq \ell\bigl(u(q)-u(p)\bigr)
		\for
		q \in U.
	\]
	For sufficiently small $s$, the diffeomorphism $h_s$ maps each of 
	the three sets
	\[
		U,
		\qquad
		\{q \in U \mid f(q)>0\},
		\qquad
		\{q\in U \mid f(q)<0\}
	\]
	onto itself; each set may have its own parameter interval. 
	Taking the minimum gives a common interval.
	Since $f(p)=0$, this gives $f(h_s(p))=0$. 
	Thus, \gzit{eq:forward-expansion} implies
	\[
		0=\lim_{s\downarrow0}\frac{f(h_s(p))-f(p)}s
		=\ell\bigl(du(p).X(p)\bigr).
	\]
	As $\ell$ is arbitrary, $du(p)X(p)=0$, and hence $X(p)=0$.
	\item Charts are admissible test functions in 
	\gzit{eq:forward-expansion}, since their values 
	lie in convenient vector spaces. 
	They determine the generator pointwise; see Proposition \ref{prop:generator-unique} below.
	Taking only $U=V=M$ would now restrict the test functions to global smooth functions. 
	Uniqueness of the generators would then require the action 
	$X \mto \Lc_X$ on $C^\infty(M, \Rz)$ to be faithful. 
	\sca{Kriegl \& Michor} \cite[Lemma 32.3]{KM} prove faithfulness for 
	the action on local smooth functions, interpreted as sheaf 
	derivations 
	according to their convention on vector fields in 
	\sca{Kriegl \& Michor} \cite[Section 32.1]{KM}. 
	For smoothly regular manifolds, 
	\sca{Kriegl \& Michor} \cite[Lemma 32.4]{KM} identify these with bounded derivations of the global function algebra 
	$C^\infty(M,\Rz)$. 
	The local definition of motions used here requires no such 
	additional smooth regularity assumption.
\end{enumerate}

\begin{examples}
\label{ex:motions}
\begin{enumerate}[label=\textup{(\roman*)}, ref=\textup{(\roman*)}]
	\item\label{itm:motions-smooth-families} Let $M$ be a smooth manifold and 
	$G \seq \Diff(M)$ a subgroup. Every map
	\[
		h \cln (-\varepsilon,\varepsilon) \to G,
		\qquad
		h_0=\id_M,
	\]
	with $\varepsilon>0$, for which the maps
	\[
		(s,p)  \mto  h_s^{\pm1}(p)
	\]
	are jointly smooth restricts to a motion on $[0,\varepsilon)$. 
	Indeed, put
	\[
	    X(p)\cq\left.\frac{d}{ds}\right|_{s=0}h_s(p).
	\]
	Joint smoothness of $(s,p)\mto h_s(p)$ makes $X$ a smooth vector field. 
	Let $U\seq M$ be open and $x\in U$.
	Since $h_0=\id_M$, both maps $(s,p)\mto h_s^{\pm1}(p)$ take $(0,x)$ into $U$. 
	By \sca{Kriegl \& Michor} \cite[Section~4.16]{KM},
	their common inverse image of $U$ contains a product neighbourhood
	$(-\delta,\delta)\*V$ of $(0,x)$. 
	This gives \gzit{eq:trapping}.
	Let $F$ be a convenient vector space. 
	For $|s|<\delta$, we put $T_s^\pm f\cq (f\circ h_s^{\pm1})|_V$ for 
	$f\in C^\infty(U,F)$.
	By smoothness of evaluation, $(s,f,p)\mto f(h_s^{\pm1}(p))$
	is smooth. 
	Applying the exponential law twice, first in $p$ and then in $f$, gives a smooth curve
	$s\mto T_s^\pm$ in $C^\infty(C^\infty(U,F),C^\infty(V,F))$
	by \sca{Kriegl \& Michor} \cite[Lemma~27.17]{KM}.
	Each $T_s^\pm$ is linear. 
	Hence, this curve takes values in the closed subspace $L(C^\infty(U,F),C^\infty(V,F))$ and is smooth there with the induced topology.
	At $s=0$, both pullbacks are restriction to $V$. Differentiating
	$h_s(h_s^{-1}(p))=p$ at $s=0$ gives
	\[
	    \left.\frac{d}{ds}\right|_{s=0}h_s^{-1}(p)=-X(p).
	\]
	Thus, evaluating the operator derivative at $f$ and then at $p$,
	we obtain
	\[
	    \left(\left.\frac{d}{ds}\right|_{s=0}T_s^\pm\right)f
	    =\pm(\Lc_Xf)|_V.
	\]
	This proves \gzit{eq:forward-expansion} with the remainder
	$o(s)$ in $L(C^\infty(U,F),C^\infty(V,F))$.
	
	The converse fails. 
	For the translation group $G$ of $\Rz$, put
	\[
		a(s) \cq s+s^{3/2},
		\qquad
		h_s(x) \cq x+a(s)
		\for 
		0 \le s< 1, ~ x \in \Rz.
	\]
	Then $h$ is a motion with generator $\partial_x$, but 
	admits no smooth extension across $s=0$. 
	Indeed,
	\[
		a'(s)=1+\frac32\sqrt{s}  \to 1,
		\qquad
		a''(s)=\frac{3}{4\sqrt{s}}  \to +\infty
		\for 
		s\downarrow0,
	\]
	so $a$ has no $C^2$-extension across zero.
\end{enumerate}
For the following examples, let $U\seq \mathbb R$ be open and 
$x\in U$. 
We choose $0<r<\tfrac12$ such that $(x-2r,x+2r)\seq  U$ and take
\[
	V=(x-r,x+r),
	\qquad 
	\delta=r^2.
\]
\begin{enumerate}[label=\textup{(\roman*)}, ref=\textup{(\roman*)}]
\setcounter{enumi}{1}
	\item\label{itm:motions-reparametrization} The family
	\[
		h_s(x) \cq x+\sqrt{s}
		\for 
		s\ge0,\ x\in\Rz
	\]
	satisfies $h_0=\mathrm{id}_{\Rz}$, is continuous at 
	zero and smooth for $s>0$. It is not a motion, since
	\[
		\frac{h_s(x)-x}{s}
		=\frac1{\sqrt{s}}  \to +\infty
		\for 
		s\downarrow0,
	\]
	so no generator exists. The reparametrized family
	\[
		\widetilde h_s(x) \cq h_{s^2}(x)=x+s
		\for 
		s\ge0
	\]
	is a motion with generator $\partial_x$.
	\item\label{itm:motions-same-generator} On $\Rz$, the motions
	\[
		h_s(x) \cq x+s,
		\qquad 
		k_s(x) \cq x+s+s^2
		\for 
		s \geq 0
	\]
	have the same generator $\partial_x$ but distinct germs at 
	zero.
	\item\label{itm:motions-zero-generator} On $\Rz$, the motion
	\[
		h_s(x) \cq x+s^2
		\for 
		s \geq 0
	\]
	has generator $X(x)=0$.
	
	This illustrates that a motion can have zero generator without being the identity motion, even as a germ: 
	$h_s \ne \id_{\Rz}$ for every $s>0$.\qedhere
\end{enumerate}
\end{examples}

\begin{proposition}
\label{prop:generator-unique}
Let $M$ be a smooth manifold, and let
$G \seq \Diff(M)$ be a subgroup. Then, every 
motion in $G$ has a unique generator.
\end{proposition}

\begin{proof}
Let $h$ be a motion with generator $X$. 
We apply it to a chart $u \cln U \to u(U) \seq F$ with 
$U \seq M$ and $F$ a convenient vector space.
The singleton $\{u\}$ is bounded, and point evaluation is 
continuous, so
\[
	\lim_{s \downarrow 0}\frac{u(h_s^{\pm1}(p))-u(p)}s
	=\pm du(p).X(p)
	\with p \in U.
\]
A second generator gives the same limits. 
Since $F$ is Hausdorff and $du(p)$ injective, the vector fields 
agree at every $p$.
\end{proof}

\begin{lemma}
\label{lem:motion-product-commutator}
Let $G$ be a subgroup of $\Diff(M)$, and let
$h,k\in\Half(G)$ with generators $X,Y$, respectively.
For sufficiently small $s\geq0$, put
\[
    a_s\cq h_s\circ k_s,
    \qquad
    c_s\cq k_{\sqrt{s}}^{-1}\circ h_{\sqrt{s}}^{-1}
           \circ k_{\sqrt{s}}\circ h_{\sqrt{s}}.
\]
Then $a$ and $c$ are motions with generators $X+Y$ and
$\Lb{X}{Y}$, respectively.
\end{lemma}

\begin{proof}
Let $U\seq M$ be open and $x\in U$.
Repeated application of \gzit{eq:trapping} and \gzit{eq:forward-expansion} gives fixed open neighbourhoods
$x \in U_4 \seq U_3 \seq U_2 \seq U_1 \seq U_0=U$
and $\delta>0$ such that, for $j=0,1,2,3$ and $0 \leq r < \delta$, we have $h_r^{\pm1}(U_{j+1})\seq U_j$ and 
$k_r^{\pm1}(U_{j+1})\seq U_j$, and the corresponding first-order expansions \gzit{eq:forward-expansion} hold on every adjacent pair. 
The neighbourhoods and $\delta$ are independent of the convenient target space $F$.
For each $j$, we define the bounded linear operators
$A_{r,j}^{\pm}\cq(h_r^{\pm1})^*$, $B_{r,j}^{\pm}\cq(k_r^{\pm1})^*$, 
and $R_jf\cq f|_{U_{j+1}}$ from $C^\infty(U_j,F)$ to $C^\infty(U_{j+1},F)$.
For $j=0,1,2$, they satisfy
\begin{equation}
\label{eq:motion-pc-restrictions}
    A_{r,j+1}^{\pm}R_j=R_{j+1}A_{r,j}^{\pm},
    \qquad
    B_{r,j+1}^{\pm}R_j=R_{j+1}B_{r,j}^{\pm},
\end{equation}
and
\begin{equation}
\label{eq:motion-pc-cancellation}
    A_{r,j+1}^{+}A_{r,j}^{-}
    =A_{r,j+1}^{-}A_{r,j}^{+}
    =R_{j+1}R_j,
\end{equation}
with analogous identities for $B_r^{\pm}$.

We henceforth suppress the domain indices and write $1$ for
the appropriate restriction operator. 
All limits are taken as $r\downarrow0$ in the respective spaces
$L(C^\infty(U_i,F),C^\infty(U_j,F))$.
Put
\[
    P_r^{\pm}\cq\frac{A_r^{\pm}-1}{r},
    \qquad
    Q_r^{\pm}\cq\frac{B_r^{\pm}-1}{r}.
\]
\gzit{eq:forward-expansion} gives
\begin{equation}
\label{eq:motion-pc-limits}
    P_r^{\pm}\to\pm\Lc_X,
    \qquad
    Q_r^{\pm}\to\pm\Lc_Y,
    \qquad
    A_r^{\pm},B_r^{\pm}\to1.
\end{equation}
These operators and their limits are bounded linear operators and commute with postcomposition by every continuous linear map 
$\ell \cln F_1 \to F_2$ between convenient vector spaces. 
For pullbacks and restrictions this follows from associativity, and for Lie derivatives from the chain rule; 
\sca{Kriegl \& Michor} \cite[Theorem~3.18]{KM}. 
This compatibility is preserved under linear combinations and compositions. 
Moreover, for every convenient vector space $F$, every $j=0,1,2,3$, and every smooth function $f\in C^\infty(U_j,F)$, the convergences in \gzit{eq:motion-pc-limits} hold after applying the operators to $f$, with convergence in $C^\infty(U_{j+1},F)$. 
This verifies assumption \ref{itm:natural-product-pointwise} of Lemma~\ref{lem:AsF}.
Since the neighbourhoods and the parameter interval are independent of $F$, Lemma~\ref{lem:AsF} applies to the products below and to their compositions.

We write $A_r\cq A_r^+$, $B_r\cq B_r^+$,
$P_r\cq P_r^+$, and $Q_r\cq Q_r^+$.
For the product family, the exact identity
\[
    \frac{B_rA_r-1}{r}=Q_rA_r+P_r
\]
and Lemma~\ref{lem:AsF} give
\begin{equation}
\label{eq:motion-pc-product}
    \frac{B_rA_r-1}{r}
    \to\Lc_Y+\Lc_X=\Lc_{X+Y}.
\end{equation}
Similarly,
\begin{equation}
\label{eq:motion-pc-product-inverse}
    \frac{A_r^-B_r^--1}{r}
    =P_r^-B_r^-+Q_r^-
    \to-\Lc_X-\Lc_Y.
\end{equation}
Since $a_r^*=B_rA_r$ and $(a_r^{-1})^*=A_r^-B_r^-$,
\gzit{eq:motion-pc-product} and
\gzit{eq:motion-pc-product-inverse}, together with \gzit{eq:trapping}, show that $a$ is a motion with generator $X+Y$.

For the commutator, \gzit{eq:motion-pc-restrictions} gives
the exact cancellation
\[
    D_r\cq\frac{A_rB_r-B_rA_r}{r^2}
    =P_rQ_r-Q_rP_r.
\]
Applying Lemma~\ref{lem:AsF} to both
products yields
\begin{equation}
\label{eq:motion-pc-bracket-limit}
    D_r\to\Lc_X\Lc_Y-\Lc_Y\Lc_X
    =\Lc_{\Lb{X}{Y}}.
\end{equation}
The same lemma gives $A_r^-B_r^- \to 1$.
By \gzit{eq:motion-pc-cancellation}, $B_rA_rA_r^-B_r^-=1$.
Consequently, another application of Lemma~\ref{lem:AsF} gives
\begin{equation}
\label{eq:motion-pc-commutator}
    \frac{A_rB_rA_r^-B_r^--1}{r^2}
    =D_rA_r^-B_r^-
    \to\Lc_{\Lb{X}{Y}}.
\end{equation}
Interchanging $h$ and $k$ in this argument gives
\begin{equation}
\label{eq:motion-pc-commutator-inverse}
    \frac{B_rA_rB_r^-A_r^--1}{r^2}
    \to\Lc_{\Lb{Y}{X}}
    =-\Lc_{\Lb{X}{Y}}.
\end{equation}
Now $c_{r^2}^*=A_rB_rA_r^-B_r^-$ and $(c_{r^2}^{-1})^*=B_rA_rB_r^-A_r^-$.
Putting $s=r^2$ in \gzit{eq:motion-pc-commutator} and
\gzit{eq:motion-pc-commutator-inverse}, we obtain
$c_s^*=1+s\Lc_{\Lb{X}{Y}}+o(s)$ and $(c_s^{-1})^*=1-s\Lc_{\Lb{X}{Y}}+o(s)$. 
Here both remainders are taken in
$L(C^\infty(U,F),C^\infty(U_4,F))$.
For $0\leq s<\delta^2$, both $c_s$ and $c_s^{-1}$ map
$U_4$ into $U$. 
Since $c_0=\id_M$, this proves that $c$ is a
motion with generator $\Lb{X}{Y}$.
\end{proof}

We now state our main result of this section.

\begin{theorem}
\label{thm:lie-alg-motions}
Let $M$ be a smooth manifold, and let $G$ be a subgroup of 
$\Diff(M)$.
Then:
\begin{enumerate}[label=\textup{(\roman*)}, ref=\textup{(\roman*)}]
	\item\label{itm:motion-germs-group} The set $\gMot(G)$ of all germs of motions forms a group.
	\item\label{itm:motion-generators-lie-algebra} The set $\half(G)$ of all generators of motions is a real Lie subalgebra of $(\Kin(M),[\cdot,\cdot])$.
\end{enumerate}
\end{theorem}

\begin{proof}
We first prove, for germs of motions $\alpha \cq \germ(h)$ and $\beta \cq \germ(k)$, with generators $X,Y$, respectively, that the following are also germs of motions:  
\begin{enumerate}[label=\textup{(\alph*)}, ref=\textup{(\alph*)}]
	\item\label{itm:motion-operation-identity} $e=\germ(s \mto \id_M)$, with generator $0$;
	\item\label{itm:motion-operation-inverse} $\alpha^{-1}=\germ(s \mto h_s^{-1})$, with generator 
	$-X$;
	\item\label{itm:motion-operation-product} $\alpha\beta=\germ(s \mto h_s \circ k_s)$, with 
	generator $X+Y$;
	\item\label{itm:motion-operation-scaling} for $\lambda \in \Rz$, let 
	$\lambda \odot \alpha \cq \germ(s \mto (\lambda \odot 
	h)_s)$, 
	where
	\[
		(\lambda \odot h)_s
		\cq 
		\begin{cases}
			h_{\lambda s}		 	& \for \lambda>0, \\
			\id_M					& \for \lambda=0, \\
			h_{|\lambda|s}^{-1}		& \for \lambda<0,
		\end{cases}
	\]
	with generator $\lambda X$;
	\item\label{itm:motion-operation-bracket} the commutator
	$[\alpha,\beta] \cq 
	\germ(s \mto k_{\sqrt{s}}^{-1} \circ h_{\sqrt{s}}^{-1}
	\circ k_{\sqrt{s}} \circ h_{\sqrt{s}}),
	$
	with generator $\Lb{X}{Y}$.
\end{enumerate}
Let $U$ be an open neighbourhood of $x\in M$. 
Repeated use of \gzit{eq:trapping} gives a neighbourhood $V \seq  U$ of $x$ and a sufficiently small parameter interval such that all 
compositions below are defined and \gzit{eq:forward-expansion} 
applies. 
We write $1$ for the identity operator. 
We understand the operators $\phantom{}^*$, $1$, $\Lc$, and $o(s)$ on $V$, see remark~\ref{rem:motion-expansion-restriction} on page~\pageref{rem:motion-expansion-restriction}, after application to an arbitrary $f \in C^\infty(U,F)$ which is omitted from the notation. 

The neighbourhoods and parameter interval are chosen independently 
of $F$ and remain fixed as $s\downarrow0$. 
All constructions depend only on the germs, since equality is preserved by inversion, composition, and the substitutions 
$s \mto \lambda s$, with $\lambda>0$, and $s \mto \sqrt{s}$. 
Every family considered takes the value $\id_M$ at zero.

We put $A_t^{\pm} \cq (h_t^{\pm 1})^*$ and 
$B_t^{\pm} \cq (k_t^{\pm 1})^*$.
By \gzit{eq:forward-expansion}, we have
\begin{equation}
\label{eq:four}
	A_t^{\pm}=1 \pm t\Lc_X+o(t) 
	\AND 
	B_t^{\pm}=1 \pm t\Lc_Y+o(t).
\end{equation}

The constant identity family has generator $0$. 
Inversion interchanges the forward and inverse expansions, hence 
changes the generator from $X$ to $-X$. 
This proves \ref{itm:motion-operation-identity} and \ref{itm:motion-operation-inverse}.
\ref{itm:motion-operation-product} follows from Lemma~\ref{lem:motion-product-commutator}.
For \ref{itm:motion-operation-scaling}, let $\lambda>0$. 
Substitution in \gzit{eq:four} gives
$A_{\lambda t}^{\pm}=1 \pm t\Lc_{\lambda X}+o(t)$.
For $\lambda<0$, we apply the same consideration to the inverse 
motion with scale $|\lambda|$, obtaining as generator 
$-|\lambda|X=\lambda X$.
The case $\lambda=0$ is \ref{itm:motion-operation-identity}.
\ref{itm:motion-operation-bracket} follows from Lemma \ref{lem:motion-product-commutator}.

Now \ref{itm:motion-germs-group} follows from \ref{itm:motion-operation-identity}, \ref{itm:motion-operation-inverse}, and \ref{itm:motion-operation-product}.
Finally, \ref{itm:motion-generators-lie-algebra} follows since, by \ref{itm:motion-operation-identity}--\ref{itm:motion-operation-bracket}, $\half(G)$ is a real vector 
subspace of $\Kin(M)$ closed under brackets, hence a real Lie 
subalgebra.
\end{proof}

\begin{corollary}
	\label{cor:motion-quotient}
	The map
	\[
		D \cln \gMot(G) \to (\half(G),+),
		\qquad
		D(\germ(h)) \cq X_h,
	\]
	is a surjective group homomorphism. 
	Its kernel
	\[
		N \cq \{\alpha \in \gMot(G) \mid D\alpha=0\}
	\]
	is normal, and
	\[
		\Phi \cln \gMot(G)/N \to \half(G),
		\qquad
		\Phi(\alpha N) \cq D\alpha,
	\]
	is an isomorphism of additive groups. 
	The operations
	\[
		(\alpha N)+(\beta N)=(\alpha\beta)N,
		\qquad
		\lambda(\alpha N)=(\lambda\odot\alpha)N,
		\AND
		[\alpha N,\beta N]=[\alpha,\beta]N
	\]
	are well-defined and make $\Phi$ an isomorphism of real Lie 
	algebras.
	The zero element is $N$, and the additive inverse of $\alpha N$ 
	is $\alpha^{-1}N$.
\end{corollary}

\sca{Hubicska \& Muzsnay} \cite{HM2019} associate a Lie algebra 
with an arbitrary subgroup of the diffeomorphism group of a compact 
finite-dimensional manifold, using smooth curves with prescribed 
leading derivative at the identity. 
The commutator formula for smooth local families on convenient 
manifolds is given in 
\sca{Mauhart \& Michor} \cite{MauhartMichor1992}. 
\sca{Kipka \& Ledyaev} \cite{KL2014} prove a 
corresponding formula on Banach manifolds under locally uniform 
operator remainder estimates. 
Here we use the 
motion condition \gzit{eq:forward-expansion}, with remainders 
measured in the operator 
topology of Subsection~\ref{subsec:convenient-calculus}. This 
condition imposes no smooth 
dependence on the parameter, and the construction requires no Lie 
group structure.

\pgph


In Example \ref{ex:motions}\ref{itm:motions-smooth-families}, the generator of the nonsmooth motion is also obtained from a smooth curve in the same group. The following example shows that a subgroup can have a nonzero motion generator even when every continuous curve in the subgroup is constant.

\begin{example}
Let $G$ be the group of rational rotations of the circle $\Rz/\Zz$. 
Every continuous curve in $G$ is constant. 
Indeed, evaluation at the origin gives a continuous curve in $\Qz/\Zz$, which is totally disconnected, and a rotation is determined by its value at the origin. 
Hence, the tangent Lie algebra of \sca{Hubicska \& Muzsnay} \cite[Definition 3.2 and Remark 3.3]{HM2019} is zero.

Nevertheless, the rotation field belongs to $\half(G)$. 
To see this, we put $h_0=\id$. 
For $0<s<1$, choose a rational number $q(s)$ between $s-s^2$ and $s$, and let $h_s$ be rotation by $q(s)$. 
Since $q(s)\leq s$, the local neighbourhoods for the ordinary rotation flow in Example \ref{ex:motions}\ref{itm:motions-smooth-families} also give \gzit{eq:trapping} for $h$. 
Since $q(s)/s \to 1$, substituting $q(s)$ for the rotation parameter in the forward and inverse pullback expansions gives \gzit{eq:forward-expansion} with the same generator.
\end{example}

\subsection{Quantization along motions}
\label{subsec:quantization-along-motions}

The preceding constructions associate operators on a common 
invariant domain with smooth coherent maps and smooth coherent vector fields. 
We now relate these constructions by differentiating quantized motions. 

\pgph


Let $G$ be a smooth coherent $*$-group. 
A motion $k \in \Half(G)$ is called a \bfi{coherent adjoint} of 
a motion $h \in \Half(G)$ if
\[
	K(h_sz,z')=K(z,k_sz')
\]
for all $z,z' \in Z$ and all sufficiently small $s \geq 0$.

\begin{lemma}
\label{lem:coherent-adjoint-motions}
Let $G$ be a smooth coherent $*$-group, and let $X$ and $Y$ be the 
generators of $h$ and a coherent adjoint $k$, respectively. 
Then the following hold:
\begin{enumerate}[ref=\textup{(\roman*)}]
	\item\label{itm:adjoint-motion-generators} $X$ is a coherent vector field with coherent adjoint $Y$.
	\item\label{itm:adjoint-motion-isometries} If $h$ preserves the coherent product, that is,
	\begin{equation}
	\label{eq:kernel-preservation}
		K(h_s z,h_s z')=K(z,z')
	\end{equation}
	for all $z,z'\in Z$ and all sufficiently small $s\geq0$, 
	then
	\[
		h^{-1} \cln s\longmapsto h_s^{-1}
	\]
	is a coherent adjoint of $h$, with generator $-X$. In 
	particular, $X$ is coherent with coherent adjoint $-X$, and
	\[
		L_XK+R_XK=0.
	\]
	\item\label{itm:adjoint-motion-conjugation} Suppose there exists a fixed 
	$J\in\operatorname{Diff}(Z)$ such that
	\[
		A^*=J A^{-1} J^{-1}
		\qquad
		\text{ for every }
		A \in G.
	\]
	Then the involution preserves motions. If 
	$h\in\operatorname{Mot}(G)$ has generator $X$, then $h^*$ 
	has generator
	\[
		-J_*X.
	\]
	In particular, if $A^*=A^{-1}$ for every $A\in G$, then the 
	involution preserves motions and $h^*$ has generator $-X$.
	\item\label{itm:adjoint-motion-involution} Suppose that the involution \bfi{preserves motions}, that is,
	\[
		h\in\operatorname{Mot}(G)
		\quad\Longrightarrow\quad
		h^*\in\operatorname{Mot}(G),
		\qquad h^*_s:=(h_s)^*.
	\]
	Then, for every $h\in\operatorname{Mot}(G)$, the motion 
	$h^*$ is a coherent adjoint of $h$. Consequently, every 
	motion in $G$ has a coherent generator, and the generator of 
	$h^*$ is a coherent adjoint of the generator of $h$.
\end{enumerate} 
\end{lemma}

\begin{proof}
To see \ref{itm:adjoint-motion-generators}, we take the right derivative at $s=0$ which gives $L_XK=R_YK$. 
To see \ref{itm:adjoint-motion-isometries}, by Theorem \ref{thm:lie-alg-motions}, $s \mto h_s^{-1}$ is a motion with generator $-X$. 
Substituting $h_s^{-1}w$ for the second argument of \gzit{eq:kernel-preservation} gives
$K(h_s z,w)=K(z,h_s^{-1}w)$.
Thus, $h^{-1}$ is a coherent adjoint of $h$. 
Part (i) now gives $L_XK=R_{-X}K=-R_XK$.
For \ref{itm:adjoint-motion-conjugation} let 
$h \in \operatorname{Mot}(G)$ have generator 
$X$, and put $k_s \cq h_s^*=J h_s^{-1} J^{-1}$.
Then $k_0=\id_Z$ and $k_s^{\pm1}=J h_s^{\mp1} J^{-1}$.
Let $U\seq  Z$ be open and $z \in U$. 
Apply the motion condition for $h$ to $J^{-1}(U)$ and $J^{-1}(z)$. It supplies an open neighbourhood $J^{-1}(z)\in W\seq  J^{-1}(U)$
and $\delta>0$ such that $h_s^{\pm1}(W)\seq  J^{-1}(U)$ for 
$0 \leq s < \delta$.
Thus, $V \cq J(W)$ is an open neighbourhood of $z$ contained in 
$U$, and $k_s^{\pm1}(V) \seq  U$ for $0\leq s<\delta$.
Let $F$ be a convenient vector space and $f \in C^\infty(U,F)$. 
Applying \gzit{eq:forward-expansion} to $f \circ J$ gives, on $V$,
\[
\begin{aligned}
	f\circ k_s^{\pm1}
	&=\bigl((f\circ J)\circ h_s^{\mp1}\bigr)\circ J^{-1}
	=f\mp s\bigl(L_X(f\circ J)\bigr)\circ J^{-1}+o(s)\\
	&=f\mp sL_{J_*X}f+o(s).
\end{aligned}
\]
The last equality follows from the chain rule. 
The remainder is $o(s)$ in the required operator topology: It is obtained from the corresponding remainder for $h$ by composition with the fixed continuous linear pullbacks induced by $J$ and $J^{-1}$.
Consequently, $k$ satisfies both motion conditions with generator 
$-J_*X$. Taking $J=\mathrm{id}_Z$ proves the final assertion.
Finally, for \ref{itm:adjoint-motion-involution}, let $h\in\operatorname{Mot}(G)$ with generator $X$. 
By assumption, $h^*$ is a motion. 
Let $Y$ be its generator. 
We have $K(h_s z,z')=K(z,(h_s)^*z')$ for $z,z' \in Z$.
Thus, $h^*$ is a coherent adjoint of $h$. 
By part \ref{itm:adjoint-motion-generators}, $L_XK=R_YK$. 
Hence, $X$ is coherent, and $Y$ is a coherent adjoint of $X$.
\end{proof}

\begin{example}
We show that the involution of a coherent $*$-group need not 
preserve motions, even when the coherent product is nondegenerate. Let $Z=s(\mathbb C)$ be the space of rapidly decreasing 
sequences, with seminorms
\[
p_m(z)=\sup_{n\ge1}n^m|z_n|
\]
and coherent product $K(z,w)=\sum_{n\ge1}\ol{z_n}w_n$. Let 
$G$ consist of the invertible finite matrices extended by the 
identity, with conjugate transpose as involution.

Choose a real-valued $\varphi\in C_c^\infty((1/2,3/2))$ with 
$\varphi(1)=1$, and put
\[
\ell_s(z)=\sum_{n\ge2}\varphi(ns)z_n,
\qquad
h_sz=z+\bigl(sz_1+\ell_s(z)\bigr)e_1.
\]
Each sum is finite for $s>0$. For every $k\ge0$ and 
sufficiently 
large $m$,
\[
|\partial_s^k\ell_s(z)|
\le C_{k,m}s^{m-k-1}p_m(z).
\]
Thus $\ell_s$, extended by zero for $s\le0$, is jointly smooth 
and flat at zero. Since
\[
h_s^{-1}z
=\frac{z_1-\ell_s(z)}{1+s}e_1+\sum_{n\ge2}z_ne_n,
\]
both forward and inverse actions are jointly smooth near zero. 
By Example \hyperref[itm:motions-smooth-families]{\ref*{ex:motions}\ref*{itm:motions-smooth-families}}, $h$ is a motion with generator $X(z)=z_1e_1$.

On the other hand,
\[
h_s^*e_1=(1+s)e_1+\sum_{n\ge2}\varphi(ns)e_n,
\qquad
p_0(h_{1/n}^*e_1-e_1)\ge1.
\]
Hence $h_s^*e_1\not\to e_1$ as $s\downarrow0$. 
Taking $U=\{z \mid p_0(z-e_1)<1/2\}$ shows that $h^*$ violates \gzit{eq:trapping}, and therefore is not a motion.
\end{example}

We record a consequence of the mapping-space topology, see 
Subsection \ref{subsec:convenient-calculus}. 
Let $M$ be a smooth manifold, and let $F$ be a convenient vector 
space. 
For fixed $z \in M$ and smooth vector fields $X_1,\ldots,X_n$ on 
$M$, the map
\[
	J_{z;X_{1:n}} \cln C^\infty(M,F) \to  F,
	\qquad
	f  \mto  (\Lc_{X_1}\cdots\Lc_{X_n}f)(z),
\]
is continuous and real-linear. 
To see continuity, we work in a chart: This evaluation is a finite 
sum of evaluations of higher differentials of $f$ at fixed vectors. 
Polarization expresses these as finite linear combinations of 
derivatives along fixed smooth chart curves. 
Such evaluations are continuous by the defining mapping-space 
topology. 

\begin{theorem}
\label{thm:orbit-derivative}
Let $G$ be a smooth coherent $*$-group, and 
let $h \in \Half(G)$ with generator $X \in \vect(Z)$. 
Then, for every $\psi \in \Qz(\wh{Z})$,
\begin{equation}
\label{eq:orbit-derivative}
	\lim_{s\downarrow0} \frac{\wh{\Gamma}(h_s)\psi-\psi}{s}
	=d\Gamma(X)\psi
	\qquad
	\text{ in }
	\ol{\Qz(Z)}.
\end{equation} 
In particular, this derivative depends only on $X$. 
For $z,z' \in Z$,
\begin{equation}
\label{eq:overlap}
	\bra*{z}d\Gamma(X)\ket*{z'}
	=R_XK(z,z').
\end{equation}
\end{theorem}

\begin{proof}
It suffices that we consider $\psi \cq (P\kappa)(z)$ and 
$P \cq \Lc_{X_1}\cdots\Lc_{X_n}$ for $z \in Z$ and 
$X_1, \ldots, X_n \in \vect(Z)$.
We choose an open neighbourhood $U \seq Z$ of $z$. 
Since $h$ is a motion, this supplies an open neighbourhood 
$z \in V \seq  U$ such that the pullback $f \mto (f \circ h_s)|_V$ has the expansion $h_s^*=\rho+s\Lc_X+o(s)$.
Evaluation at $\kappa|_U$ is continuous. 
Consequently,
\[
	F_s \cq 
	\frac{(\kappa \circ h_s)|_V-\kappa|_V}{s}-(\Lc_X\kappa)|_V
	=s^{-1}R_s(\kappa|_U)
	\to 0
	\qquad
	\text{ in } C^\infty(V,\ol{\Qz(Z)}).
\]
Applying the continuous evaluation $J_{z;X_{1:n}}$ gives
$(PF_s)(z) \to 0$ in $\ol{\Qz(Z)}$.
By Theorems~\ref{thm:common-domain-finite-quantization} and
\ref{thm:intrinsic-representation}, and the chain rule, we get
$\wh{\Gamma}(h_s)\psi=\bigl(P(\kappa\circ h_s)\bigr)(z)$ and $d\Gamma(X)\psi=(P\Lc_X\kappa)(z)$.
Thus, \gzit{eq:orbit-derivative} holds for $\psi$. 
Finite linear combinations prove it on $\Qz(\wh{Z})$. 
Finally, for $z' \in Z$, we get equation \gzit{eq:overlap} from
$\bra*{z}d\Gamma(X)\ket*{z'}=\bra*{z}\Lc_X\kappa(z')=R_XK(z,z')$.
\end{proof}

\begin{corollary}
Let $G$ be a smooth coherent $*$-group whose involution preserves motions. 
Let $h \in \Half(G)$ with generator $X$. 
Then $X$ is coherent, and for $\psi \in \Qz(\wh{Z})$,
\begin{equation}
\label{eq:strong-limit-star-group}
	\lim_{s\downarrow0}\frac{\wh{\Gamma}(h_s)\psi-\psi}{s}
	=d\Gamma(X)\psi
	\qquad
	\text{ in }
	\ol{\Qz(Z)}.
\end{equation}
Moreover, the generator of $h^*$ is a coherent adjoint of $X$.
\end{corollary}

\begin{proof}
By Lemma~\hyperref[itm:adjoint-motion-involution]{\ref*{lem:coherent-adjoint-motions}\ref*{itm:adjoint-motion-involution}}, $X$ is coherent, and the generator of 
$h^*$ is a coherent adjoint of $X$. Theorem~\ref{thm:orbit-derivative} therefore gives
\[
	\lim_{s\downarrow0}\frac{\wh{\Gamma}(h_s)\psi-\psi}{s}
	=d\Gamma(X)\psi
	\qquad
	\text{for every }\psi\in\Qz(\wh Z),
\]
with convergence in $\ol{\Qz(Z)}$. 
This proves \gzit{eq:strong-limit-star-group}.
\end{proof}

\begin{corollary}
\label{cor:unitary-motion-derivative}
Let $G$ be a smooth coherent $*$-group of $Z$ consisting of unitary coherent maps, and let $h \in \Half(G)$ with generator $X$. 
Then $-X$ is a coherent adjoint of $X$. 
Consequently, $d\Gamma(X)$ is skew-symmetric on $\Qz(\wh Z)$, and $i\,d\Gamma(X)$ is symmetric there. 
Both operators are closable. Moreover, for every $\psi \in \Qz(\wh Z)$,
\[
	\lim_{s\downarrow0}
	\frac{U_{h_s}\psi-\psi}{s}
	=d\Gamma(X)\psi
	\qquad\text{ in }\ol{\Qz(Z)}.
\]
\end{corollary}

\begin{proof}
By Lemma~\hyperref[itm:adjoint-motion-isometries]{\ref*{lem:coherent-adjoint-motions}\ref*{itm:adjoint-motion-isometries}}, $-X$ is a coherent adjoint of $X$. 
Theorem~\ref{thm:intrinsic-representation} and real linearity therefore give
$
(d\Gamma(X)\phi)^*\psi
=\phi^*d\Gamma(-X)\psi
=-\phi^*d\Gamma(X)\psi
$
for $\phi,\psi\in\Qz(\wh Z)$. 
Thus, $d\Gamma(X)$ is skew-symmetric and $i\,d\Gamma(X)$ is symmetric. 
Theorem~\ref{thm:intrinsic-representation} gives closability of $d\Gamma(X)$, and consequently also of $i\,d\Gamma(X)$.
By Corollary~\ref{cor:unitary},
$U_{h_s}|_{\Qz(\wh Z)}=\wh\Gamma(h_s)$.
The asserted limit now follows from Theorem~\ref{thm:orbit-derivative}.
\end{proof}


\section{Coherent constructions in quantum theory}
\label{sec:applications}

We now apply the preceding theory to quantum dynamics and bosonic 
quantization. 
The coherent product supplies the inner products of coherent 
states, while coherent maps and coherent vector fields supply 
quantum operators. 
The common invariant domain constructed in Section 
\ref{sec:coherent-quantization} allows these operators to be 
composed, and their coherent shadows allow matrix elements to be 
calculated by differentiation and substitution. 
Thus, the geometric constructions of the preceding sections lead 
directly to computations in quantum theory.

In Subsection~\ref{subsec:schrodinger-equation}, we 
express dynamical and spectral questions in terms of coherent 
matrix elements of the propagator and the resolvent. 
If the quantum evolution is induced by a unitary coherent motion, 
the evolution of coherent states is determined by trajectories on 
the coherent manifold. 
More precisely, if the propagator carries $\ket{z'}$ to $\ket{z(t)}$, its matrix element between $\ket z$ and $\ket{z'}$ is $K(z,z(t))$.
The propagation kernel is then obtained by evaluating the coherent 
product along these trajectories. 
When the exact evolution does not preserve the chosen family of 
coherent states, one may seek an approximate evolution within this 
family. 

Subsection~\ref{subsec:variational-principles} obtains the coherent action by restricting the quantum action to coherent states. 
If the tangent images of the coherent-state map are invariant under multiplication by $i$, stationarity is equivalent to the Dirac–Frenkel condition.

In Subsection~\ref{subsec:klauder}, we apply the 
general construction to Klauder spaces. 
Their coherent transformations and vector fields give the familiar 
bosonic operators. 
The canonical commutation relations follow from the corresponding 
relations between vector fields, and operator products are defined 
on the common invariant domain. 
Normal ordering expresses these operators through their coherent 
shadows. 
Creation and annihilation operators, Segal fields, and Weyl 
operators thus become concrete instances of the coherent 
quantization developed in Section \ref{sec:coherent-quantization}.

\subsection{The Schr\"odinger equation}
\label{subsec:schrodinger-equation}

In this section, we show how Schr\"odinger equations on a completed quantum space can be studied in terms of the coherent product. 
In Subsubsection~\ref{subsubsec:exactly-solvable-case}, a smooth
one-parameter group of unitary coherent maps gives the exact
evolution of coherent states when its generator $H$ satisfies
$\H\ket z=i\hbar\,d\Gamma(H)\ket z$ for every $z\in Z$.
Theorem~\ref{thm:hamIVP} gives the corresponding result for a
time-dependent Hamiltonian. 
Many more concrete applications to quantum mechanics can be found in the survey by \sca{Zhang} et al. \cite{ZhaFG}.

\subsubsection{The time-independent Schr\"odinger equation}
The dynamics of a conservative physical system is traditionally given by a Hamiltonian, a self-adjoint operator
\[
 \H \cln \Dom \H \seq \ol{\Qz(Z)} \lto \ol{\Qz(Z)}
\]
whose spectrum is bounded from below. 
The time-independent Schr\"odinger equation
\begin{equation}
\label{eq:tiSchr}
	\H\psi=E\psi,
	~~~ 
	\psi \in \Dom \H,
\end{equation}
determines the stationary states of the quantum system described by the Hamiltonian $\H$. 
The kernel of $E-\H$ is called the eigenspace of $E$. 
Its nonzero elements are the eigenvectors corresponding to $E$, and can be normalized to have norm one. 
The spectrum $\spec \H$ consists of those $E \in \Cz$ for which $E-\H$ has no everywhere defined bounded inverse on $\ol{\Qz(Z)}$. 
It is real and bounded from below. 
Spectral values need not be eigenvalues; the continuous spectrum is described by the spectral measure of $\H$.

We write
\[
	\iota \cq \frac{i}{\hbar}
	~~~
	\text{ for }
	\hbar>0,
\]
and denote the spectral measure of $\H$ by $P$. We use the spectral theorem and its bounded functional calculus; see Teschl~\cite[Theorems 3.1 and 3.6]{Teschl2014}.

\begin{theorem}
\label{thm:eig}
Let $\H$ be a time-independent Hamiltonian on $\ol{\Qz(Z)}$.
\begin{enumerate}[label=(\roman*), ref=\textup{(\roman*)}]
	\item\label{itm:spectral-mean-projection} For $\psi \in \ol{\Qz(Z)}$ and $E \in \Rz$,
	\begin{equation}
	\label{eq:Pi}
		 \Pi(E)\psi \cq \lim_{T\to\infty}\frac1{2T}\int_{-T}^{T}
		 e^{\iota t(E-\H)}\psi\,d t=P(\{E\})\psi
	\end{equation}
	is the orthogonal projection of $\psi$ onto the eigenspace of $\H$ corresponding to $E$. The integral is a Hilbert-space integral, and the limit is in the Hilbert norm. In particular, for $z \in Z$,
	\begin{equation}
	\label{eq:zE}
		 \ket z_E \cq \lim_{T\to\infty}\frac1{2T}\int_{-T}^{T}
		 e^{\iota t(E-\H)}\ket z\,d t
		 \in \Qz(Z)^\*
	\end{equation}
	is either zero or an eigenvector of $\H$ corresponding to $E$.
	\item\label{itm:spectral-coherent-totality} The vectors $\ket z_E$ have dense complex linear span in this eigenspace.
\end{enumerate}
\end{theorem}

\begin{proof}
For fixed $E$, the spectral multiplier of the average in~\gzit{eq:Pi} is
\[
	 a_T(\lambda)=
	 \begin{cases}
 \displaystyle\frac{\sin(T(E-\lambda)/\hbar)}{T(E-\lambda)/\hbar},&\lambda\ne E,\\[4pt]
	 1,&\lambda=E.
	 \end{cases}
\]
It satisfies $|a_T(\lambda)|\le1$ and converges pointwise to $\mathbf1_{\{E\}}(\lambda)$. 
Thus, by the spectral theorem and dominated convergence,
\[
	 \left\|\bigl(a_T(\H)-P(\{E\})\bigr)\psi\right\|^2
	 =\int_{\Rz}|a_T(\lambda)-\mathbf1_{\{E\}}(\lambda)|^2
	 \,d\langle\psi,P(\lambda)\psi\rangle \lto 0.
\]
This proves \ref{itm:spectral-mean-projection}. 
For \ref{itm:spectral-coherent-totality}, apply the continuous projection $P(\{E\})$ to the dense coherent span $\Qz(Z)$.
\end{proof}
In particular,
\begin{equation}
\label{eq:deltaH2}
	 \Pi(E)=
	 \begin{cases}
		 \ket E\bra E,&E\text{ is a simple eigenvalue of }\H,\\
		 0,&E\text{ is not in the spectrum of }\H,
	 \end{cases}
\end{equation}
where $\ket E$ is normalized. 
At a continuous-spectrum value without an eigenvalue, the averages~\gzit{eq:zE} therefore vanish.

\subsubsection{The resolvent}
The resolvent of the self-adjoint Hamiltonian $\H$ is the bounded operator
\begin{equation}
\label{eq:resolv}
	 G(E) \cq (E-\H)^{-1}
	 ~~~
	 \text{ for }
	 E\notin\spec \H.
\end{equation}
To express its coherent matrix elements, we define
\begin{equation}
\label{eq:Kt}
	 K_t(z,z') \cq \bra z e^{-\iota t\H}\ket{z'}
\end{equation}
and write
\begin{equation}
\label{eq:KtE}
	 K_t(z,z')=\int_{\Rz}e^{-\iota tE}\,d\mu_{z,z'}(E),
	 ~~~ 
	 \mu_{z,z'}(B) \cq \bra z P(B)\ket{z'}
\end{equation}
for Borel sets $B \seq \Rz$. 
The measure $\mu_{z,z'}$ is a finite complex measure. 
We call its support, meaning the support of its total variation, the classical spectrum of $K_t(z,z')$.

\begin{theorem}
\label{thm:res}
The resolvent satisfies
\begin{equation}
\label{eq:Gbar}
	 G(\ol E)=G(E)^*
	 ~~~
	 \text{ for }
	 E\notin\spec \H.
\end{equation}
For $\im E>0$, it is given by
\begin{equation}
\label{eq:resolv2}
 	\begin{aligned}
 		G(E)\psi&=-\iota\int_0^\infty e^{\iota t(E-\H)}\psi\,d t
		 && \for \psi\in\ol{\Qz(Z)},\\
 		\bra z G(E)\ket{z'}&=\int_{\Rz}(E-E')^{-1}\,d\mu_{z,z'}(E').
 	\end{aligned}
\end{equation}
In particular,
\begin{equation}
\label{eq:GE}
	 G(E)(z,z') \cq \bra z G(E)\ket{z'}
 	=-\iota\int_0^\infty e^{\iota tE}K_t(z,z')\,d t.
\end{equation}
Moreover, if all zeros $E_1,\ldots,E_n$ of a polynomial $A$ are simple and lie outside $\spec \H$, and a polynomial $B$ has smaller degree than $A$, then the bounded operator $r(\H)$, where 
$r \cq B/A$, satisfies
\begin{equation}
\label{eq:ratex}
	 \bra z r(\H)\ket{z'}
	 =-\sum_{j=1}^n\frac{B(E_j)}{A'(E_j)}G(E_j)(z,z').
\end{equation}
\end{theorem}
\begin{proof}
Equation~\gzit{eq:Gbar} follows from self-adjointness. For $\im E>0$,
\[
	 \|e^{\iota t(E-\H)}\psi\|=e^{-t\im E/\hbar}\|\psi\|.
\]
Hence the Hilbert-space integral in~\gzit{eq:resolv2} exists. 
The spectral theorem and
\[
	 -\iota\int_0^\infty e^{\iota t(E-E')}\,d t=(E-E')^{-1}
\]
give~\gzit{eq:resolv2}. Taking coherent matrix elements gives~\gzit{eq:GE}. Finally, the partial fraction decomposition
\[
	 \frac{B(x)}{A(x)}=\sum_{j=1}^n\frac{B(E_j)}{A'(E_j)}(x-E_j)^{-1}
\]
and the bounded functional calculus give~\gzit{eq:ratex}.
\end{proof}
The operator $r(\H)$ is the bounded extension of $A(\H)^{-1}B(\H)$ from $\Dom B(\H)$.

For a bounded operator $T$, put $\im T \cq (T-T^*)/(2i)$. 
We use the notation
\begin{equation}
\label{eq:deltaH}
 	\delta(E-\H) \cq -\frac1\pi\lim_{\eps \downarrow0}
 	\im G(E+i\eps )
\end{equation}
in the distributional sense: After multiplication by a smooth compactly supported function of $E$ and integration, the limit exists strongly. 
It represents the spectral measure, so that, for bounded Borel functions $f$,
\begin{equation}
\label{eq:specM}
	 f(\H)=\int_{\Rz}f(E)\,d P(E)
   	 =\int_{\Rz}f(E)\delta(E-\H)\,d E.
\end{equation}

More explicitly, for every bounded continuous 
$f \cln \Rz\to\Cz$,
\begin{equation}
\label{eq:specMe}
	 \bra z f(\H)\ket{z'}=-\frac1\pi\lim_{\eps \downarrow0}
	 \int_{\Rz}f(E)\bra z\im G(E+i\eps )\ket{z'}\,d E.
\end{equation}
Indeed,
\[
	 -\frac1\pi\bra z\im G(E+i\eps )\ket{z'}
	 =\int_{\Rz}\frac{\eps }{\pi((E-\lambda)^2+\eps ^2)}
	 \,d\mu_{z,z'}(\lambda).
\]
Fubini's theorem applies because $\mu_{z,z'}$ has finite total variation. 
The integral of $f$ against the Poisson kernel is bounded by $\|f\|_\infty$ and converges to $f(\lambda)$, so dominated convergence proves~\gzit{eq:specMe}. 
Notice that the matrix element of the operator imaginary part is
\[
	 \bra z\im G(E+i\eps )\ket{z'}
	 =\frac{G(E+i\eps )(z,z')-\ol{G(E+i\eps )(z',z)}}{2i}.
\]
In particular, for $f(E)=e^{-\iota tE}$,
\begin{equation}
\label{eq:specKt}
	 K_t(z,z')=-\frac1\pi\lim_{\eps \downarrow0}
	 \int_{\Rz}e^{-\iota tE}
	 \bra z\im G(E+i\eps )\ket{z'}\,d E.
\end{equation}
These formulas are a form of spectral recovery from the resolvent; compare Teschl~\cite[Section 4.1]{Teschl2014}.

If $\H$ is bounded, and $f$ is holomorphic on a neighbourhood of the closed region enclosed by a positively oriented simple contour $C$ in the resolvent set containing $\spec \H$ in its interior, then
\begin{equation}
\label{eq:fH2}
	 f(\H)=\frac1{2\pi i}\int_C f(E)G(E)\,d E.
\end{equation}
This follows by applying the scalar Cauchy formula in the spectral representation. 
For unbounded $\H$, formulas~\gzit{eq:specM}--\gzit{eq:specKt} require no finite contour surrounding its spectrum.

The classical spectrum of each $K_t(z,z')$ is contained in 
$\spec \H$. 
Thus, the classical spectrum of each $K_t(z,z')$ gives information about the quantum spectrum, although an individual matrix element may not detect every spectral component. 
Its atoms give eigenvalue information. 
For related numerical methods for the calculation of quantum spectra, see Mandelshtam \& Taylor~\cite{ManT2} and Neumaier \& Mandelshtam~\cite{NeuM}.

\subsubsection{An exactly solvable case}
\label{subsubsec:exactly-solvable-case}
When using the above approach to determine the spectrum of $\H$, the main work is to find $K_t(z,z')$. 
Exact formulas are possible when the quantum evolution is induced by a unitary one-parameter group of coherent maps of $Z$ generated by a classical Hamiltonian $H$.

Thus, suppose that $U_t$ is such a group, that $(t,z) \mto  U_tz$ is smooth, and that its real generator $H$ satisfies
\[
 	\ket z\in\Dom \H,
 	~~~ 
 	\H\ket z
 	=i\hbar\,d\Gamma(H)\ket z
    =i\hbar\,d\kappa(z)H(z)
   	\for 
   	z \in Z.
\]
For $z'\in Z$, put $z(t) \cq U_tz'$. 
Then $\dot z(t)=H(z(t))$, and
\[
	 i\hbar\frac{d}{d t}\ket{z(t)}
	 =i\hbar\,d\kappa(z(t))H(z(t))=\H\ket{z(t)}.
\]
Uniqueness of the strong Schr\"odinger solution therefore gives
\[
	 e^{-\iota t\H}\ket{z'}=\ket{U_tz'},
\]
and hence
\begin{equation}
\label{eq:Ktw}
	 K_t(z,z')=K(z,U_tz')=K(z,z(t)).
\end{equation}
Thus, $K_t$ is computable from the coherent product and the label trajectories. Moreover, the projected coherent vector~\gzit{eq:zE} takes the form
\begin{equation}
\label{eq:zdE}
 	\ket{z'}_E
 	=\lim_{T\to\infty}\frac1{2T}\int_{-T}^{T}
 	e^{\iota tE}\ket{U_tz'}\,d t,
\end{equation}
with convergence in the Hilbert norm. When the product and the trajectories are explicit, these formulas give direct access to the propagation kernel and the components of coherent states in Hilbert eigenspaces.

\subsubsection{The Schwinger--Dyson equation}
\label{subsubsec:schwinger-dyson}
We now assume that $\H$ is the self-adjoint closure of an operator $\H_0\in\Az(G)$ on $\Qz(\wh{Z})$, where $G$ is a smooth coherent $*$-group. To apply the coherent shadow calculus to the resolvent and propagator, let $\Kc_b$ be the space of kernels
\[
	 S_B(z,z') \cq \bra z B\ket{z'},
	 \with 
	 B \text{ bounded on } \ol{\Qz(Z)}.
\]
Coherent totality determines $B$ uniquely. We extend the notation $\Oc(\H_0)$ to this space by
\[
	 \bigl(\Oc(\H_0)S_B\bigr)(z,z') \cq \bra z B\H_0\ket{z'}.
\]
This is a smooth kernel because $\kappa$ and $\H_0\kappa$ are smooth and $B$ is bounded. It need not belong to $\Kc_b$. The definition agrees with the right shadow action of Subsection~\ref{subsec:enveloping-shadow-operators} wherever both are defined. If $e\in\Exp(G)$ represents $\H_0$, boundedness of $B$ permits differentiation of its matrix elements, and the substitution and differentiation rules of Corollary~\ref{cor:formal-shadow-action} give
\[
	 \Oc(\H_0)S_B=\wt{\Oc}(e)S_B.
\]
Consequently, these rules apply independently of the chosen representative.

\begin{theorem}
\label{thm:SD}
\leavevmode
	\begin{enumerate}[label=(\roman*), ref=\textup{(\roman*)}]
		\item\label{itm:shadow-resolvent-equation} For $E\notin\spec \H$, the resolvent kernel $G(E)(z,z')$ is the unique element of $\Kc_b$ satisfying
		\begin{equation}
		\label{eq:Res}
		 	(E-\Oc(\H_0))G(E)=K.
		\end{equation}
		In particular, one may write
		\begin{equation}
		\label{eq:cohRes}
		 	G(E)=(E-\Oc(\H_0))^{-1}K,
		\end{equation}
		where the inverse means the inverse on the range of the injective map
		\[
			E-\Oc(\H_0) \cln \Kc_b\to C^\infty(Z \*  Z,\Cz).
		\]
		\item\label{itm:shadow-propagator-equation} The propagation kernel satisfies
		\begin{equation}\label{eq:SD}
		 	i\hbar\partial_tK_t=\Oc(\H_0)K_t,
		 	~~~ 
		 	K_0=K.
		\end{equation}
	\end{enumerate}
\end{theorem}

\begin{proof}
First, $\Qz(Z)$ is a core for $\H$. 
Indeed, the smooth map $z \mto (\kappa(z),\H_0\kappa(z))$ takes values in the closure of the graph of $\H_0|_{\Qz(Z)}$. 
This closed linear subspace contains its iterated coherent derivatives. 
Their first components span $\Qz(\wh{Z})$, and closability identifies their second components with $\H_0$ applied to the first. Hence, $\H_0$ and $\H_0|_{\Qz(Z)}$ have the same closure.

For \ref{itm:shadow-resolvent-equation}, the definition of the extended shadow action gives
\[
	 \bigl((E-\Oc(\H_0))G(E)\bigr)(z,z')
	 =\bra z G(E)(E-\H)\ket{z'}=K(z,z').
\]
If $(E-\Oc(\H_0))S_B=0$ with $B$ bounded, coherent totality gives
$B(E-\H)\psi=0$ for $\psi\in\Qz(Z)$. 
The core property extends this identity to $\Dom \H$. 
Since $E-\H$ maps this domain onto $\ol{\Qz(Z)}$, it follows that $B=0$. 
This proves injectivity and the uniqueness assertion, and gives the stated meaning of~\gzit{eq:cohRes}.

For \ref{itm:shadow-propagator-equation}, coherent vectors belong to $\Dom \H$. Therefore,
\[
	 i\hbar\partial_tK_t(z,z')
	 =\bra z e^{-\iota t\H}\H\ket{z'}
	 =\bigl(\Oc(\H_0)K_t\bigr)(z,z').
\]
The initial condition follows at $t=0$.
\end{proof}
We call~\gzit{eq:SD} the Schwinger--Dyson equation in the present coherent setting. 
It is the coherent matrix-element form of the Schr\"odinger equation.

\subsubsection{The time-dependent Schr\"odinger equation}
\label{subsubsec:time-dependent-schrodinger}
The dynamics of a conservative physical system subject to time-dependent external forces is traditionally given by a family of self-adjoint Hamiltonians $\H(t)$ on $\ol{\Qz(Z)}$, each bounded from below. 
The time-dependent Schr\"odinger equation
\begin{equation}\label{eq:tdSchr}
	 i\hbar\partial_t\psi(t)=\H(t)\psi(t)
\end{equation}
is understood as an equality in $\ol{\Qz(Z)}$, with 
$\psi(t)\in\Dom \H(t)$. 
A strong solution is a continuously differentiable Hilbert-space curve satisfying this equation.

In the time-independent case, $\psi(t)=e^{-\iota tE}\psi_0$ is a solution if and only if $\H\psi_0=E\psi_0$. 
For time-dependent Hamiltonians, a single spectral representation no longer describes the evolution. 
The treatment of the exactly solvable case by trajectories on $Z$, however, extends to the time-dependent situation.

Let $H_t$ be a smooth time-dependent real coherent vector field on $Z$. 
The corresponding compatibility with the quantum Hamiltonian is
\begin{equation}
\label{eq:dissip}
	 \Qz(\wh{Z}) \seq \Dom \H(t),~~~
	 \H(t)|_{\Qz(\wh{Z})}=i\hbar\,d\Gamma(H_t).
\end{equation}
The following result reduces the evolution of coherent states to a differential equation on $Z$; for Glauber coherent states, see also Mehta et al.~\cite{MehCSV}.

\begin{theorem}
\label{thm:hamIVP}

Let $I$ be a closed interval containing $0$, and suppose that \gzit{eq:dissip} holds for every $t\in I$. 
Assume that the equation $\dot z(t)=H_t(z(t))$ has a unique solution on all of $I$ for every initial point and initial time, with jointly smooth transition maps $U(t,s)z$.
Then, for every $z_0 \in Z$, the strong solution of
\begin{equation}
\label{eq:coherent-schrodinger-ivp}
	 i\hbar\partial_t\psi_t=\H(t)\psi_t,
	 ~~~
	 \psi_0=\ket{z_0},
\end{equation}
is unique and has the form $\psi_t=\ket{z(t)}$ for all $t\in I$, where
\[
 	\dot z(t)=H_t(z(t)),
 	~~~ 
 	z(0)=z_0.
\]
\end{theorem}

\begin{proof}
By \gzit{eq:dissip} and Theorem \ref{thm:intrinsic-representation}, $H(t)\kappa(z)=i\hbar\,d\kappa(z)H_t(z)$. 
Since $H(t)$ is symmetric on coherent vectors, we obtain
\[
	0
	=\langle H(t)\kappa(z),\kappa(w)\rangle
	  -\langle\kappa(z),H(t)\kappa(w)\rangle
	=-i\hbar\,(L_{H_t}K+R_{H_t}K)(z,w).
\]
Thus, $L_{H_t}K+R_{H_t}K=0$.
For uniqueness, it suffices by linearity to prove
\begin{equation}
\label{eq:uniSol}
 	i\hbar\partial_t\eta_t=\H(t)\eta_t,
 	~~~
 	\eta_0=0
	~~~ \Rightarrow ~~~
	\eta_t=0
	\for 
	t \in I.
\end{equation}
Self-adjointness gives
\[
	 \frac{d}{d t}\|\eta_t\|^2
	=2\re \left\langle\eta_t,-\frac{i}{\hbar}\H(t)\eta_t\right\rangle=0.
\]
Thus, the norm is constant, and its initial value is zero. 
This proves uniqueness.

For existence, put $z(t)=U(t,0)z_0$ and $\psi_t=\kappa(z(t))$. Smoothness of the evolution and of $\kappa$ gives a smooth Hilbert-space curve. 
By~\gzit{eq:dissip} and Theorem~\ref{thm:intrinsic-representation},
\[
	 i\hbar\partial_t\psi_t
	 =i\hbar\,d\kappa(z(t))H_t(z(t))
	 =i\hbar\,d\Gamma(H_t)\ket{z(t)}
	 =\H(t)\psi_t.
\]
Its initial value is $\ket{z_0}$, completing the proof.
\end{proof}

By Proposition~\ref{prop:quantized-unitary-evolution}, the quantizations of $U(t,s)$ form a strongly continuous unitary evolution $V(t,s)$ preserving $\Qz(\wh{Z})$. Hence, for $\psi_0\in\Qz(\wh{Z})$, the curve $\psi_t=V(t,0)\psi_0$ is the unique strong solution of
\begin{equation}
\label{eq:general-schrodinger-ivp}
	 i\hbar\partial_t\psi_t=\H(t)\psi_t,~~~\psi(0)=\psi_0.
\end{equation}
For arbitrary $\psi_0\in\ol{\Qz(Z)}$, it defines a continuous norm-preserving evolution, obtained by approximation with finite linear combinations of coherent vectors. 
Strong differentiability requires the corresponding domain conditions.

If an initial vector admits a Bochner-integrable representation
\[
	 \psi_0=\int_Z\ket z\,d\mu(z),
\]
then boundedness of $V(t,0)$ gives
\[
	 \psi_t=\int_Z\ket{U(t,0)z}\,d\mu(z).
\]
For example, this applies if $z \mto \ket z$ is strongly measurable and$\int_Z\sqrt{K(z,z)}\,d|\mu|(z)<\infty$.

\subsection{Variational principles}
\label{subsec:variational-principles}

\subsubsection{The coherent action principle}
\label{subsubsec:coherent-action-principle}

Variational principles play an important role in modern physics, and they are fruitfully applied in classical and quantum mechanics. In this subsection, we first give an overview of Lagrangian systems on smooth manifolds, where we restrict our considerations to smooth solutions of the Euler--Lagrange equations. 

\pgph


Let $I=[a,b]$, where $a<b$, and let $M$ be a smooth manifold. A Lagrangian on $TM$ is a smooth function 
$L \cln I \*  TM \to \Cz$. 
In local coordinates, $D_L$ and $D_R$ denote differentiation with respect to position and velocity, with $t$ fixed. 
The Lagrangian $1$-form $\theta_t^L$ on $TM$ is
\[
	 (\theta_t^L)_{(p,v)}(\xi)
	  \cq D_RL(t,p,v)\bigl(d\pi(p,v)\xi\bigr)
	 \for 
	 p\in M,\ v\in T_pM, ~ \xi\in T_{(p,v)}TM,
\]
where $\pi \cln TM \to M$ is the bundle projection. 
The Lagrangian $2$-form is
\[
	 \omega_t^L \cq -d\theta_t^L,
\]
where the exterior derivative is taken with $t$ fixed.

The action functional $S$ is defined on smooth paths 
$c \cln I \to M$ by
\[
	 S[c] \cq \int_I L(t,c(t),\dot c(t))\,d t.
\]
A smooth path $c \cln I \to M$ is stationary if
\[
	 \left.\frac{d}{d s}S[c_s]\right|_{s=0}=0
\]
for every smooth variation $c_s$ with fixed endpoints. 
Integration by parts in local coordinates gives the Euler--Lagrange equations
\begin{equation}
\label{eq:Euler-Lagrange}
 	\frac{d}{d t}D_RL(t,c(t),\dot c(t))
 	=D_LL(t,c(t),\dot c(t)).
\end{equation}

Let $Z$ be a coherent manifold, and let $\hbar>0$. The coherent action associated with a smooth time-dependent Hamiltonian $H \cln I \*  Z \to \Cz$, $(t,z) \mto  H_t(z)$, is
\begin{equation}
\label{eq:coherent-action}
	 S_H[z] \cq \int_I L_H(t,z(t),\dot z(t))\,d t
\end{equation}
for every smooth path $z:I\to Z$, with the coherent Lagrangian
\begin{equation}
\label{eq:coherent-Lagrangian}
	 L_H(t,z,v) \cq -\hbar\theta_z(v)-H_t(z)
 	=i\hbar D_RK(z,z).v-H_t(z)
\end{equation}
for $z\in Z$ and $v \in T_zZ$. Applying~\gzit{eq:Euler-Lagrange} to~\gzit{eq:coherent-Lagrangian} gives the coherent Euler--Lagrange equations
\[
 	\hbar\omega_{z(t)}(\dot z(t),v)=d H_t(z(t)).v
 	\for 
	 v \in T_{z(t)}Z.
\]
Furthermore, the Lagrangian $1$-form and Lagrangian $2$-form for the coherent $1$-form $\theta$ and the coherent $2$-form $\omega$ read
\[
	 \begin{aligned}
	 \theta^{L_H}_{(z,v)}(\xi)
	 &=i\hbar D_RK(z,z)\bigl(d\pi(z,v)\xi\bigr),\\
	 \omega^{L_H}_{(z,v)}(\xi,\eta)
	 &=2\hbar\operatorname{Im}D_LD_RK(z,z)
	     \bigl(d\pi(z,v)\xi,d\pi(z,v)\eta\bigr),
	 \end{aligned}
\]
for $z \in Z$, $v \in T_zZ$, and $\xi,\eta \in T_{(z,v)}TZ$. 

\begin{theorem}
\label{thm:coherent-action-stationarity}
Let $H \in C^\infty(I\* Z,\Cz)$ and $\hbar>0$. 
A smooth curve $z \cln I \to Z$ is stationary for the coherent action $S_H$ \Iff
\begin{equation}
\label{eq:stationary-euler-lagrange}
	 \hbar\omega_{z(t)}(\dot z(t),u)=d H_t(z(t)).u
	 ~~~
	 \text{ for every }
	 u \in T_{z(t)}Z.
\end{equation}
\end{theorem}
\begin{proof}
Let $z_s$ be a smooth variation of $z \cln I \to Z$ with fixed endpoints, and put
\[
 	\eta(t) \cq \left.\frac{\partial}{\partial s}z_s(t)\right|_{s=0}.
\]
Since $\omega=d\theta$,
\[
	 \left.\frac{\partial}{\partial s}
	 \bigl[\theta_{z_s(t)}(\partial_tz_s(t))\bigr]\right|_{s=0}
	 =\omega_{z(t)}(\eta(t),\dot z(t))
 		 +\frac{d}{d t}\bigl[\theta_{z(t)}(\eta(t))\bigr].
\]
The boundary term vanishes because $\eta(a)=\eta(b)=0$. 
Thus,
\[
	 \begin{aligned}
		 \left.\frac{d}{d s}S_H[z_s]\right|_{s=0}
		 &=\int_I\bigl[-\hbar\omega_{z(t)}(\eta(t),\dot z(t))
		 	-d H_t(z(t)).\eta(t)\bigr]\,d t\\
		 &=\int_I\bigl[\hbar\omega_{z(t)}(\dot z(t),\eta(t))
	     	-d H_t(z(t)).\eta(t)\bigr]\,d t.
	 \end{aligned}
\]
If~\gzit{eq:stationary-euler-lagrange} holds, the first variation vanishes. Conversely, suppose that $z$ is stationary. 
In a chart around $z(t_0)$, with $a<t_0<b$, choose variations whose coordinate variation is $\rho(t)v$, where $v$ is an arbitrary model-space vector and $\rho$ is a smooth real function supported in a sufficiently small interval around $t_0$. 
These variations agree with $z$ outside that interval and are defined for sufficiently small $s$. 
Applying the fundamental lemma of the calculus of variations to the real and imaginary parts gives
$\hbar\omega_{z(t)}(\dot z(t),u)-d H_t(z(t)).u=0$ for every tangent vector $u$ at every interior time. 
The identity extends to the endpoints by continuity.
\end{proof}

\begin{corollary}
\label{cor:coherent-hamiltonian-curves}
Let $H \in C^\infty(I \*  Z,\Cz)$ and $\hbar>0$. 
If $H_t \in \Ez(\omega)$ and $X_{H_t}$ are associated Hamiltonian vector fields, stationarity is equivalent to
\[
	 \dot z(t)-\hbar^{-1}X_{H_t}(z(t))\in\ker_{\Cz}\omega_{z(t)}.
\]
If the chosen $X_{H_t}$ are real and depend smoothly on $t$, every integral curve of $\hbar^{-1}X_{H_t}$ on $I$ is stationary.
\end{corollary}

\begin{proof}
By~\gzit{eq:dfhom}, $d H_t=X_{H_t}\omega$. Hence, stationarity is equivalent to
$\omega_{z(t)}\bigl(\dot z(t)-\hbar^{-1}X_{H_t}(z(t)),u\bigr)=0$
for every real tangent vector $u$. Complex multilinearity gives the stated kernel condition.
\end{proof}

For complex-valued $H$, stationarity also requires 
$d(\im H_t)(z(t))=0$, since $\omega$ is real.

\subsubsection{The Dirac--Frenkel variational principle}
\label{subsubsec:dirac-frenkel}

Let $\H(t)$ be a family of self-adjoint Hamiltonians on $\ol{\Qz(Z)}$. We consider smooth paths $\psi(t) \in \Dom \H(t)$ and fixed-endpoint variations $\psi_s(t)$ for which
\[
 (s,t) \mto \bigl(\psi_s(t),\H(t)\psi_s(t)\bigr)
\]
is smooth as a map into $\ol{\Qz(Z)} \* \ol{\Qz(Z)}$. The quantum action is
\begin{equation}
\label{eq:schrAction}
	 S[\psi] \cq \int_I L(t,\psi(t),\dot\psi(t))\,d t,
	 \qquad
	 L(t,\psi,\dot\psi) \cq \psi^*(i\hbar\dot\psi-\H(t)\psi).
\end{equation}
Writing $\eta=\left.\partial_s\psi_s\right|_{s=0}$ and $r=i\hbar\dot\psi-\H(t)\psi$, integration by parts and self-adjointness give
\[
 	\left.\frac{d}{d s}S[\psi_s]\right|_{s=0}
 	=2\operatorname{Re}\int_I\langle\eta(t),r(t)\rangle\,d t.
\]
Here $\eta(t)\in\Dom \H(t)$ and $\left.\partial_s(\H(t)\psi_s(t))\right|_{s=0}=\H(t)\eta(t)$ by closedness of $\H(t)$. In particular, every solution of the Schr\"odinger equation in this class is stationary. The action remains complex-valued; its imaginary part is the endpoint term
\[
	 \operatorname{Im}S[\psi]
 	=\frac{\hbar}{2}\bigl(\|\psi(b)\|^2-\|\psi(a)\|^2\bigr).
\]
To find approximate solutions of the Schr\"odinger equation, \sca{Dirac}~\cite{Dir.var} and \sca{Frenkel}~\cite{Fre} proposed variational equations for a parametrized family of states. 
The resulting Dirac--Frenkel variational principle plays an important role in approximation schemes for the dynamics of quantum systems. 
Its identification with stationarity of the restricted quantum action requires a condition on the tangent spaces, stated below.

In many cases, a viable approximation is obtained by restricting the state vectors $\psi(t)$ to a family of easily manageable states $\ket z$ smoothly parametrized by parameters $z$ in a manifold $Z$, which can often be given a physical meaning. 
Here ``easily manageable'' means that the variational equations can be written down explicitly without integrals. 
In particular, the inner products $K(z,z') \cq \bra z\ket{z'}$ should be explicitly computable and smooth, making $Z$ a coherent manifold. 
Conversely, given a coherent manifold, one can use its coherent states as the $\ket z$.

Thus, we assume now that $Z$ is a coherent manifold, that 
$\kappa(z) \in \Dom \H(t)$ for all $(t,z) \in I \*  Z$, and that
\[
 	(t,z)\mto \H(t)\kappa(z)
\]
is smooth with values in $\ol{\Qz(Z)}$. We restrict $\psi(t)$ to the coherent states $\ket{z(t)}$ and obtain variational equations on $Z$. When these equations admit a solution, the resulting states provide an evolution within the chosen family of coherent states.

Inserting the ansatz $\psi(t)=\ket{z(t)}$ into \gzit{eq:schrAction} gives the restricted quantum action
\[
	S_{\mathrm{res}}[z] \cq \int_I
 	\bra{z(t)}\bigl(i\hbar\partial_t-\H(t)\bigr)\ket{z(t)}\,d t.
\]
On a general real trial manifold, stationarity of this action is the Kramer--Saraceno variational principle. 
A geometric treatment is given in \sca{Kramer \& Saraceno}~\cite{KraS}; its relation to Dirac--Frenkel is discussed in \sca{Lasser \& Su}~\cite[Sections 1.1 and 3]{LS2022}. 
For applications in quantum chemistry, see \sca{Lubich}~\cite{Lubich}. An important application is time-dependent Hartree--Fock, obtained by restricting to Slater determinants. 
Their rays are parametrized by a Grassmann manifold,
the manifold of all $k$-dimensional subspaces of a vector space. 
The corresponding group-theoretic coherent states are discussed in \sca{Zhang} et al.~\cite{ZhaFG}. 
Thus, a state-vector parametrization also retains the phase. 
Enlarging the trial family can improve the approximation, but its accuracy depends on the system and the chosen family.

\begin{proposition}
\label{prop:coherent-restricted-action}
Under the above assumptions, the coherent action is the restriction of the quantum action to coherent states, with
\[
	 H(t,z) \cq \bra z\H(t)\ket z.
\]
For a smooth path $z \cln I \to Z$, put
\[
	 r_z(t) \cq i\hbar\,d\kappa(z(t))\dot z(t)-\H(t)\kappa(z(t)).
\]
The path is stationary \Iff
\[
	 \operatorname{Re}\langle d\kappa(z(t))u,r_z(t)\rangle=0
	 \qquad\text{for every }u\in T_{z(t)}Z.
\]
If $d\kappa(z)T_zZ$ is invariant under multiplication by $i$ for every $z \in Z$, this is equivalent to the Dirac--Frenkel condition
\[
	 \langle d\kappa(z(t))u,r_z(t)\rangle=0
	 \qquad\text{for every }u\in T_{z(t)}Z.
\]
\end{proposition}
\begin{proof}
The function $H$ is smooth by the assumed smoothness of $\kappa$ and $\H(t)\kappa(z)$, and is real-valued by self-adjointness. For $\psi(t) \cq \ket{z(t)}$, we have
\[
	 \begin{aligned}
		 L_H(t,z(t),\dot z(t))
	 	 &=i\hbar D_RK(z(t),z(t)).\dot z(t)-\bra{z(t)}\H(t)\ket{z(t)}
		 =\bra{z(t)}(i\hbar\partial_t-\H(t))\ket{z(t)}\\
		 &=\psi(t)^*(i\hbar\dot\psi(t)-\H(t)\psi(t)),
	 \end{aligned}
\]
identifying $S_H$ with $S_{\mathrm{res}}$. 
Every smooth variation of $z$ gives an admissible variation of $\psi$ by the assumed smoothness of $(t,z) \mto  \H(t)\kappa(z)$. 
The first-variation formula and the local variation argument in Theorem~\ref{thm:coherent-action-stationarity} therefore give the real-part condition. 
If the tangent image is invariant under multiplication by $i$, applying that condition also to $i\,d\kappa(z(t))u$ makes the imaginary part vanish. 
This gives the Dirac--Frenkel condition, whose converse is immediate.
\end{proof}

The resolvent equation~\gzit{eq:Res} and the Schwinger--Dyson equation~\gzit{eq:SD} need not admit closed-form solutions. 
However, we may use these equations to determine the parameters in an approximate closed-form ansatz for $K_t(z,z')$ or $G(E)(z,z')$, with kernels in the space $\Kc_b$ of Subsubsection~\ref{subsubsec:schwinger-dyson}. 
The propagator ansatz is required to be differentiable in $t$. 
In a collocation approach, the ansatz is substituted into~\gzit{eq:SD} or~\gzit{eq:Res}. 
The residual is evaluated at sufficiently many points, and the parameters are chosen to minimize a weighted sum or integral of its squared modulus.

\subsection{Bosonic quantization of Klauder spaces}
\label{subsec:klauder}

Let $\Hz$ be a complex Hermitian space, with Hermitian form
$\beta \cln \Hz \* \Hz \to \Cz$. 
Thus, $p^*q \cq \beta(p,q)$ is antilinear in $p$ and linear in $q$.

We equip $\Hz$ with a Hausdorff locally convex topology for which it
is convenient and $\beta$ is bounded on products of bounded sets.
Thus, $\beta$ is smooth by 
\sca{Kriegl \& Michor} \cite[Lemma~5.5]{KM}. 

The topology need not be induced by $\|v\|_\beta=(v^*v)^{1/2}$.

We write $L_{\Cz}(\Hz)$ for the space of bounded complex-linear 
maps $\Hz \to \Hz$ and put
\[
	\Lc_\beta(\Hz) \cq 
	\{C \in L_{\Cz}(\Hz) \mid \text{ there is } C^*\in L_{\Cz}(\Hz)
	\text{ with }
	(Cp)^*q=p^*C^*q,\ p,q \in \Hz\}.
\]
The adjoint $C^*$ is unique. 
The usual sum and product rules turn $\Lc_\beta(\Hz)$ into a unital 
complex $*$-algebra. 
Its elements need not be bounded for $\|\cdot\|_\beta$. 

\pgph


We apply the constructions of Section~\ref{sec:coherent-quantization}
to the \bfi{Klauder space} 
\[
	\Kl(\Hz) \cq \Cz \*  \Hz,
\] 
with points $z \cq [z_0,\z]$, $z' \cq [z'_0,\z']$ and coherent 
product
\begin{equation}
\label{eq:klauder-kernel}
	K(z,z') \cq \exp(\ol z_0+z'_0+\z^*\z').
\end{equation}
This coherent space and its oscillator transformations are 
described by 
\sca{Neumaier \& Ghaani Farashahi} 
\cite[Sections~6.1--6.2]{NGF2022}. We put their results into the present context and give new proofs for some of them.

\pgph


We regard $\Kl(\Hz)$ as a real convenient manifold with its given 
complex structure.
The coherent product is smooth, so the preceding results apply. 

From now on, we set
\[
	Z \cq \Kl(\Hz)
\]
if not stated explicitly otherwise.

\subsubsection{The Klauder kernel and oscillator transformations}

For $\z \in \Hz$, we abbreviate $\ket*{[0,\z]}$ by $\ket*{\z}$. 
Thus,
\begin{equation}
\label{eq:klauder-reduced-vectors}
	\ket*{[z_0,\z]}=e^{z_0}\ket*{\z},
	~~~
	\braket*{\z}{\z'}=e^{\z^*\z'}.
\end{equation}
In particular, the vectors $\ket*{\z}$, $\z \in \Hz$, span $\Qz(Z)$.

\begin{proposition}
\label{prop:klauder-kernel}
The coherent-state map
$\kappa \cln \Kl(\Hz) \to \ol{\Qz(\Kl(\Hz))}$ is holomorphic and 
tangent-injective.
\end{proposition}

\begin{proof}
Proposition~\ref{prop:smooth-kappa} gives smoothness. 
For $z=[z_0,\z] \in \Kl(\Hz)$, $(u_0,u) \in T_z \Kl(\Hz)$, and 
$\z'\in \Hz$,
differentiation of \gzit{eq:klauder-reduced-vectors} gives
\begin{equation}
\label{eq:klauder-tangent-pairing}
	\bra*{\z'}d\kappa(z)(u_0,u)
	=e^{z_0+(\z')^*\z}(u_0+(\z')^*u).
\end{equation}
The right-hand side is complex-linear in $(u_0,u)$.
Since the reduced coherent vectors are total in 
$\ol{\Qz(\Kl(\Hz))}$,
$d\kappa(z)$ is complex-linear. 
Hence, $\kappa$ is holomorphic by 
\sca{Kriegl \& Michor} \cite[Theorem~7.19(8)]{KM}.
If $d\kappa(z)(u_0,u)=0$, then \gzit{eq:klauder-tangent-pairing}
with $\z'=0$ gives $u_0=0$. 
Taking $\z'=u$ then gives $u^*u=0$, and positive definiteness gives 
$u=0$.
\end{proof}

We write $\Osc(\Hz)$ for the set of maps $[\rho,p,q,A]$ with 
$\rho \in \Cz$, $p,q \in \Hz$, and $A \in \Lc_\beta(\Hz)$ defined by
\begin{equation}
\label{eq:klauder-semigroup-action}
	[\rho,p,q,A][z_0,\z] \cq [z_0+\rho+p^*\z,A\z+q]
	\for 
	[z_0,\z] \in \Kl(\Hz).
\end{equation}

\begin{proposition}[%
	\sca{Neumaier \& Ghaani Farashahi}
	{\cite[Section~6.2, Propositions~6.3 and~6.4(i)]{NGF2022}}%
]\label{prop:klauder-semigroup-laws}
$\Osc(\Hz)$ is an involutive semigroup of coherent maps, with
identity $[0,0,0,1]$ and
\begin{align*}
	[\rho,p,q,A][\eta,p',q',B]
	&=[\rho+\eta+p^*q',\,p'+B^*p,\,q+Aq',\,AB],
\\
	[\rho,p,q,A]^*&=[\ol\rho,q,p,A^*]
	\for 
	B \in \Lc_{\beta}(\Hz).
\end{align*}
Its group of units is the \bfi{linear oscillator group}
$\operatorname{LOsc}(\Hz) \cq \{[\rho,p,q,A] \mid A \in
\Lc_\beta(\Hz)^\* \}$, 
with
\begin{equation}
\label{eq:klauder-semigroup-inverse}
	[\rho,p,q,A]^{-1}
	=[p^*A^{-1}q-\rho,\,-(A^{-1})^*p,\,-A^{-1}q,\,A^{-1}].
\end{equation}
The unitary elements form the subgroup
\begin{equation}
\label{eq:klauder-unitary-group}
	\begin{split}
		\operatorname{UOsc(\Hz)}
		& \cq \{A_{\alpha,q,U} \mid \alpha \in \Rz,\ q \in \Hz,\ 
		U \in \Lc_{\beta}(\Hz), \
		U^*U=UU^*=1\}, \with \\
		A_{\alpha,q,U}& \cq [i\alpha-q^*q/2,\,-U^*q,\,q,\,U].
	\end{split}
\end{equation}
\end{proposition}

Note that when $\Hz$ is a Hilbert space with its norm topology,
$ \Lc_\beta(\Hz)= \Lc_{\Cz}(\Hz)$. 
The parameter space $\Cz \* \Hz^* \* \Hz \* \GL(\Hz)$,
with the operator norm on the last factor, then turns
$\operatorname{LOsc}(\Hz)$ into a complex Banach--Lie group, where $\Hz^*$ is the complex-conjugate space of $\Hz$.

\subsubsection{Coherent vector fields and the common domain}

Vector fields on $\Kl(\Hz)$ are smooth sections of its real tangent 
bundle.
Since $T_z \Kl(\Hz)\cong(\Cz \* \Hz)_{\Rz}$, the complex structure is
\[
	J_z[\eta_0,\bm{\eta}]=[i\eta_0,i\bm{\eta}]
	\for
	[\eta_0,\bm{\eta}] \in T_z \Kl(\Hz).
\]
We use $iX \cq JX$ and
$\lambda X \cq (\re \lambda)X+(\im \lambda)JX$ for pointwise 
complex multiplication.
Thus, $\lambda X$ is again a real tangent vector field.

The \bfi{oscillator Lie algebra} $\osc(\Hz)$ is the set of
block matrices
\[
X_{\gamma,p,q,A} \cq 
	\begin{pmatrix}0&p^*&\gamma\\0&A&q\\0&0&0\end{pmatrix}
	\with
	\gamma \in \Cz, ~ p,q \in \Hz, ~ A \in \Lc_\beta(\Hz),
\]
with Lie product
\[
    X_{\gamma,p,q,A} \lp X_{\gamma',p',q',A'}
    \cq -[X_{\gamma,p,q,A},X_{\gamma',p',q',A'}]
    \for 
    X_{\gamma,p,q,A},X_{\gamma',p',q',A'} \in \osc(\Hz),
\]
where we use \gzit{eq:brackets}.

The affine oscillator action and its matrix bracket are given by \sca{Neumaier \& Ghaani Farashahi} 
\cite[(51)--(52) and (66)]{NGF2022}. 
The following proposition characterizes the smooth coherent vector fields on $\Kl(\Hz)$.

\begin{proposition}[%
	cf. \sca{Neumaier \& Ghaani Farashahi}
	{\cite[Proposition~6.3 and (66)]{NGF2022}}%
	]\label{prop:klauder-fields}
Let $X$ be a smooth real vector field on $Kl(\Hz)$.
Then $X$ is a coherent vector field \Iff it has the form
\begin{equation}
\label{eq:klauder-affine-fields}
	X=X_{\gamma,p,q,A},~~~
	X_{\gamma,p,q,A}[z_0,\z]=[\gamma+p^*\z,A\z+q],
\end{equation}
with $\gamma \in \Cz$, $p,q \in \Hz$, and $A \in \Lc_\beta(\Hz)$.
Its coefficient map is holomorphic. 
Its parameters and coherent adjoints are unique, and, for 
$\lambda \in \Cz$,
\begin{equation}
\label{eq:klauder-field-star}
	X_{\gamma,p,q,A}^*=X_{\ol\gamma,q,p,A^*},
	~~~
	\lambda X_{\gamma,p,q,A}
	=X_{\lambda\gamma,\ol\lambda p,\lambda q,\lambda A}.
\end{equation}
For $Y \cq X_{\delta,p',q',D}$,
\begin{equation}
\label{eq:klauder-affine-bracket}
	X_{\gamma,p,q,A} \lp X_{\delta,p',q',D}
	=X_{p^*q'-p'^*q,\,D^*p-A^*p',\,Aq'-Dq,\,AD-DA}.
\end{equation}
With pointwise complex multiplication, these fields realize the 
complex Lie $*$-algebra $\osc(\Hz)$.
In particular, $(\lambda X)^*=\ol\lambda X^*$.
\end{proposition}

\begin{proof}
Let $Y$ be a coherent adjoint of $X$.
Since $T_zZ \cong (\Cz\*\Hz)_{\Rz}$, we write
$X(z) \cq [f(z),F(z)]$ and $Y(z) \cq [g(z),G(z)]$,
where $f,g \cln Z \to \Cz$ and $F,G \cln Z \to \Hz$ are smooth and
$z \cq [z_0,\z], z' \cq [z'_0,\z'] \in \Kl(\Hz)$.
Differentiating \gzit{eq:klauder-kernel} and dividing by $K(z,z')$
shows that \gzit{eq:vector-fields-coherent-adjoint} is equivalent to
$\ol{f(z)}+F(z)^*\z'=g(z')+\z^*G(z')$.
We put $o \cq [0,0]$, $\gamma \cq f(o)$, $p \cq G(o)$, and 
$q \cq F(o)$.
Taking $z=z'=o$ gives $g(o)=\ol\gamma$. 
Taking first $z'=o$ and then $z=o$ gives
$f(z)=\gamma+p^*\z$ and $g(z')=\ol\gamma+q^*\z'$.
Hence, $(F(z)-q)^*\z'=\z^*(G(z')-p)$.
The right-hand side is independent of $z_0$. 
Since $\z'$ is arbitrary, nondegeneracy of $\beta$ shows that $F(z)$ is independent of $z_0$. 
Similarly, $G(z')$ is independent of $z'_0$.
Thus, putting $Au\cq F([0,u])-q$ and $Bv\cq G([0,v])-p$, we have
$(Au)^*v=u^*Bv$ for $u,v \in \Hz$.
The right-hand side is antilinear in $u$ and the left-hand side is
linear in $v$. 
By nondegeneracy, $A$ and $B$ are complex-linear.
Both are smooth by restriction, hence bounded. 
Thus, $A \in \Lc_\beta(\Hz)$ and $B=A^*$.
This gives \gzit{eq:klauder-affine-fields} and
$Y=X_{\ol\gamma,q,p,A^*}$. 

Conversely, for any parameters as in
\gzit{eq:klauder-affine-fields}, these two fields satisfy \gzit{eq:vector-fields-coherent-adjoint} by substitution. 
The value $X(o)$ determines $\gamma$ and $q$, and
$X([0,u])-X(o)=[p^*u,Au]$ for $u \in \Hz$ determines $p$ and $A$. 
Thus, the parameters and the coherent adjoint are unique. 
The coefficient map is holomorphic since it is affine with bounded complex-linear part. 
Since $(\ol\lambda p)^*u=\lambda p^*u$, pointwise multiplication gives the scalar rule in \gzit{eq:klauder-field-star}, and hence
$(\lambda X)^*=\ol\lambda X^*$.

Finally, for $X=X_{\gamma,p,q,A}$ and $Y=X_{\delta,p',q',D}$,
$ dX(z).Y(z)=[p^*(D\z+q'),A(D\z+q')]$.
Subtracting the same expression with $X$ and $Y$ interchanged
gives $X\lp Y=-[X,Y]$ in \gzit{eq:klauder-affine-bracket}.
\end{proof}

\begin{proposition}
\label{prop:klauder-half}
\leavevmode

\begin{enumerate}[ref=\textup{(\roman*)}]
	\item\label{itm:klauder-unitary-generators} Write $Y_{\tau,q,A} \cq X_{i\tau,-q,q,A}$ and 
	$u(\Hz) \cq \{A \in \Lc_\beta(\Hz) \mid A^*=-A\}$.
	Then
	\begin{equation}
	\label{eq:klauder-half-equality}
		\half(\operatorname{UOsc(\Hz)})\seq \uosc(\Hz)
		\cq \{Y_{\tau,q,A} \mid \tau \in \Rz,\ q \in \Hz,\ A \in 
		u(\Hz)\},
	\end{equation}
	and $\uosc(\Hz)$ is a real form of $\osc(\Hz)$: Pointwise 
	multiplication induces the isomorphism
	\begin{equation}
	\label{eq:klauder-complexification}
		\uosc(\Hz) \otimes_{\Rz} \Cz \to \osc(\Hz),
		~~~ 
		Y \otimes \lambda \mto \lambda Y.
	\end{equation}
	\item\label{itm:klauder-unitary-generators-equality} Suppose that every 
	$A \in u(\Hz)$ admits unitary elements 
	$U_A(t) \in \Lc_\beta(\Hz)$ for $|t|<\eps_A$, with
	\[
		U_A(0)=1,
		~~~
		\left.\frac{d}{dt}\right|_{t=0}U_A(t)v=Av
		\for 
		v \in \Hz,
	\]
	and jointly smooth evaluations
	$(t,v) \mto U_A(t)v$ and $(t,v) \mto U_A(t)^{-1}v$.
	Then
	\[
		\half(\operatorname{UOsc(\Hz)})=\uosc(\Hz).
	\]
\end{enumerate}
\end{proposition}

\begin{proof}
\ref{itm:klauder-unitary-generators} Proposition~\ref{prop:klauder-fields} gives the displayed
anti-Hermitian fields. They form a real Lie algebra, and
\[
	X=\re X +i \im X
	\with 
	\re X \cq \frac{X-X^*}{2},
	~
	\im X \cq \frac{X+X^*}{2i}
\]
is the unique decomposition into an anti-Hermitian field plus 
$i$
times an anti-Hermitian field. This proves the real-form 
assertion.
A motion in $\operatorname{UOsc(\Hz)}$ preserves $K$. 
Differentiation at zero gives $L_XK+R_XK=0$, so $X^*=-X$ and this proves the inclusion.

\ref{itm:klauder-unitary-generators-equality} Put $h_t=A_{t\alpha,tq,U_A(t)}$. 
Proposition~\ref{prop:klauder-semigroup-laws} gives smooth forward and inverse evaluations, 
$h_0=\id_Z$,
and
\[
	\left.\frac{d}{dt}\right|_{t=0}h_t[z_0,v]
	=[i\alpha-q^*v,Av+q]=Y_{\alpha,q,A}[z_0,v].
\]
Example~\hyperref[itm:motions-smooth-families]{\ref*{ex:motions}\ref*{itm:motions-smooth-families}} makes the restriction to nonnegative
times a motion. 
This proves the reverse inclusion.
\end{proof}

We put 
\begin{equation}
\label{eq:PvQq}
	P_v \cq X_{0,0,v,0},
	~~~
	Q_q \cq X_{0,q,0,0}, 
	~~~
	L_A \cq X_{0,0,0,A},
	~~~
	S_\gamma \cq X_{\gamma,0,0,0}
\end{equation}
for $\gamma \in \Cz$, $v,q \in \Hz$, and $A \in \Lc_{\beta}(\Hz)$.
For $\z,v_1,\ldots,v_r \in \Hz$, we write
\begin{equation}
\label{eq:klauder-jet-definition}
	\ket*{\z;P_{v_1},\ldots,P_{v_r}}
	\cq \ket*{[0,\z];P_{v_1},\ldots,P_{v_r}}
	=\left.\partial_{t_1}\cdots\partial_{t_r}
	\ket*{\z+t_1v_1+\cdots+t_rv_r}\right|_{t=0}.
\end{equation}
Here $\ket*{\z;\emptyset} \cq \ket*{\z}$, and the derivatives are
taken in $\ol{\Qz(Z)}$. 
Holomorphy of $\kappa$ makes 
\gzit{eq:klauder-jet-definition} 
complex-multilinear and symmetric in the constant fields $P_{v_j}$.
We claim that
\begin{equation}
\label{eq:klauder-domain-equality}
	\Qz(\wh{Z})
	=\Span_{\Cz}\{\ket*{\z;P_{v_1},\ldots,P_{v_r}} \mid r\geq0,\ 
	\z,v_j\in \Hz\}.
\end{equation}
Each $P_v=X_{0,0,v,0}$ is coherent by
Proposition~\ref{prop:klauder-fields}. 
Hence, every vector on the right-hand side of \gzit{eq:klauder-domain-equality} belongs to $\Qz(\wh Z)$.

Conversely, consider an iterated Lie derivative of $\kappa$ at
$[z_0,\z]$. 
By the chain and product rules, it is a finite sum of
ordinary derivatives of $\kappa$ at that point, evaluated on
tangent vectors. 
Indeed, at each differentiation one differentiates
either $\kappa$ once more or one of the vector-valued arguments
of a derivative already present.

By multilinearity, we may split each tangent vector as
$[a,v]=[a,0]+[0,v]$. 
Since $\kappa([z_0,\z])=e^{z_0}\ket*{\z}$, each differentiation in a
direction $[a,0]$ contributes a factor $a$. 
The remaining directions $[0,v_1],\ldots,[0,v_r]$ give
$e^{z_0}\ket*{\z;P_{v_1},\ldots,P_{v_r}}$
by \gzit{eq:klauder-jet-definition}, including $r=0$.
Thus, every generator of $\Qz(\wh Z)$ belongs to the displayed
span. 
This proves \gzit{eq:klauder-domain-equality}.
Differentiating \gzit{eq:klauder-reduced-vectors} also gives
\begin{equation}
\label{eq:klauder-jet-pairing}
	\braket*{\z'}{\z;P_{v_1},\ldots,P_{v_r}}
	=e^{(\z')^*\z}\prod_{j=1}^r (\z')^*v_j
	\for 
	\z' \in \Hz.
\end{equation}

Theorem~\ref{thm:finite-quantization} gives, for every
$S=[\rho,p,q,A]\in\Osc(\Hz)$,
\begin{equation}\label{eq:klauder-finite-coherent-action}
	\Gamma(S)\ket*{\z}=e^{\rho+p^*\z}\ket*{A\z+q}.
\end{equation}
If $S\in\operatorname{LOsc}(\Hz)$, both $S$ and $S^*$ are smooth
coherent diffeomorphisms. 
Theorem~\ref{thm:common-domain-finite-quantization}
therefore supplies $\Gamma(S)$ on $\Qz(\wh{Z})$, with its
multiplication and formal-adjoint identities, and, 
applied to \gzit{eq:klauder-finite-coherent-action}, gives
\begin{equation}
\label{eq:klauder-finite-jet-action}
	\Gamma(S)\ket*{\z;P_{v_1},\ldots,P_{v_r}}
	=e^{\rho+p^*\z}
	\sum_{J\seq [r]}
	\Big(\prod_{j\notin J}p^*v_j\Big)
	\ket*{A\z+q;(P_{Av_j})_{j\in J}},
\end{equation}
where $[r] \cq \{1,\ldots,r\}$. 
For $S \in \operatorname{UOsc(\Hz)}$, we identify
$\Gamma(S)$ with its unique unitary extension to $\ol{\Qz(Z)}$,
supplied by Corollary~\ref{cor:unitary}.

\subsubsection{Creation and annihilation operators}
\label{subsubsec:klauder-ladders}

Proposition~\ref{prop:klauder-fields} gives
$Q_q^*=P_q$ and $L_B^*=L_{B^*}$. 
Hence, Theorem~\ref{thm:intrinsic-representation} supplies the
\bfi{annihilation operator} and \bfi{creation operator}
\begin{equation}\label{eq:klauder-ladder-definition}
	q^*\a \cq d\Gamma(Q_q) 
	\AND
	\a^*q \cq d\Gamma(P_q).
\end{equation}
We also put $N(A) \cq d\Gamma(L_A)$ for $A \in \Lc_\beta(\Hz)$. 
Here $\a$ and $\a^*$ are symbols; the expressions in 
\gzit{eq:klauder-ladder-definition} denote the operators. 
The \bfi{vacuum vector} is $\ket*{0} \cq \ket*{[0,0]}$.

\begin{proposition}[%
	cf. \sca{Neumaier \& Ghaani Farashahi} 
	{\cite[pp.~44--45]{NGF2022}} %
]\label{prop:klauder-infinitesimal}
Let $p,q,\z,u_1,\ldots,u_r \in \Hz$ and 
$B,C \in \Lc_\beta(\Hz)$, and put $\P=(P_{u_1},\ldots,P_{u_r})$. 
Then
\begin{align}
	(\a^*q)\ket*{\z;\P}
	&=\ket*{\z;\P,P_q},
\label{eq:klauder-creation-jets} \\
	(q^*\a)\ket*{\z;\P}
	&=q^*\z\ket*{\z;\P}
	+\sum_{j=1}^r q^*u_j
	\ket*{\z;P_{u_1},\ldots,\wh{P_{u_j}},\ldots,P_{u_r}},
\label{eq:klauder-annihilation-jets} \\
	N(B)\ket*{\z;\P}
	&=\ket*{\z;\P,P_{B\z}}
	+\sum_{j=1}^r
	\ket*{\z;P_{u_1},\ldots,P_{Bu_j},\ldots,P_{u_r}},
\label{eq:klauder-linear-jets}
\end{align}
where the hat $\;\wh{\phantom{t}}\;$ denotes omission. 
Consequently,
\begin{equation}
\label{eq:klauder-affine-quantization}
	d\Gamma(X_{\gamma,p,q,C})
	=\gamma I+p^*\a+\a^*q+N(C).
\end{equation}
The maps $q \mto \a^*q$ and $B \mto  N(B)$ are complex-linear,
and $q \mto  q^*\a$ is conjugate-linear. 
On $\Qz(\wh{Z})$, we have the \bfi{canonical commutation relations} 
\begin{equation}
\label{eq:klauder-ccr}
	p^*\a \lp \a^*q=p^*q,
	~~~
	p^*\a \lp q^*\a=\a^*p \lp \a^*q=0.
\end{equation}
Moreover,
\begin{align}
	N(B) \lp \a^*q&=\a^*(Bq),&
	N(B) \lp q^*\a&=-(B^*q)^*\a,
\label{eq:klauder-linear-commutators} \\
	N(B) \lp N(C)&=N(BC-CB),&
	(q^*\a)^*&=\a^*q,
	~~~
	N(B)^*=N(B^*).
\label{eq:klauder-formal-adjoints}
\end{align}
All these operators preserve $\Qz(\wh{Z})$ and are closable in
$\ol{\Qz(Z)}$. 
Every displayed product is a composition on $\Qz(\wh{Z})$. 
The vacuum state satisfies
\[
	\braket*{0}{0}=1
	\AND
	(q^*\a)\ket*{0}=N(B)\ket*{0}=0.
\]
\end{proposition}

\begin{proof}
Differentiating \gzit{eq:klauder-reduced-vectors} along
$X=X_{\gamma,p,q,C}$ gives
$d\Gamma(X)\ket*{\z}=(\gamma+p^*\z)\ket*{\z}+\ket*{\z;P_{C\z+q}}$. 
In particular, $(q^*\a)\ket*{\z}=(q^*\z)\ket*{\z}$, $N(B)\ket*{\z}=\ket*{\z;P_{B\z}}$, and
$d\Gamma(S_\gamma)\ket*{\z}=\gamma\ket*{\z}$. 
Theorem~\ref{thm:intrinsic-representation} gives 
\gzit{eq:klauder-creation-jets} directly. 
For \gzit{eq:klauder-annihilation-jets}, the product rule 
differentiates either the coherent vector or the affine factor 
$q^*\z$; the latter contributes $q^*u_j$ and has no second 
derivatives. 
For \gzit{eq:klauder-linear-jets}, a derivative either 
differentiates the coherent vector or replaces $B\z$ by $Bu_j$; a 
second derivative of $B\z$ vanishes.
Since $\kappa$ is holomorphic, its derivatives are complex-linear. 
Therefore, 
\gzit{eq:klauder-creation-jets}--\gzit{eq:klauder-linear-jets} are 
linear.
Theorem \ref{thm:intrinsic-representation} gives
\gzit{eq:klauder-affine-quantization}, first on coherent vectors
and then on $\Qz(\wh{Z})$ by injectivity of $\Sh$ in 
Proposition~\ref{prop:shadow-algebra}. 
The same argument gives $d\Gamma(S_\gamma)=\gamma$.
By Proposition \ref{prop:klauder-fields},
$Q_p \lp P_q=S_{p^*q}$ and $Q_p \lp Q_q=P_p \lp P_q=0$
as well as $L_B \lp P_q=P_{Bq}$, $L_B \lp Q_q=-Q_{B^*q}$, and
$L_B \lp L_C=L_{BC-CB}$. 
The commutator identity of
Theorem~\ref{thm:intrinsic-representation} proves all the
commutation relations. 
Its adjoint identity, together with $Q_q^*=P_q$ and $L_B^*=L_{B^*}$, proves the formal-adjoint
identities. 
The same theorem supplies invariance and closability.
Finally, \gzit{eq:klauder-reduced-vectors} gives the vacuum 
norm,
and the coherent-vector formulas at $\z=0$ give the vacuum 
identities.
\end{proof}

The right-hand side of the mixed commutation relation in
\sca{Neumaier \& Ghaani Farashahi}
\cite[Theorem~6.6(iv), (65)]{NGF2022} should read $p^*q$.
Indeed, replacing $p,q$ by $sp,tq$ with $s,t \in \Rz$ in the Weyl
relation of the same theorem gives the scalar factor $e^{stp^*q}$.
Comparison of the coefficients of $st$ gives
$[p^*\a,\a^*q]=p^*q$, as in \gzit{eq:klauder-ccr}.

\subsubsection{Normal ordering}
\label{subsec:klauder-normal-ordering}

By \sca{Neumaier \& Ghaani Farashahi} 
\cite[Theorem~6.6(i--ii)]{NGF2022}, every function $F \cln \Hz \* 
\Hz \to \Cz$ determines a unique \bfi{normal ordering}
$\NO(F) \cln \Qz(Z) \to \Qz(Z)^\*$ with
$\bra*{\z}\NO(F)\ket*{\z'}=e^{\z^*\z'}F(\z,\z')$.
We use the  notation
\[
	\no{F(\a^*,\a)} \cq \NO(F).
\]
Conversely, every linear map $T \cln \Qz(Kl(\Hz)) \to 
\Qz(Kl(\Hz))^\*$ has the 
unique normal symbol
\begin{equation}
\label{eq:klauder-normal-symbol}
	F_T(\z,\z')=e^{-\z^*\z'}\bra*{\z}T\ket*{\z'}.
\end{equation}
The map $F \mto \NO(F)$ is complex-linear,
$\NO(1)=\inj$, with $\inj$ here viewed as the canonical 
embedding $\Qz(Z)\to\Qz(Z)^ \* $, and
$\NO(F)^*=\NO(F^*)$, where
$F^*(\z,\z') \cq \ol{F(\z',\z)}$ and the adjoint is understood
in the sense of sesquilinear forms. 

\pgph


For arbitrary $F$, $\NO(F)$ is only a map into the antidual.
To obtain an operator on $\Qz(\wh{Z})$, let $T \in \Az(\LOsc(\Hz))$
and define $F_T$ by \gzit{eq:klauder-normal-symbol}.
Since $\ket*{[z_0,\z]}=e^{z_0}\ket*{\z}$,
\begin{equation}
\label{eq:klauder-shadow-normalization}
	(\Sh T)([z_0,\z],[z'_0,\z'])
	=K([z_0,\z],[z'_0,\z'])F_T(\z,\z').
\end{equation}
The coherent product never vanishes. 
Thus, Proposition~\ref{prop:shadow-algebra} makes $T \mto  F_T$
injective. We say that $F$ is \bfi{realized in $\Az(\LOsc(\Hz))$}
if $F=F_T$ for some $T \in \Az(\LOsc(\Hz))$. 
The realizing operator is unique, and in this case we also write
\[
	\no{F(\a^*,\a)}
	= T
	~~~
	\text{ on }
	\Qz(\wh{Z}).
\]
Its restriction to $\Qz(Z)$, viewed in the antidual, is the 
normal ordering. 
By Proposition~\ref{prop:expolynomial-representation},
it preserves $\Qz(\wh{Z})$ and is closable in 
$\ol{\Qz(Z)}$.

\begin{proposition}
\label{prop:klauder-normal-polynomials}
Let $p_1,\ldots,p_n,q_1,\ldots,q_m\in \Hz$, with $m,n \geq 0$, and 
put
\[
	F(\z,\z') \cq \prod_{j=1}^m\z^*q_j
	\prod_{k=1}^n p_k^*\z'.
\]
Then $F$ is realized in $\Az(\LOsc(\Hz))$, and
\begin{equation}
\label{eq:klauder-normal-monomial}
	\no{F(\a^*,\a)}
	=(\a^*q_1)\cdots(\a^*q_m)(p_1^*\a)\cdots(p_n^*\a)
	~~~
	\text{ on }
	\Qz(\wh{Z}).
\end{equation}
Empty products denote $1$ for functions and $I$ for operators.
For $B \in \Lc_\beta(\Hz)$, the reduced normal symbol of
$N(B)$ is $\z^*B\z'$. 
Thus, the notation
\begin{equation}
\label{eq:klauder-quadratic-order}
	\a^*B\a \cq N(B),
	~~~
	\no{\a^*B\a} = N(B)
\end{equation}
is consistent with normal ordering on the common domain. 
If $B=\sum_{j=1}^m q_jp_j^*$, then
\[
	N(B)=\sum_{j=1}^m(\a^*q_j)(p_j^*\a)
	~~~
	\text{ on }
	\Qz(\wh{Z}).
\]
\end{proposition}

\begin{proof}
First we have
$(p^*\a)\ket*{\z'}=(p^*\z')\ket*{\z'}$ and 
$\bra*{\z}(\a^*q)=(\z^*q)\bra*{\z}$.
The second equality is an identity of functionals on
$\Qz(\wh{Z})$. 
Applying these identities successively shows that
the product on the right of \gzit{eq:klauder-normal-monomial}
has matrix element $e^{\z^*\z'}F(\z,\z')$.
It belongs to $\Az(\LOsc(\Hz))$, so it realizes $F$.

By \gzit{eq:klauder-linear-jets} with $r=0$ and
\gzit{eq:klauder-jet-pairing},
$
\bra*{\z}N(B)\ket*{\z'}
=\braket*{\z}{\z';P_{B\z'}}
=e^{\z^*\z'}\z^*B\z'
$.
This proves \gzit{eq:klauder-quadratic-order}.
For $B=\sum_jq_jp_j^*$, the stated finite sum has the same
normal symbol as $N(B)$. 
Injectivity of $T \mto  F_T$ proves their equality on the common 
domain.
\end{proof}

For general $B\in\mathcal L_\beta(\Hz)$,
\gzit{eq:klauder-quadratic-order} denotes the already constructed
operator $N(B)$ and requires no infinite operator sum.

For $p,q\in \Hz$, we define the \bfi{complex Weyl operator} on $\Qz(\wh{Z})$ by
\[
	W(p,q) \cq \Gamma([-p^*q/2,-ip,iq,1]).
\]
Since $[-p^*q/2,-ip,iq,1]$ belongs to 
$\operatorname{LOsc}(\Hz)$, we have 
$W(p,q) \in \Az(\LOsc(\Hz))$. 
If $p=q$, it belongs to $\operatorname{UOsc(\Hz)}$ by 
Proposition~\ref{prop:klauder-semigroup-laws}. 
Corollary~\ref{cor:unitary} then gives a unitary extension of
$W(q,q)$ to $\ol{\Qz(\Kl(\Hz))}$. 

\begin{proposition}
\label{prop:klauder-normal-exponential}
Let $S=[\rho,p,q,A] \in \operatorname{LOsc}(\Hz)$. 
Then
\begin{equation}
\label{eq:klauder-oscillator-normal-symbol}
	F_{\Gamma(S)}(\z,\z')
	=\exp\bigl(\rho+p^*\z'+\z^*q+\z^*(A-1)\z'\bigr).
\end{equation}
Consequently,
\begin{equation}
\label{eq:klauder-normal-oscillator}
	\Gamma(S)
	=\no{\exp\bigl(\rho+p^*\a+\a^*q+\a^*(A-1)\a\bigr)}
	~~~
	\text{ on }
	\Qz(\wh{Z}).
\end{equation}
In particular,
\begin{equation}
\label{eq:klauder-normal-weyl}
	W(p,q)=e^{-p^*q/2} \! \no{e^{i(p^*\a+\a^*q)}}
	~~~
	\text{ on }
	\Qz(\wh{Z}).
\end{equation}
\end{proposition}

\begin{proof}
By \gzit{eq:klauder-finite-coherent-action},
$\bra*{\z}\Gamma(S)\ket*{\z'}
=\exp\bigl(\rho+p^*\z'+\z^*(A\z'+q)\bigr)$.
Division by $e^{\z^*\z'}$ proves
\gzit{eq:klauder-oscillator-normal-symbol}. 
Since $\Gamma(S) \in \Az(\LOsc(\Hz))$, uniqueness of realization gives \gzit{eq:klauder-normal-oscillator}. 
Substitution of $S=[-p^*q/2,-ip,iq,1]$ proves 
\gzit{eq:klauder-normal-weyl}.
\end{proof}

This is the common-domain form of the normal-ordering formula
given in \sca{Neumaier \& Ghaani Farashahi} 
\cite[Theorem 6.6(iii)]{NGF2022}. 
The same calculation gives a weak identity for $\Gamma(S)$ on
$\Qz(Z)$ whenever $S \in \Osc(\Hz)$. 
The colon expression specifies the operator through its normal symbol; it does not require a power-series definition of an operator exponential.

Injectivity defines a product on realized symbols by 
$F_T\star F_U \cq F_{TU}$ for $T,U\in\Az(\LOsc(\Hz))$.
For example, the canonical commutation relations give
\begin{equation}
\label{eq:klauder-normal-contraction}
	F_{(p^*\a)(\a^*q)}(\z,\z')
	=(p^*\z')(\z^*q)+p^*q.
\end{equation}
Indeed, $(p^*\a)(\a^*q)=(\a^*q)(p^*\a)+p^*q$,
and the first term is normally ordered by
\gzit{eq:klauder-normal-monomial}. 

\pgph


We define the \bfi{Segal fields} on $\Qz(\wh{Z})$ by
\[
	\Phi_s(q) \cq q^*\a+\a^*q.
\]
We have $\Phi_s(q)^*=\Phi_s(q)$ on $\Qz(\wh{Z})$. 
We put $\phi_q(\z,\z') \cq q^*\z'+\z^*q$ and define
\[
	\no{\Phi_s(p)\Phi_s(q)} \cq\ \no{(\phi_p\phi_q)(\a^*,\a)}.
\]
The symbol $\phi_p\phi_q$ is realized by a linear combination
of the normal monomials in 
Proposition~\ref{prop:klauder-normal-polynomials}.
Thus, on $\Qz(\wh{Z})$,
\begin{equation}
\label{eq:klauder-wick-two-fields}
	\no{\Phi_s(p)\Phi_s(q)}
	=\Phi_s(p)\Phi_s(q)-p^*q,
	~~~
	\bra*{0}\Phi_s(p)\Phi_s(q)\ket*{0}=p^*q.
\end{equation}
To verify these identities, expand the product into its four
creation--annihilation terms. 
Only $(p^*\a)(\a^*q)$ requires a commutation to put the creator on 
the left, producing $p^*q$.
Each remaining quadratic term has zero vacuum expectation:
An annihilator on the right removes $\ket*{0}$, and a creator on
the left kills $\bra*{0}$ by the formal-adjoint identity.


\PrintProjectIndices

\printbibliography[heading=bibintoc]

\end{document}